%% file: main.tex
\documentclass[preprint,12pt,authoryear]{elsarticle}
\usepackage[margin=1in]{geometry}
\usepackage{setspace}
\usepackage{geometry}
\usepackage{amsmath}
\usepackage{amssymb}
\usepackage{graphicx} 
\usepackage{longtable}
\usepackage{url}
\usepackage{array}
\usepackage{subcaption}
\usepackage{multirow}
\usepackage{tikz}
\usepackage[colorlinks=false,linkcolor=blue,citecolor=blue]{hyperref}
\usepackage{natbib}
\usepackage{mathrsfs}
\usepackage{color}
\usepackage{amsfonts}
\usepackage{bbm}
\usepackage{stackengine}
\usepackage{lineno}
\usepackage{setspace}
\usepackage{algorithm}
\usepackage{algpseudocode}
\usepackage{kantlipsum}
\usepackage{mwe}
\usepackage{diagbox}
\usepackage{color, colortbl}
\usepackage{placeins}
\usepackage{xcolor} 
\usepackage{amsthm}
\usepackage{appendix}

\newcolumntype{C}[1]{>{\centering\arraybackslash}p{#1}}

\DeclareMathAlphabet{\mathcal}{OMS}{cmsy}{m}{n}
\usetikzlibrary{shapes,arrows}

\newcommand{\subtitle}[1]{%
  \posttitle{%
    \par\end{center}
    \begin{center}\large#1\end{center}
    \vskip0.5em}%
}

\journal{Transport Policy}

\begin{document}

\begin{frontmatter}

\title{The impacts of remote work on travel: insights from nearly three years of monthly surveys}

\author[mitcee]{Nicholas S. Caros}
\ead{caros@mit.edu}
\author[mitcee]{Xiaotong Guo}
\ead{xtguo@mit.edu}
\author[mitcee]{Yunhan Zheng}
\ead{yunhan@mit.edu}
\author[mitdusp]{Jinhua Zhao\corref{cor}}
\ead{jinhua@mit.edu}

\address[mitcee]{Department of Civil and Environmental Engineering, Massachusetts Institute of Technology, Cambridge, MA 02139, USA}
\address[mitdusp]{Department of Urban Studies and Planning, Massachusetts Institute of Technology, Cambridge, MA 02139, USA}

\begin{abstract}
 
% \linenumbers

Remote work has expanded dramatically since 2020, upending longstanding travel patterns and behavior. 
More fundamentally, the flexibility for remote workers to choose when and where to work has created much stronger connections between travel behavior and organizational behavior. 
This paper uses a large and comprehensive monthly longitudinal survey over nearly three years to identify new trends in work location choice, mode choice and departure time of remote workers. 
The travel behavior of remote workers is found to be highly associated with employer characteristics, task characteristics, employer remote work policies, coordination between colleagues and attitudes towards remote work. 
Approximately one third of all remote work hours are shown to take place outside of the home, accounting for over one third of all commuting trips. 
These commutes to ``third places'' are shorter, less likely to occur during peak periods, and more likely to use sustainable travel modes than commutes to an employer's primary workplace. 
Hybrid work arrangements are also associated with a greater number of non-work trips than fully remote and fully in-person arrangements.
Implications of this research for policy makers, shared mobility provides and land use planning are discussed.

\end{abstract} 


\end{frontmatter}

% \linenumbers

\section{Introduction}

% Background - Rise in remote work: magnitude, timing
In 2020, a slow and steady rise in remote working over decades was dramatically and irreversibly accelerated by the onset of the COVID-19 pandemic.
In the United States, employers are planning for about 32\% of all worked hours to take place remotely in the long term, a nearly seven-fold increase in remote work shares relative to 2018 \citep{barrero2021working}.
The average employee would prefer even higher levels of remote work than their employers is planning, suggesting that there is room for remote work to grow if business practices, technology and infrastructure evolve to ease existing remote work constraints. 

% Motivation - What has changed / is uncertain with regards to travel patterns? 
This sudden rise in remote work has resulted in the biggest shock to urban travel patterns in generations. 
The primary effect of increased remote work is that many commuting trips can now be replaced by working at home or close to home.
The substitution of some commuting trips for remote work also has narrow secondary effects on travel behavior, including changes to mode choice, departure time choice, home location choice and destinations for non-work travel. 
New survey data is needed to inform future travel demand forecasting models that accurately reflect these primary and secondary effects of widespread remote work.
Yet there has also been a more fundamental change to how travel decisions are made. 
Remote work offers the freedom to choose a work location, whether that is at home, at the employer's workplace, or somewhere else entirely. 
Moreover, it forces employees and employers to consider coordination with colleagues in their decisions about where and when to work remotely. 
The outcomes of these decisions have implications for travel demand, but also for productivity, personal well-being and local retail spending. 
In the past, when the vast majority of the workforce commuted to the same location on a regular schedule, the connections between travel behavior and employment attributes could be largely ignored as employees did not have the agency to act on their preferences.
In the remote work era, however, these connections can no longer be neglected.
It is now essential to understand how employment attributes and inter-organizational coordination impact travel choices for remote workers. 

% Introduce the concept of the paper and how it resolves existing issues 
To quantify emerging trends in remote work travel behavior, including connections between travel and employment, this paper first introduces an important source of remote work data: the monthly U.S. Survey of Working Arrangements and Attitudes (SWAA) \citep{barrero2021working}. 
Initially designed to understand experiences and attitudes towards remote work from early 2020 onward, the SWAA was updated in 2021 to include questions about the travel behavior of remote workers. 
This paper presents the results of each travel behavior-related question and explores how choices differ based on demographic status, lifestyle choices, geography and critically, job type, employer attributes and employer policies.
Policy-relevant insights for remote workers, employers, transportation services and urban planners are highlighted.  
Lastly, new open-source tools for conducting further analysis of the SWAA data are presented. 

% Contribution - Areas of novelty
% 1: Breadth - comprehensive data on all aspects of travel, demographics and behavior
% 4: Entirely new areas of insight that are critical! Third places, economics
% 2: Detail - disaggregate home and work location data 
% 3: Duration - monthly surveys since onset of remote work
The primary contribution of this paper is to explore the factors that contribute to the choice of work location for remote work, and the secondary effects of work location choices on mode choice, travel time and departure time. 
For the first time in literature, these choices are disaggregated by employment-related factors, thus demonstrating how travel behavior has become highly correlated with employer decisions, job types and attitudes towards work and collaboration. 
While previous surveys have investigated changes in travel behavior since the rapid rise of remote work in 2020, the SWAA is unique in its scope, duration and level of detail. 
The analyses presented in this paper are based on comprehensive monthly surveys of 5,000 or more respondents spanning nearly three years. 
Over 400 unique questions have been presented to respondents across the dozens of survey waves, enabling identification of several unexpected trends and travel behavior patterns with relevance for urban transportation and land use policy.

\section{Literature Review}

% Start with pre-COVID surveys
Transportation scholars have been interested in the topic of remote work (often referred to as ``teleworking'') since well before the boom occurred in 2020.
Much of the effort in modeling flexible work decisions has focused on the frequency and duration of flexible work, rather than the location and the choice to co-locate with others \citep{Handy_Mokhtarian_1996b, mokhtarian1998synthetic}.
This is partly due to implicit assumptions that flexible workers are making a binary choice: work at an office or work from home. 
\citet{bagley1997analyzing} and \citet{stanek1998developing} conducted surveys of workers in California to elicit preferences for working from home and from a flexible work center.
\citet{Mokhtarian_Salomon_1997} and \citet{Vana_Bhat_Mokhtarian_2008} found that attitudes towards work, family and commuting are more important than socio-demographic factors in determining preferences towards flexible work.
\citet{Pouri_Bhat_2003} includes several occupational factors in a flexible work choice model, finding that part time workers and employees of private companies are more likely to choose flexible work, while those requiring daily face-to-face interactions are less likely to choose flexible work. 
\citet{Sener_Bhat_2011} also included work characteristics in estimating a copula-based sample selection model using household travel survey data from Chicago. 
Even recent comprehensive frameworks that include duration of flexible work do not consider location choice or the impact of employer policies \citep{asgari2014choice, Asgari_Jin_2015, paleti2016generalized}. 
In a very interesting study, \citet{stiles2021working} reviews the travel patterns of remote workers from 2003 to 2017 based on their choice of work location using data from the American Time Use Survey. 
Leading up to 2020, remote work remained a niche working arrangement in the United States, restricted to specific industries and occupations.
As a result, surveys were limited to small panels and often focused on employees within a single firm or employment sector. % TODO: add some citations here from the above

% Then early COVID surveys: Salon et al, others
Several new travel and remote work surveys were issued in 2020 and early 2021 to capture new behavioral patterns resulting from pandemic-related restrictions on mobility and social activities.
This was a period of very high remote work (over 60\% of worked hours in the United States in May 2020) and considerable uncertainty about working arrangements in the medium and long term. 
\citet{beck2020insights} and \citet{echaniz2021behavioural}  used surveys during the early days of the pandemic to identify significant shifts in travel behavior due to mobility restrictions in Australia and Spain, respectively.
\citet{dianat2022assessing} surveyed 1,000 travelers in the Toronto, Canada area about changes to their activity scheduling and mode choice during and immediately after the lifting of pandemic-related restrictions. 
\citet{currie2021evidence} and \citet{jain2022covid} use a survey distributed in Melbourne, Australia during the summer of 2020 to estimate the total impact of remote work on future travel demand and identify contributing factors, respectively.
They find that access to remote work technology and employer support were significant factors in determining the likelihood of continuing remote work, and that attitudes were not a significant driver of long-term remote work preferences. 
\citet{balbontin2021impact} compared differences across countries with respect to remote work preferences in late 2020, also finding that employer support has a strong positive effect. 
In a widely cited paper, \citet{salon2021potential} explored potential changes to a wide range of travel behavior among remote workers, including shopping, restaurant patronage, air travel and home relocation. 
The number of published studies investigating remote work trends in this period  is substantial and continues to grow. 

% Then recent COVID surveys: Mahmassani, etc.
After restrictions were lifted and perceived public health threat subsided, working arrangements slowly began to stabilize.
New survey instruments were introduced to gain an understanding of the future of remote work and travel behavior. 
\citet{nayak2021potential} uses logistic regression to estimate preferences for remote work among a small sample of Indian commuters in March 2021. 
The authors several household characteristics, including poor internet connectivity and distractions caused by other household members are predictive of preferences for less remote work. 
\citet{asmussen2022modeling} moves beyond the home-office paradigm in a stated preference survey of hypothetical work location choices for Texas residents in early 2022. 
The study finds that workplace environment is at least as important as geographic location when choosing from working at home, at the employer's business premises or a ``third place'' (e.g. caf\'e, library or community center). 
Using a longitudinal survey from December 2020 to March 2022, \citet{tahlyan2022longitudinal} tracks changes in attitudes towards remote work, shopping and travel for the same group of respondents over time.

% Then economics-focused surveys, including Barrero and any census/ATUS
Remote work was not only a topic of interest to transportation scholars. 
Many social scientists, including the SWAA founders, conducted surveys to explore the economic and psychological impacts of widespread remote work.
\citet{barrero2021working} is the original working paper exploring the responses to non-travel questions on the SWAA, finding significant differences in preferences for remote work across demographic and employment categories.
Other large online surveys investigating remote work trends were conducted by \citet{bick2020work} and \citet{brynjolfsson2020covid} around the same period with similar results.  
Using a small sample of U.S. residents, \citet{tahlyan2022whom} find that middle-aged workers experienced greater remote work satisfaction than young and older workers. 
Similarly, \citet{shi2020factors} explores the factors that contributed to greater productivity while working at home among employees in the Seattle, United States area.
These surveys are not intended to elicit travel preferences and therefore do not provide a substantive travel behavior component. 
To the authors' knowledge, this is the first detailed analysis of survey results that connect remote work location choice, travel behavior and employment characteristics during the widespread remote work era.

\section{Survey methodology} \label{sec:methods}

% Note the availability of the data at wfhresearch.com
The insights in this paper are generated from the U.S. Survey of Working Arrangements and Attitudes (SWAA).
It is a comprehensive monthly survey designed and administered by the WFH Research team. 
The entire catalog of past survey questions and the cumulative response data are available online at \url{https://wfhresearch.com}. 

% Describe our specific contribution: adding and revising transport questions
The authors of this paper began to collaborate with the WFH Research team in 2021 to add over 20 travel-related questions to the survey during different survey waves.
The travel-related questions cover many different aspects of travel behavior: destination choice, mode choice, travel time, departure time, non-work trips and more. 
The responses to the travel-related questions are the basis of the insights presented in Section~\ref{sec:findings}. 
The travel-related questions are cross-tabulated with the comprehensive set of existing SWAA questions on demographics, geography, household characteristics, work characteristics and attitudes to provide new insights into the factors that impact travel behavior during the remote work era. 

% Waves, sample sizes
The first wave of the survey was distributed in May 2020, shortly after the onset of the COVID-19 pandemic in the United States.
The survey was continued in July 2020 and has been distributed monthly ever since. 
All survey waves are restricted to U.S. residents over the age of 19. 
Minor changes to the sample size and sampling methodology have been made since the survey began. 
The first travel-related questions were added in November 2021.
At that time, the monthly sample was 5,000 respondents and restricted to people who had earned at least \$10,000 USD in 2019.
In early 2022, the sample was increased to 10,000 respondents per month who had earned at least \$10,000 USD in the previous full calendar year.

% Distribution method
The SWAA is a panel survey distributed online by commercial survey providers who recruit respondents through a variety of sources. 
Respondents are not recruited for the SWAA specifically, rather they are recruited for online surveys and then provided with a link to the SWAA questionnaire. 
Each questionnaire includes approximately 40 to 60 questions, with a typical response time of 8 to 10 minutes.
No identifying information is provided and the survey administrators do not interact directly with the respondents. 

% Scaling and attention checks
Attention check questions are used to filter out a small number of low-quality responses from each survey wave.
Then, individual survey responses are then re-weighted to match the United States Current Population Survey (CPS) shares with respect to age, sex, education and income. 
The re-weighted sample is also very similar to the CPS shares for census division (region of the country) and industry. 
The sample is not weighted by race, therefore race is not used as an independent variable for the analyses in this paper.
All of the results in this paper are drawn from the re-weighted sample. 
The aggregate results of the SWAA have been shown to be consistent with the results of other similar remote work surveys \citep{wfhresearch_bench}. 

% Summarize the results of the steps above (valid samples for each month?)
After survey administration, data cleaning and scaling, the result is a high-quality, representative sample of thousands of working-age U.S. residents per month across three years and a total sample size of more than 148,000 responses. 
Questions have been added and removed over time, with over 400 different questions included in at least one survey wave. 
This rich dataset can then be used to generate new insights that connect remote work, travel behavior and organizational behavior.
The findings are presenting in the next section and the implications are discussion in Section~\ref{sec:discussion}. 

\section{Survey findings} \label{sec:findings}

% Introduce the overall section here
This section presents the travel-related findings from the SWAA survey, cross-tabulated with other variables of interest. 
The wide-ranging findings are organized into five subsections:
\begin{enumerate}
    \item Work location choice
    \item Mode choice
    \item Departure time
    \item Productivity and travel
\end{enumerate}
Each subsection is further divided among different topics of interest, such as non-work trips, trip duration, and so on.
The abundance of data and questions collected over three years make it impossible to fit an exhaustive analysis within the constraints of a single journal paper. 
We have therefore provided only a selection of the most interesting and policy-relevant findings herein.
To encourage further analysis by interested readers, we have developed tools for data cleaning, organization and visualization and shared them in a public GitHub repository: \url{https://github.com/jtl-transit/swaa}. 
These tools can also be used to replicate the results and charts below using the public survey data. 

% Establish a basic set of independent variables that will be used in remaining sections
The variation in responses to travel-related questions across standard demographic, household, employment and attitudinal groups are analysed below.
The standard demographic variables are age, gender, income and education. 
The standard household characteristic variables are number of children, internet quality, home office availability, and population density of the home ZIP code.
The standard work characteristics include industry, occupation, company size, population density of the work ZIP code, percentage of tasks requiring a computer, employment type, and percentage of tasks that can be done remotely.
Standard attitudinal variables include attitudes towards commuting, socializing at work, efficiency at home and risk of infection on certain travel modes.
All independent variables are categorical unless otherwise noted. 

% Discussion about the timing of the COVID-19 lockdowns, etc.
The travel-related questions were added to the SWAA questionnaire in November 2021, after COVID-19 vaccines were widely available in the United States and most pandemic-related restrictions on social gatherings and public activities were lifted. 
As context for the results in this section, it is worth noting that the Omicron variant of COVID-19 emerged in November 2021 leading to dramatic increase in the number of reported COVID-19 cases in the United States from December 2021 to March 2022.
Survey responses from that period may therefore reflect greater concerns about the public health risks of in-person work and shared transportation than responses from subsequent survey waves. 
Attitudes towards managing the risks of the COVID-19 pandemic varied significantly across the United States, however, so it is difficult to make generalizations about the country as a whole.

Note that not all questions were asked on all survey waves nor presented to all respondents on the same survey wave due to survey logic, so the number of respondents and timing varies by question.
In particular, certain questions about attitudes and behaviors related to remote work were only asked to respondents who were participating in some amount of remote work. 
Sample sizes and time periods are indicated below each figure.

\subsection{Work location choice} 

Despite the common misconception that remote work is synonymous with ``working from home'', remote work is actually the flexibility to choose a work location from a set of possible alternatives.
These alternatives typically include the traditional workplace and home, but, as we show in the subsections that follow, often include third places or client's workplaces as well.
The rapid rise in flexibility with regards to work location choice across much of the workforce is a major disruption to urban transportation systems, which have largely been designed and operated to serve stable commuting patterns for decades.
It is also a significant concern for commercial real estate providers, downtown retail businesses and employers, who are navigating these changes with very little information.
This section briefly describes aggregate preferences for remote work, evolution of remote work preferences over time, and some of the constraints that prevent additional remote work.
Then, the focus of this section is describing the dynamics of work location choice between home, the employer's business premises and other possible locations, which has a profound impact on both the demand for transportation.
The impact of work location choice on trip duration and total commuting time is also investigated in detail. 

\subsubsection{Work location trends}

\begin{figure*}[t!]
    \centering
    \begin{subfigure}[b]{0.45\textwidth}
        \centering
        \includegraphics[height=2.5in]{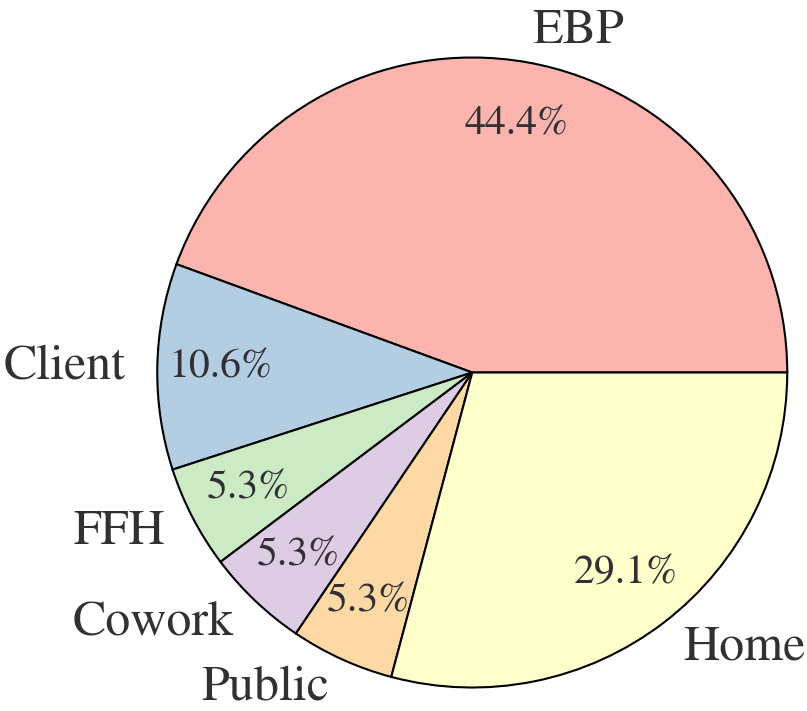}
        \caption{By worked hours}
    \end{subfigure}
    ~ 
    \begin{subfigure}[b]{0.45\textwidth}
        \centering
        \includegraphics[height=2.5in]{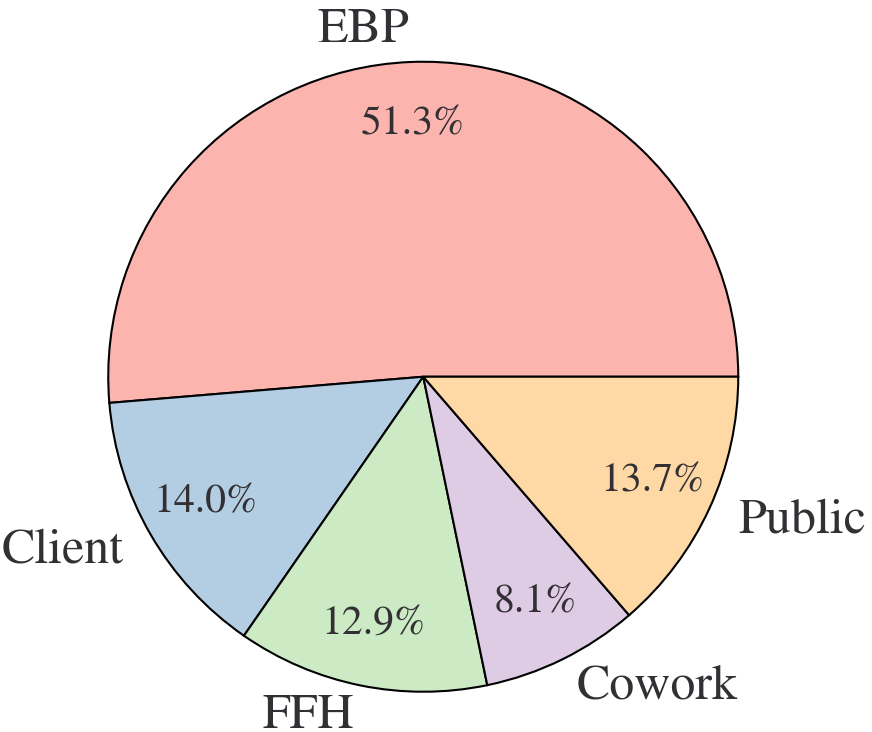}
        \caption{By number of work trips}
    \end{subfigure}
    \begin{minipage}{0.74\textwidth}
    \footnotesize (a): Nov 2021 - Jun 2022, N = 27,364; (b): Nov 2022 - Jan 2023, N = 13,091. 
    \end{minipage}
    \caption{Work location split by worked hours (a) and number of work trips (b)}
    \label{fig:thirdplace_split}
\end{figure*}

One of the primary contributions of this research is the extended investigation into the use of third places for remote work. 
The SWAA considers three types of ``third places'': public spaces (e.g. caf\'es, libraries, community centers), co-working spaces (that are not provided by the employer) and the homes of friends and family members (FFH).  
Breaking down all worked hours by location, we find that about one-third of all remote work takes place outside of the home, with a relatively even split between public spaces, coworking spaces and the homes of friends and family members.
The remaining remote work hours take place at home, while in-person work is split unevenly between the employer's business premises (EBP) and a client's workplace. 
The results are shown in Figure~\ref{fig:thirdplace_split}(a). 
After weighting by respondent earnings, we find that 17.3\% of all income in the United States is earned while working at a non-home, non-work location.

Using a question about work trips rather than worked hours, we find that 34.7\% of all work-related trips are to public spaces (13.7\%), coworking spaces (8.1\%), and FFH (12.9\%), as shown in Figure~\ref{fig:thirdplace_split}(b).
The higher share of third place trips relative to hours is because working at home does not induce a work trip. 
These results have remarkable implications; overlooking third places as a possible work location for remote workers results in over a third of all work trips being ignored. 
Third place trips are different than traditional commute trips to the employer's business premises (EBP). 
Over 43\% of respondents in the November 2021 survey had spent at least some time working at a third place during the previous week.

\begin{figure}[ht!]
    \centering
    \includegraphics[width=\textwidth]{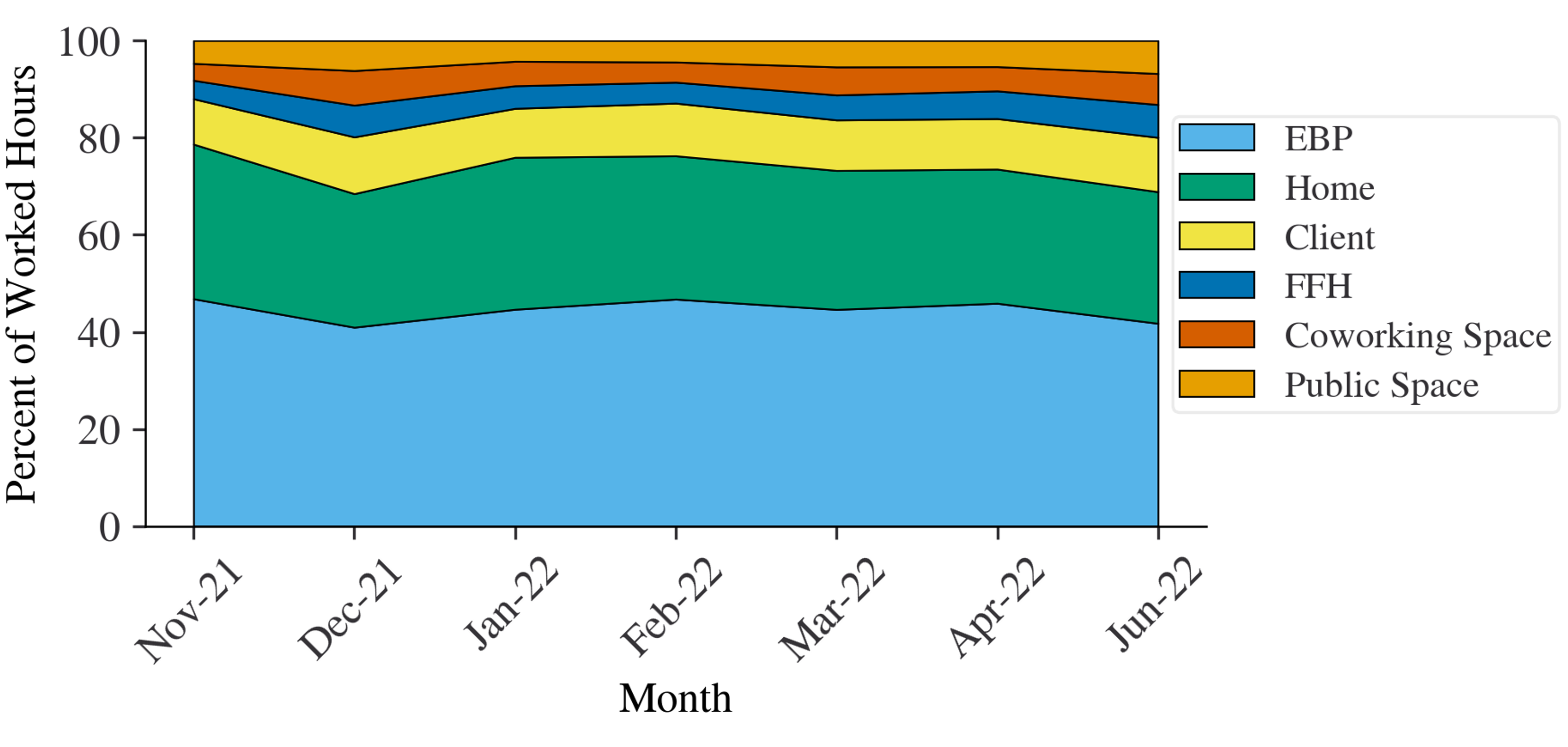}
    \begin{minipage}{0.84\textwidth}
    \footnotesize Nov 2021 - Jun 2022, N = 27,364.
    \end{minipage}
    \caption{Third place use over time}
    \label{fig:third_time}
\end{figure}

The repeated nature of the SWAA also allows us to track the use of third places over time, as shown in Figure~\ref{fig:third_time}. 
Third place use was lowest during January and February 2022, presumably due to the sharp rise in COVID cases across the United States during the same timeframe. 
By May and June of 2022, working at EBP was on the decline and working from FFH and coworking spaces was rising. 

One of the advantages of the SWAA relative to other remote work and travel surveys is the breadth of questions relating to employment and attitudes. 
Use of third places can be compared across demographic groups, household characteristics, employer characteristics, job characteristics, attitudes towards coordinating with colleagues, and general attitudes towards remote work.
Figure 
Those who are younger, male and have higher educational attainment are much more likely to make trips to third places for remote work.
The gender identity and age associations are particularly strong; people under 30 are three times more likely to be using third places for remote work than those 50 and over.

\begin{figure}[ht!]
    \centering
    \includegraphics[width=\textwidth]{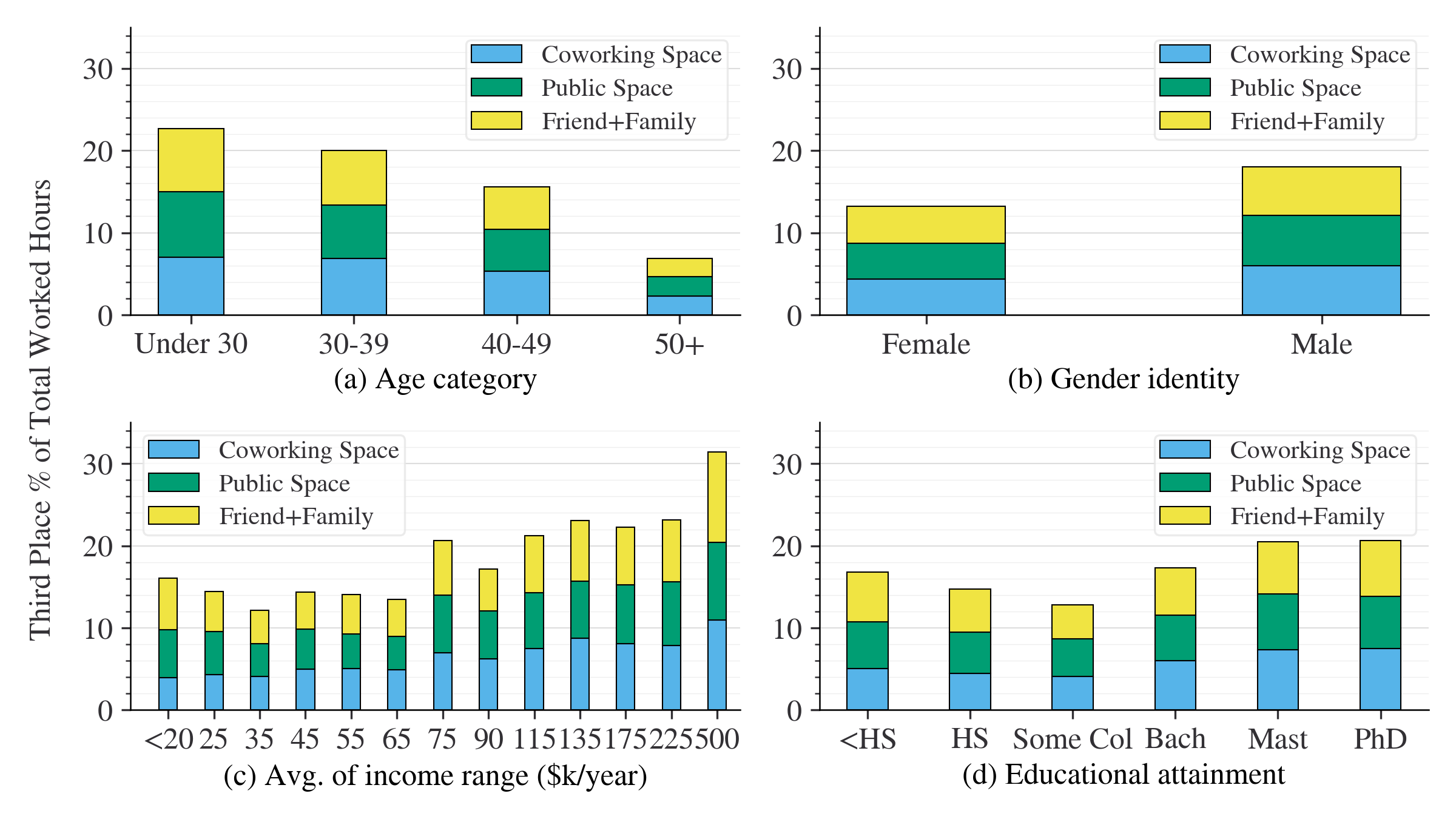}
    \begin{minipage}{0.84\textwidth}
    \footnotesize All: Nov 2021 - Jun 2022, N = 27,364.
    \end{minipage}
    \caption{Third place use by demographic group}
    \label{fig:third_demographics}
\end{figure}

\begin{figure}[ht!]
    \centering
    \includegraphics[width=\textwidth]{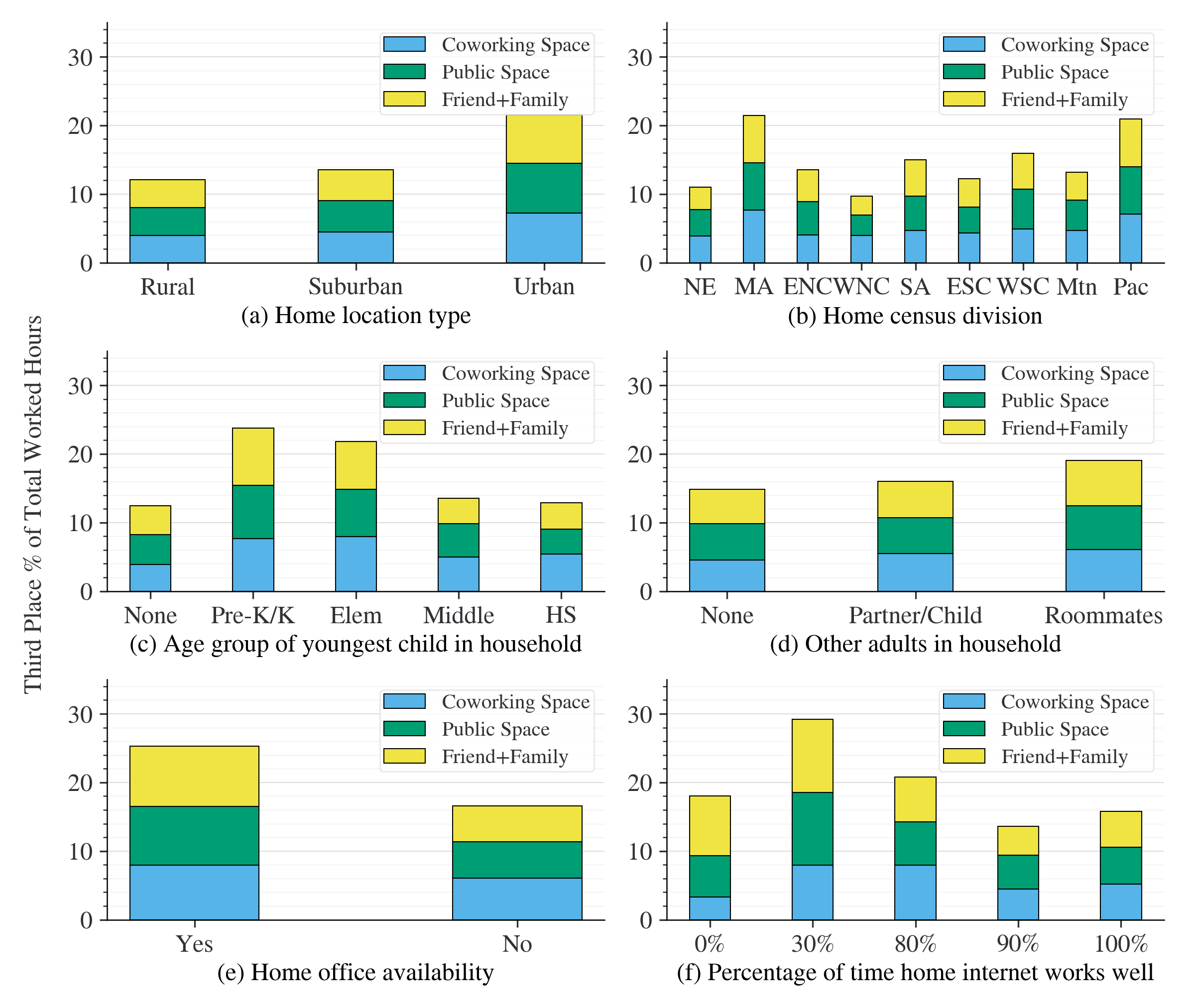}
    \begin{minipage}{0.84\textwidth}
    \footnotesize (a)-(d): Nov 2021 - Jun 2022, N = 27,364; (e)- (f): Nov 2021 - Dec 2021, N = 7,585. 
    \end{minipage}
    \caption{Third place use by household characteristics}
    \label{fig:third_household}
\end{figure}

The analysis for different household characteristics is shown in Figure~\ref{fig:third_household}. 
People living or working in an urban area, those with roommates, those with a poor internet connections and people living in the Mid-Atlantic or Pacific census divisions are much more likely to work from third places. 
This is in contrast with the results for remote work preferences, which found little association between having roommates and preferences for remote work.
This suggests that people who live with roommates have similar remote work preferences as others, but are more likely to conduct that remote work outside of the home.
Urban dwellers are more likely to have third places near their home, so it would seem reasonable that they are more likely to use third places.
Poor internet quality was found to be predictive of more third place hours, but not necessarily more third place \emph{trips}, suggesting that people with poor internet travel to third places at similar rates to others but stay longer during each visit. 
The New England and West North Central census divisions were the least likely to make trips to third places, which may be a result of colder weather in those regions during the winter months when this question was included in the questionnaire.

\begin{figure}[ht!]
    \centering
    \includegraphics[width=\textwidth]{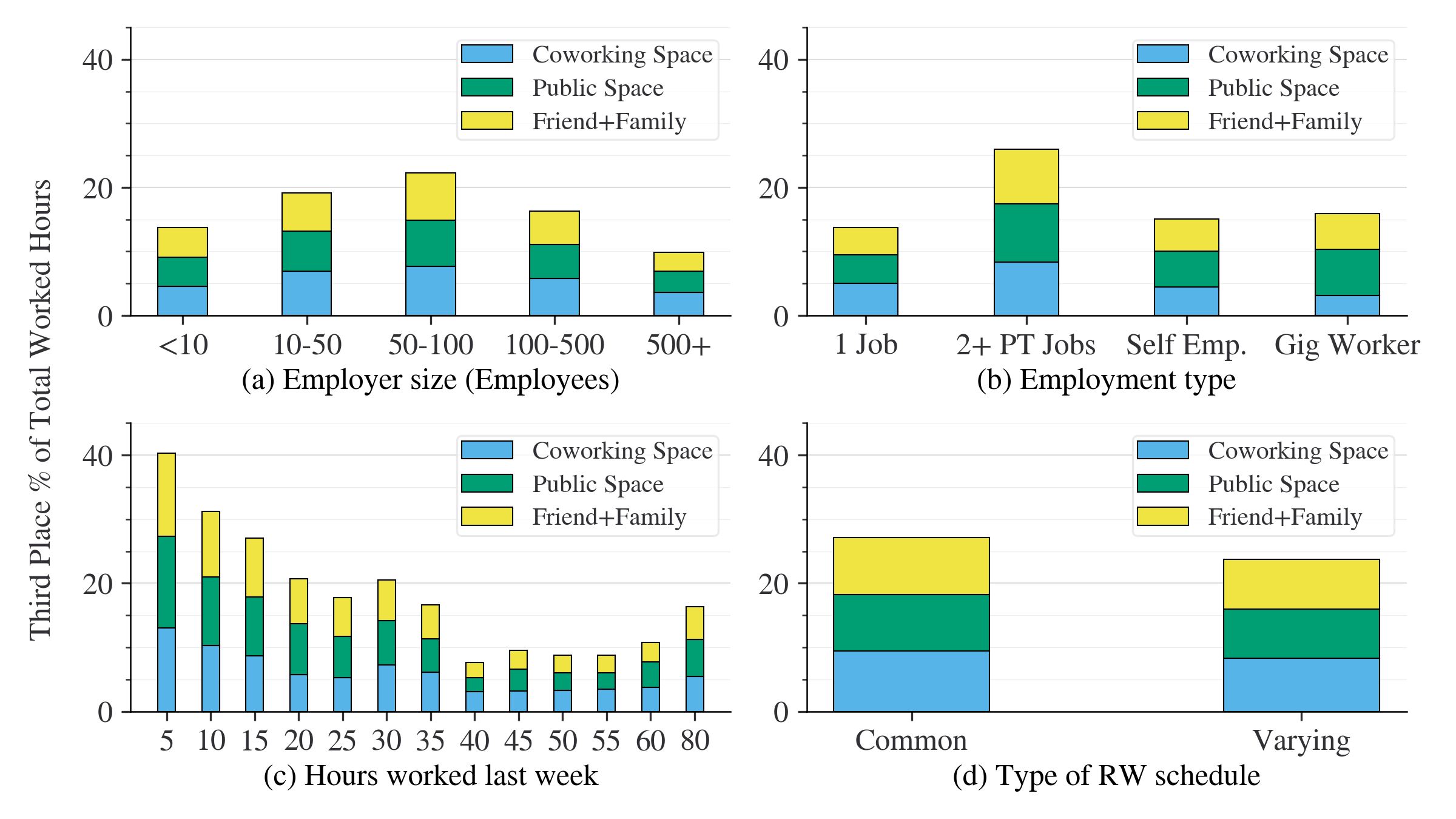}
    \begin{minipage}{0.84\textwidth}
    \footnotesize (a): Apr 2022 - Jun 2022, N = 6,583; (b): Nov 2021 - Jun 2022, N = 20,623; (c): Jun 2022, N = 4,768; (d): Mar 2022 - Apr 2022, N = 6,323. 
    \end{minipage}
    \caption{Third place use by employment characteristics}
    \label{fig:third_employment}
\end{figure}

Third place use also has strong associations with employment characteristics, as shown in Figure~\ref{fig:third_employment}. 
People who work fewer than 40 hours per week are much more likely to do that work at a third place (FFH in particular), as are people who work 2 or more part time (PT) jobs. 
Employees of medium-sized companies are also much more likely than those working for very large or small companies to use third places.
The type of remote work (RW) schedule does not have a strong bearing on third place use. 
These results underscore the need to include employment factors in travel demand models to capture commutes to third places. 

\begin{figure}[ht!]
    \centering
    \includegraphics[width=\textwidth]{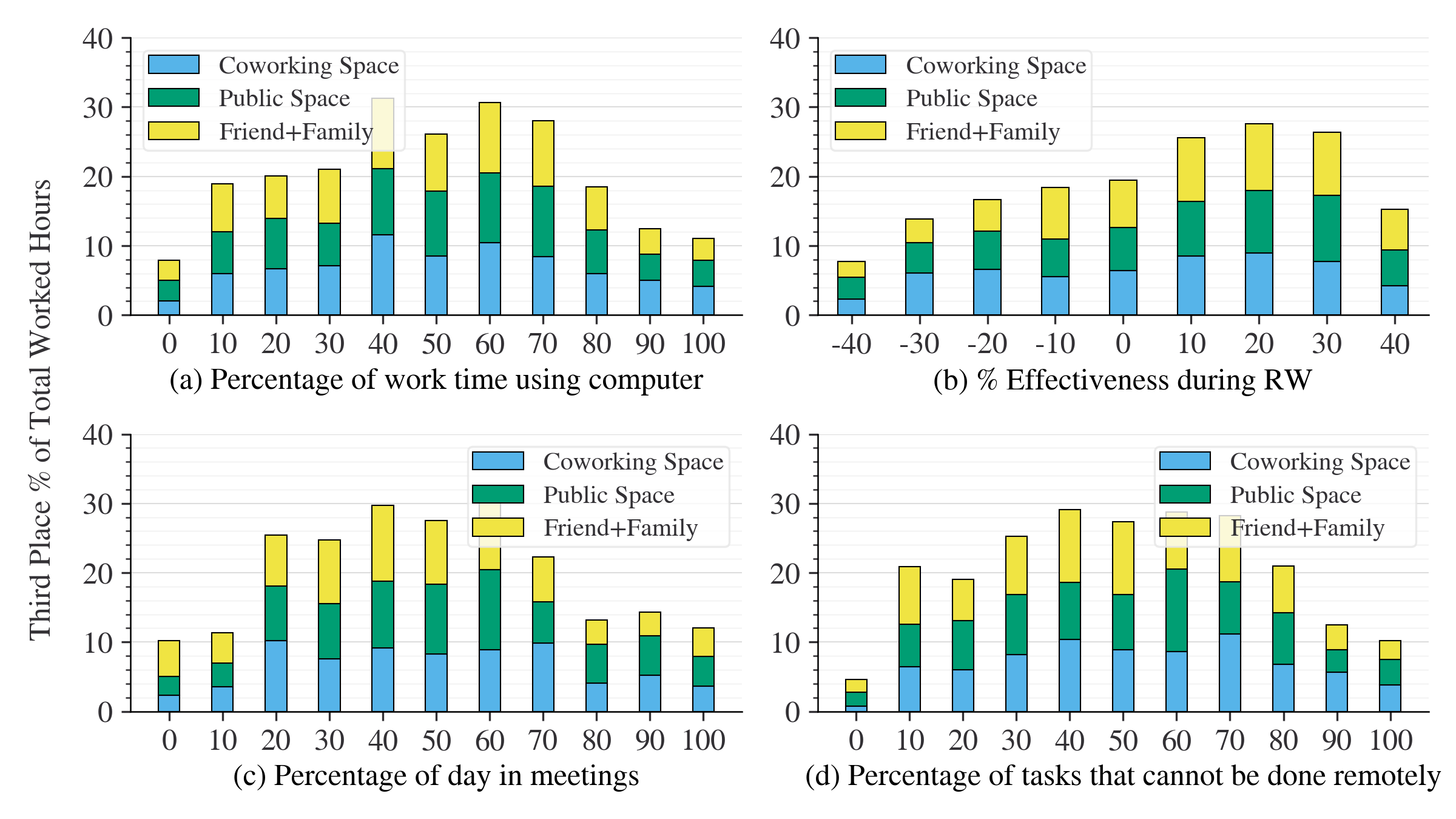}
    \begin{minipage}{0.84\textwidth}
    \footnotesize (a): Aug 2021 - Jan 2023, N = 69,436; (b): Sep 2021 - Oct 2021, N = 6,439; (c): Feb 2021 - Jan 2023, N = 96,001; (d): Jul 2020 - Jan 2023, N = 106,513. 
    \end{minipage}
    \caption{Third place use by task characteristics}
    \label{fig:third_tasks}
\end{figure}

Figure~\ref{fig:third_tasks} shows that types of tasks that a person does at work are also strongly correlated with third place use.
People who spend between 20 and 70 percent of their day meeting with others, and between 40 and 70 percent of their time using a computer are the most frequent third place users. 
Interestingly, people who find themselves to be more effective during remote work are also more likely to use third places. 
This effect could be bi-directional, which would suggest that using third places can make people feel more effective during remote work. 
Lastly, people who can do some but not all of their tasks remotely are most likely to use third places. 

\begin{figure}[ht!]
    \centering
    \includegraphics[width=\textwidth]{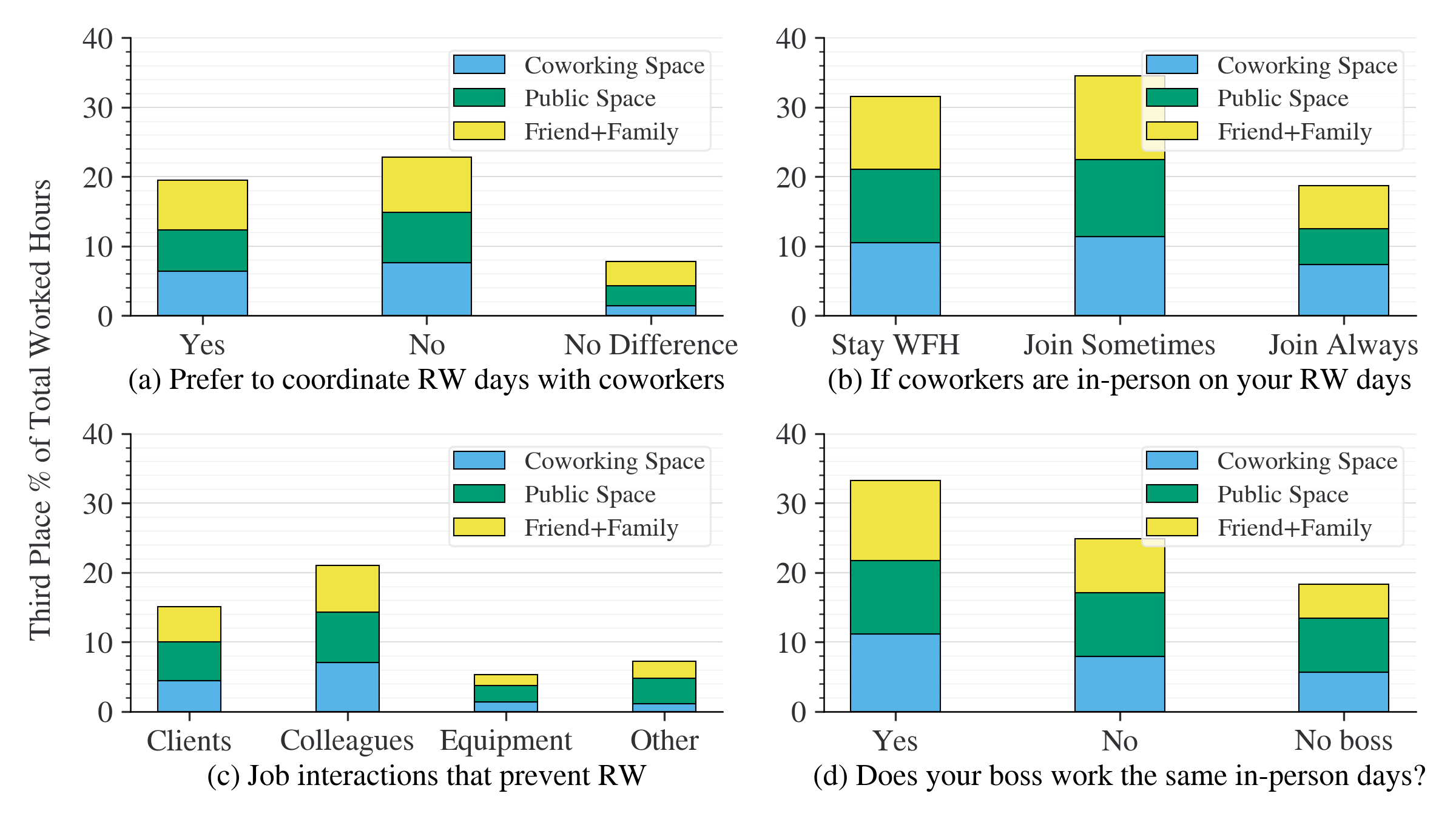}
    \begin{minipage}{0.84\textwidth}
    \footnotesize (a): Feb 2022, N = 2,925; (b): Nov 2021 - Dec 2021, N = 2,971; (c): Nov 2021, N = 3,038; (d): same as (b). 
    \end{minipage}
    \caption{Third place use by attitudes towards coordinating with colleagues}
    \label{fig:third_coordination}
\end{figure}

Connecting third place use to attitudes around coordination with colleagues is also important, as shown in Figure~\ref{fig:third_coordination}. 
People who rarely or sometimes prefer to be co-located with colleagues are more likely to use third places than those who always prefer co-location.
It can be inferred that people who always co-locate with colleagues are more likely to work at their employer's business premises.
People who claim that collaborating with colleagues is the primary barrier preventing them from additional remote work are also most likely to work at third places, possibly implying that third places are perceived as suitable for collaborative work between colleagues.
illustrates the need to include employment and task-related factors into travel demand models. 
People who need to interact with specialized equipment use third places for remote work for only about 5\% of their total working hours on average. 
Those who are able to coordinate in-person days with their boss also feel more comfortable using third places than those who do not. 

\begin{figure}[ht!]
    \centering
    \includegraphics[width=\textwidth]{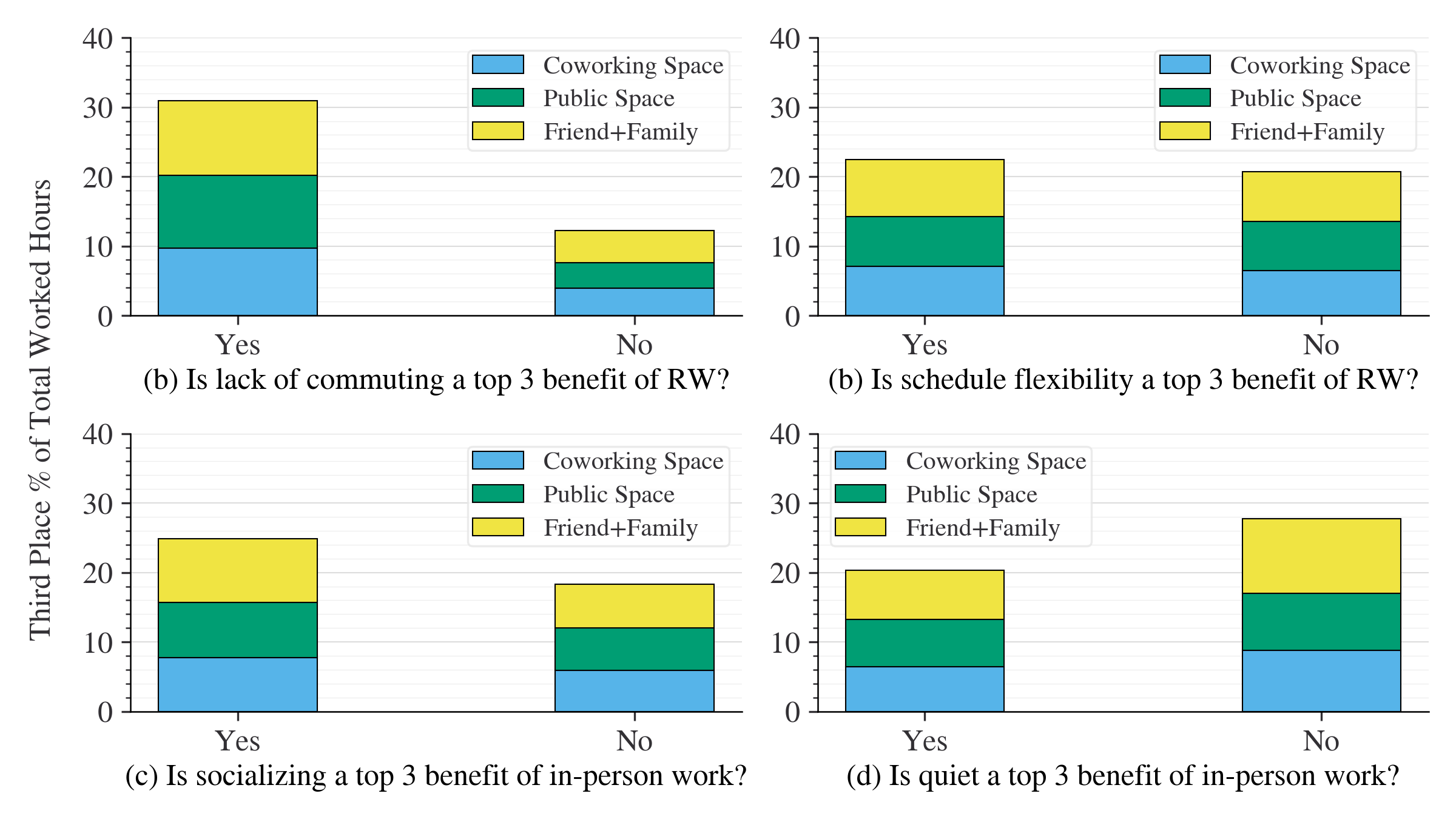}
    \begin{minipage}{0.84\textwidth}
    \footnotesize All: Feb 2022 - Jun 2022, N = 11,307.
    \end{minipage}
    \caption{Third place use by perceived benefits of remote work}
    \label{fig:third_benefits}
\end{figure}

Unsurprisingly, people who consider commuting time savings as a top benefit of remote work spend about 60\% less time at third places than those who are less concerned about commuting. 
Figure~\ref{fig:third_benefits} presents the results for third place use by different perceived benefits of remote work. 
Another interesting finding is that people who enjoy socializing during in-person work are about 7 percentage points more likely to work at third places, suggesting that third places are perceived as social environments. 

\subsubsection{Third place trip duration}

\begin{figure}[ht!]
    \centering
    \includegraphics[width=0.70\textwidth]{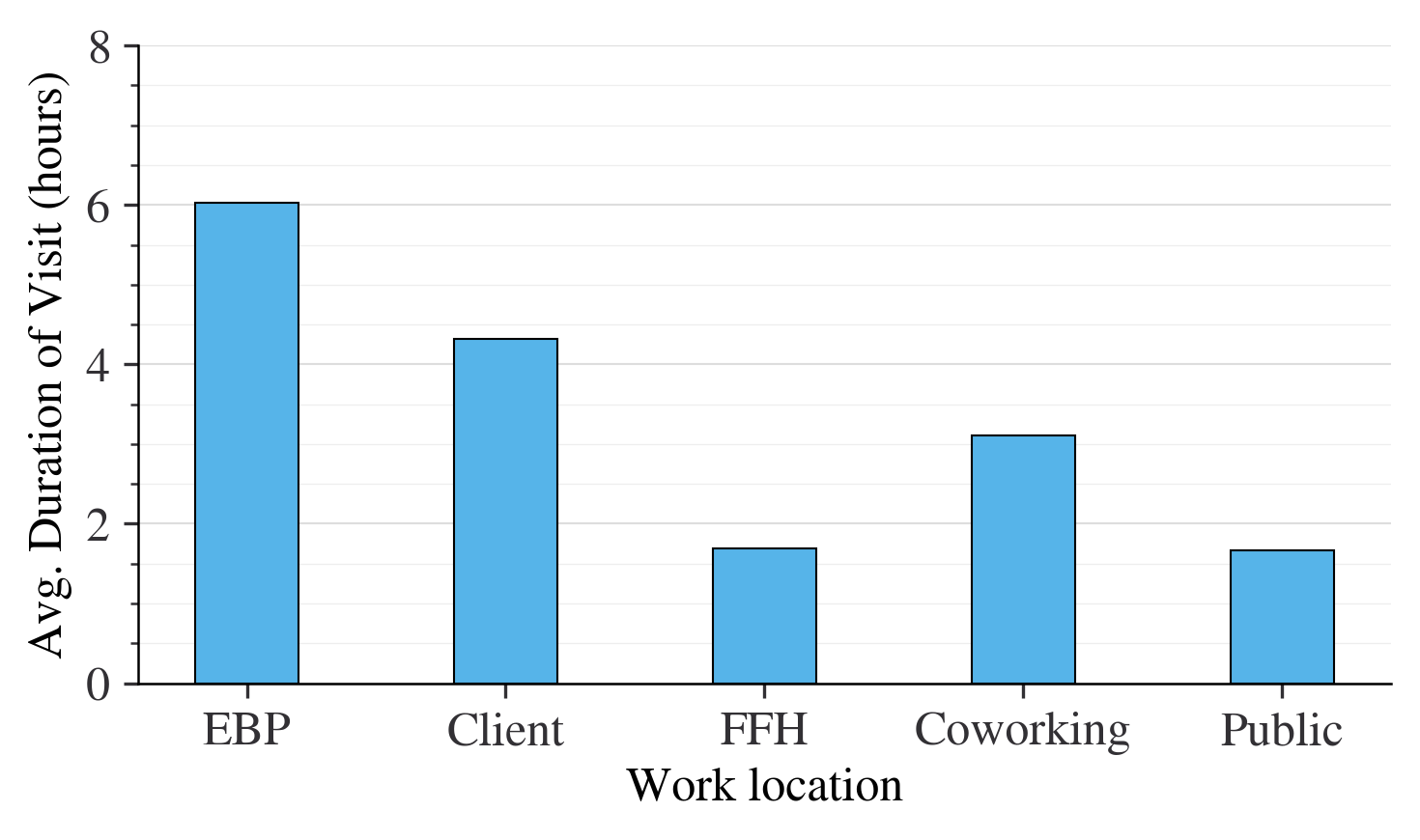}
    \begin{minipage}{0.52\textwidth}
    \footnotesize November 2022 - January 2023, N = 13,091.
    \end{minipage}
    \caption{Trip duration by work location}
    \label{fig:third_duration}
\end{figure}

Because the SWAA includes questions about time spent at workplaces and the number of trips to different workplaces, we can infer the length of time spent at each work location type by comparing the two. 
Trips to the employer's business premises, a client's workplace and coworking spaces have the longest average duration, as shown in Figure~\ref{fig:third_duration}. 
This indicates that trips to coworking places follow a similar activity schedule as traditional work trips and might anchor the daily schedule in the same way. 
Trips to work at FFH and public spaces, on the other hand, typically last for less than two hours, implying that these are often secondary work locations used for a specific task or for a temporary change of environment rather than for a full work day. 

\subsubsection{Third place travel times}

\begin{figure}[ht!]
    \centering
    \includegraphics[width=\textwidth]{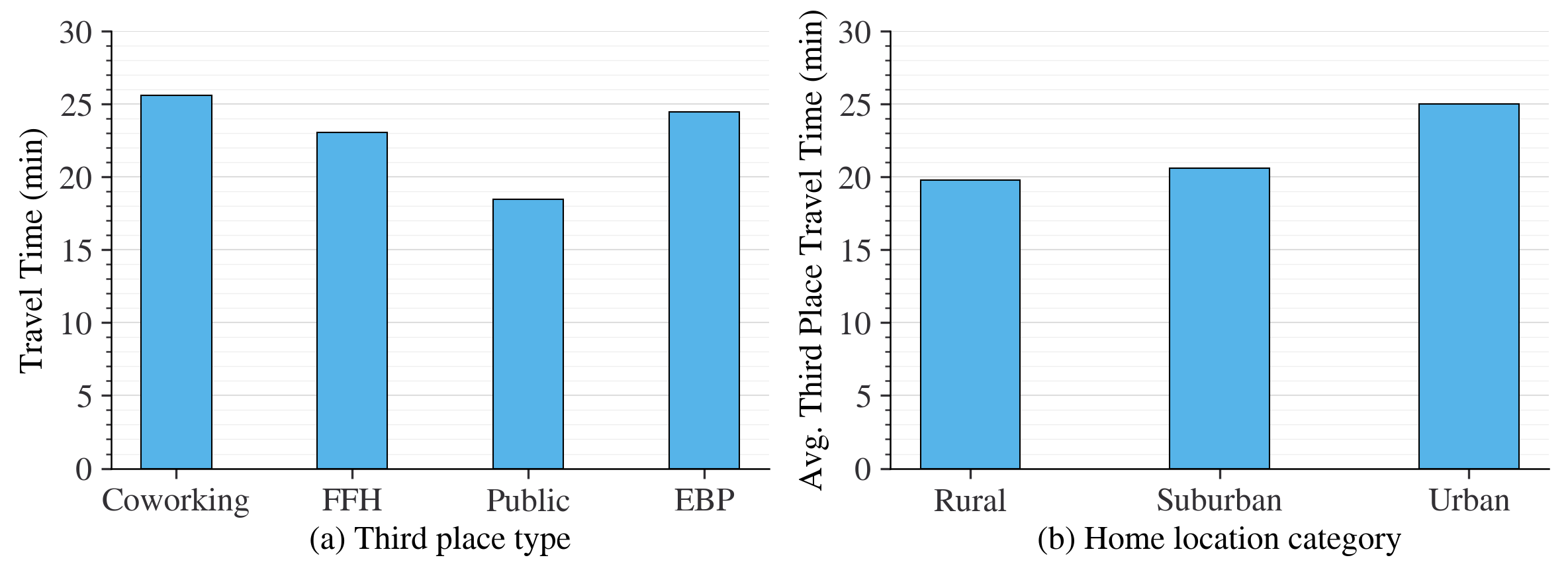}
    \begin{minipage}{0.84\textwidth}
    \footnotesize November 2022 - January 2023, N = 13,091.
    \end{minipage}
    \caption{Average travel times by third place type (a) and home location (b)}
    \label{fig:third_traveltime}
\end{figure}

Respondents who used third places were also asked to provide the travel time needed to reach the third place that they visited most recently.
This is an important concern for travel demand forecasting, given the popularity of working remotely from third places.
The initial expectation that third place commutes would be shorter that commutes to the employer's primary place were confirmed, as shown in Figure~\ref{fig:third_traveltime}.
The average one-way commuting times to coworking spaces and FFH are 26 and 22 minutes, respectively. 
Public space trips are shorter, at 18 minutes on average.
Note that these are average travel times across all modes, and third place commutes are more likely to involve slower modes such as walking or cycling, as discussed in Section~\Ref{sec:mode_choice}.

Travel times to third places were also found to vary depending on the home location of the respondent, as shown in Figure~\ref{fig:third_traveltime}(b).
Those living in urban areas were likely to travel for longer than those in suburban and rural areas, which is somewhat counter-intuitive given that there is typically a greater density of third places in urban areas. 
However, urban dwellers are more likely to use modes with low average travel speeds, such as walking and public transport, and are more likely to encounter congestion.

\subsubsection{Remote work constraints}

\begin{figure}[ht!]
    \centering
    \includegraphics[width=\textwidth]{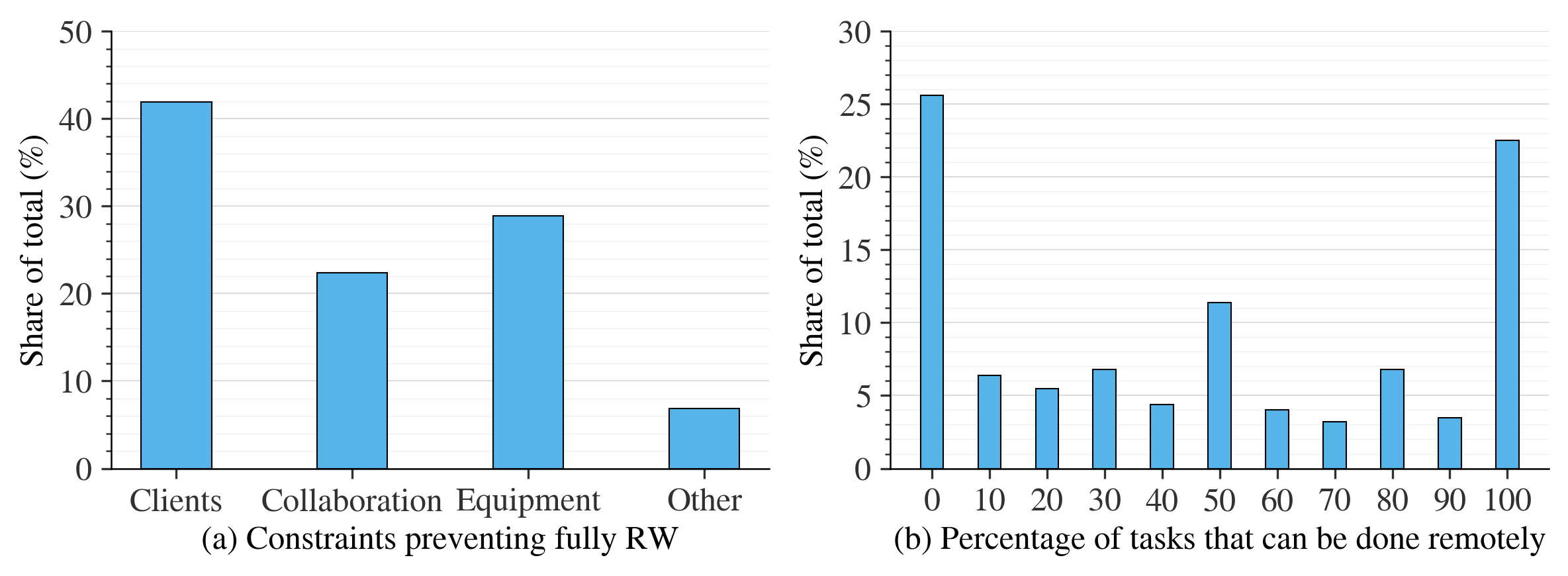}
    \begin{minipage}{0.84\textwidth}
    \footnotesize (a) Nov 2021, N = 3,038; (b): Nov 2021 - Feb 2022, N = 16,273. 
    \end{minipage}
    \caption{Constraints preventing additional remote work (a) and percentage of tasks that can be done remotely (b)}
    \label{fig:constraints}
\end{figure}

Another new question added to the SWAA in late 2021 relates to the task-related constraints that prevent additional remote work. 
The purpose of the question is to understand the potential for additional remote work in the future if some constraints are eased through improvements in communication technology. 
The results, shown in Figure~\ref{fig:constraints}, indicate that interaction with clients and equipment make up about 70\% of all constraints preventing additional remote work.
People who interact with clients in person would be those involved in retail sales, auto maintenance, and other customer-facing roles which could be difficult to do remotely.
Interaction with specialized equipment is similarly challenging to do remotely, although automation and virtual reality technology may improve remote control of equipment in the future.
Collaborative interactions with colleagues might be done remotely, however, given appropriate digital tools and technology. 
This would open up a further 22\% of roles to fully remote work. 

From the results in Figure~\ref{fig:constraints}(b), about a quarter of the workforce could be working a fully remote job, and another 23\% cannot work remotely at all.
The remaining 52\% have a job that would support hybrid work.
Put differently, about three quarters of the workforce requires some amount of in-person work in order to do their job, suggesting that a large majority of people will still need to base their home location around access to their primary work location going forward unless tasks are re-allocated between jobs. 

\subsubsection{Remote work preferences}

The SWAA survey includes three different questions about remote work: the share of remote work as a percentage of total work, the respondent's preferred share of remote work in the future, and employer plans for remote work in the future. 
This allows us to determine an upper and lower bound on future levels of remote work in the near term, assuming that preferences and plans remain stable. 
The breakdown of remote work shares by demographic and employment groups is the primary focus of other literature (e.g. \cite{barrero2021working}) and is therefore not included here.
Results for actual, planned and desired amount of remote work from the SWAA, broken down by factors related to demographics, household characteristics, employment, job tasks, remote work policies, remote work attitudes, perceived remote work benefits, attitudes towards coordination with colleagues and overall life priorities are provided in the appendix for interested readers. 

Important findings include the fact that people who work for larger employers, those who work fewer hours, and gig workers are more likely to prefer additional remote work. 
From a task breakdown perspective, people whose work is primarily done on a computer, people who speed less time in meetings and people who spend less time collaborating are also more likely to prefer additional remote work. 
People who are interested in coordinating remote work days with their boss and colleagues prefer less remote work overall, as do people who feel less stressed and more effective on remote work days. 
Interestingly, people who want to work hard to ensure their organization's success and those who consider work to be the most important priority in life anticipate that their employer will plan for a higher share of remote work than those who are less enthusiastic about their work. 

\subsection{Mode choice} \label{sec:mode_choice}

The choice of work location interacts with other travel decisions such as mode choice, departure time and even household location. 
This section investigates the choice of travel mode for commutes to the employer's business premises, commutes to third places and for non-work trips.
The responses allow us to understand how remote workers are traveling when they choose to conduct remote work outside of the home and how non-work trips differ from commuting trips. 

\subsubsection{Commuting mode choice}

\begin{figure}[ht!]
    \centering
    \includegraphics[width=0.9\textwidth]{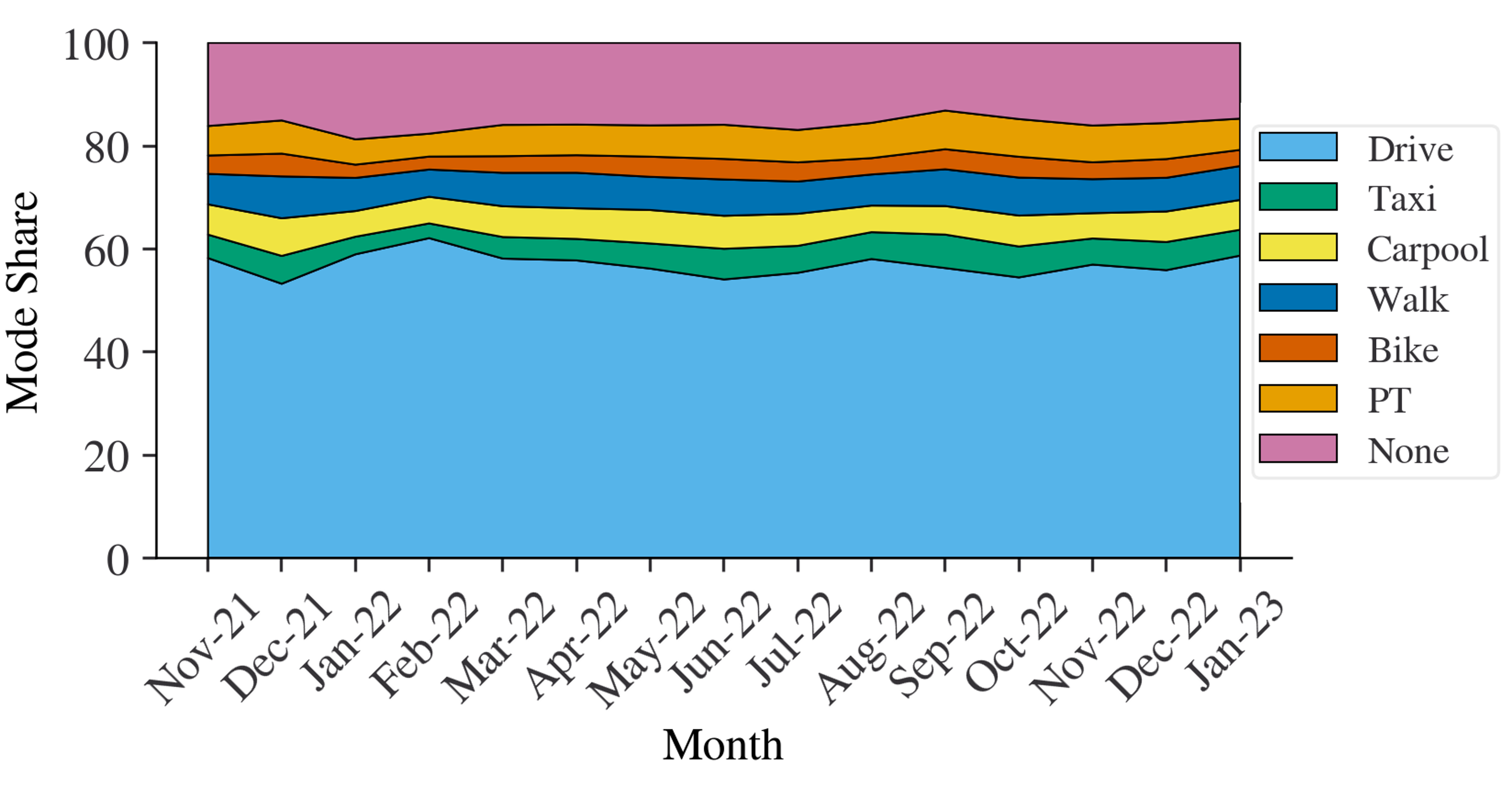}
    \begin{minipage}{0.74\textwidth}
    \footnotesize Nov 2021 - Jan 2023, N=70,029. 
    \end{minipage}
    \caption{Commuting mode shares from November 2021 to January 2023}
    \label{fig:mode_time}
\end{figure}

Questions about mode choice for commuting in general have been included in the SWAA questionnaire since November 2021.
These questions ask about current mode choices and mode choices in 2019, allowing us to determine how the disruption of remote work has affected mode choice and whether the trend has changed over time. 
The time series results are shown in Figure~\ref{fig:mode_time}. 
We find that commuting modes have been relatively constant over time, with public transit and walking increasing somewhat in the summer of 2022.
The response ``None'' indicates that the respondent did not commute during the week prior to responding to the survey.

\begin{figure}[ht!]
    \centering
    \includegraphics[width=0.9\textwidth]{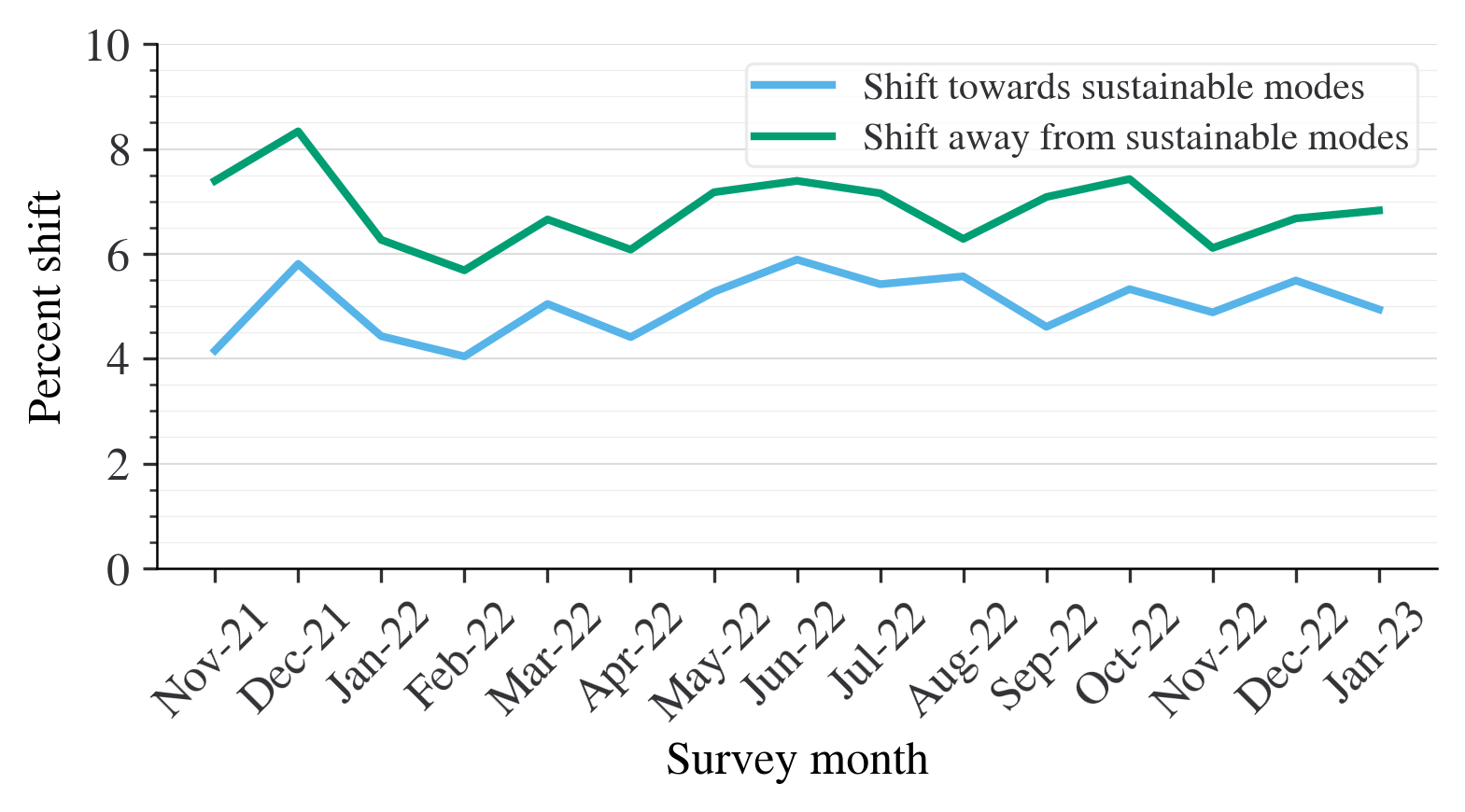}
    \begin{minipage}{0.74\textwidth}
    \footnotesize Nov 2021 - Jan 2023, N=70,029. 
    \end{minipage}
    \caption{Reported transitions to and from sustainable commuting modes since 2019, by survey wave}
    \label{fig:mode_switch}
\end{figure}

We can also look at whether people have changed commuting modes over time.
Of particular interest is the transition to and from ``sustainable'' travel modes.
For the purpose of this analysis, we will consider public transit, walking, cycling, and carpooling as sustainable modes.
Every survey wave has found that a greater share of people have transitioned \emph{away} from sustainable commuting modes since 2019 than have transitioned \emph{towards} sustainable commuting modes, as shown in Figure~\ref{fig:mode_switch}
The average across survey waves is that 6.8\% of people have switched from sustainable modes in 2019 to driving or using a taxi, while only 5.0\% have done the opposite.
These high-level results suggest that even if commuting frequency has decreased as a result of remote work, the carbon intensity of each commuting mile travelled is likely to have risen, although it should be noted that these questions do not capture important factors such as vehicle occupancy or fuel efficiency. 

\subsubsection{Third place commutes}

\begin{figure}[ht!]
    \centering
    \includegraphics[width=\textwidth]{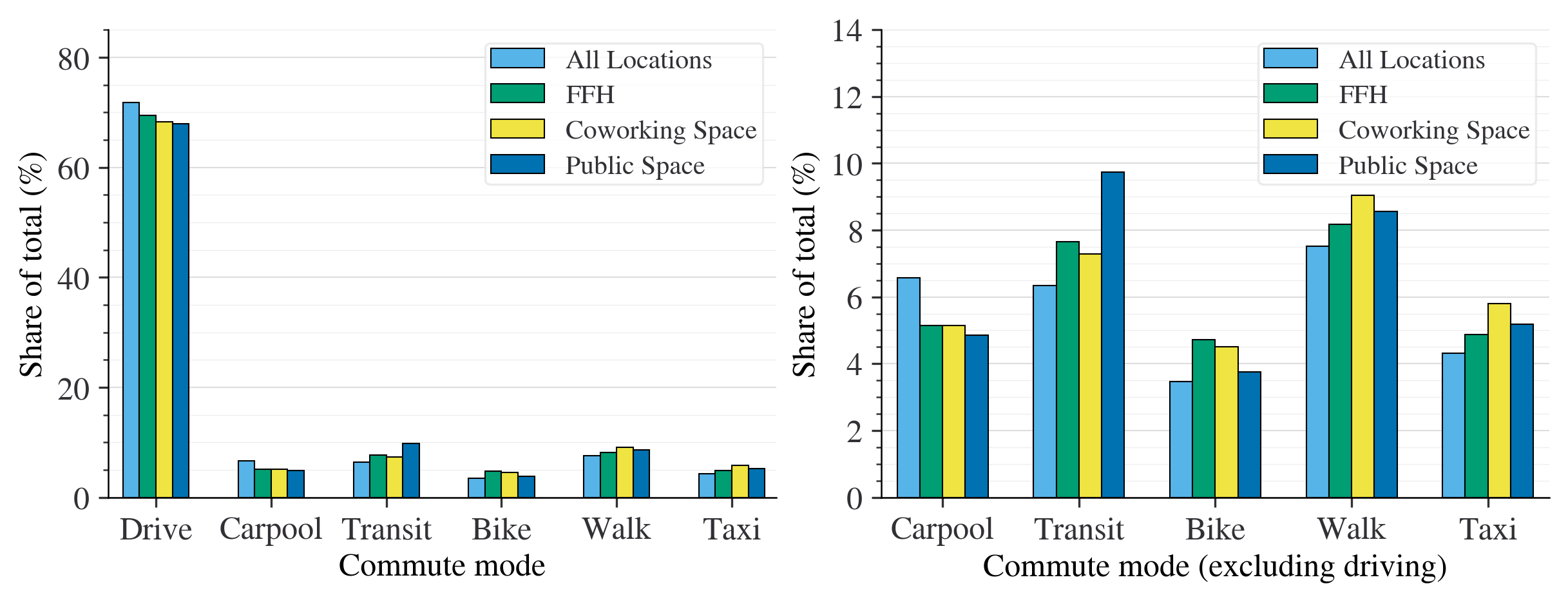}
    \begin{minipage}{0.84\textwidth}
    \footnotesize Jan 2022 - Apr 2022, N = 2,235. 
    \end{minipage}
    \caption{Mode choice by work location}
    \label{fig:third_mode}
\end{figure}

Mode choice distributions also differ for third places relative to overall commute mode shares and mode shares for trips to the employer's business premises.
Figure~\ref{fig:third_mode}(a) shows the breakdown of mode choice by work location, while Figure~\ref{fig:third_mode}(b) shows the same data with driving excluded. 
The results for third place mode choices are quite remarkable, despite appearing similar to the average commuting mode in Figure~\ref{fig:third_mode}(a).
Looking at Figure~\ref{fig:third_mode}(b), we can see that at a national level, the mode shares for cycling, walking and transit are all higher than average commuting mode shares.
Transit mode shares are much higher for remote work trips to third places, especially public spaces (54\% greater than the average mode share) suggesting that third place use is associated with greater patronage of public transit systems. 

\subsubsection{Non-work trips}

\begin{figure}[ht!]
    \centering
    \includegraphics[width=\textwidth]{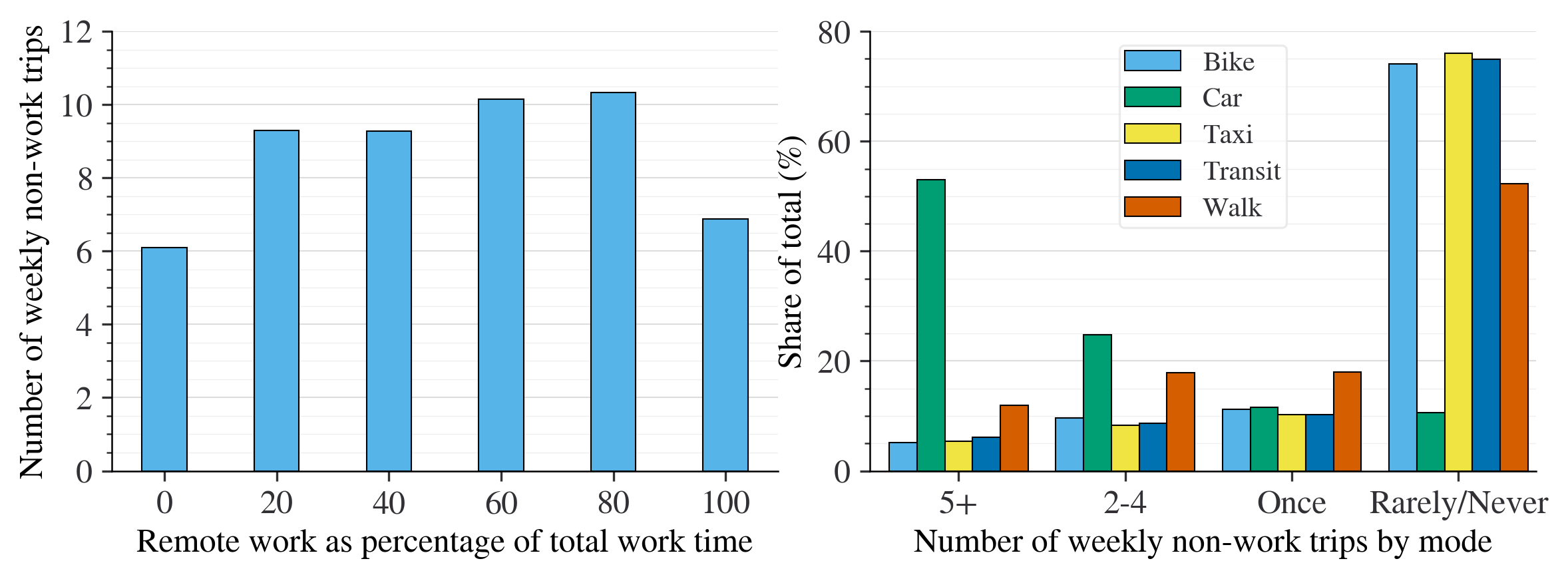}
    \begin{minipage}{0.84\textwidth}
    \footnotesize Mar 2022 - Apr 2022, N = 5,681. 
    \end{minipage}
    \caption{Non-work trips by remote work share (a) and frequency of non-work trips by mode (b)}
    \label{fig:nonwork}
\end{figure}

As remote work has grown, non-work trips have become an increasingly important contributor to overall travel demand. 
Respondents were asked to provide their frequency of trips by mode for non-work trips during consecutive survey waves in the spring of 2021. 
By plotting the number of weekly trips against remote work share in Figure~\ref{fig:nonwork}, we can observe that there are three separate groups: those who work entirely remote, those with hybrid work arrangements, and those who do not work remotely at all.
Interestingly, hybrid workers conduct the most non-work trips, and the balance of remote and non-remote work within the hybrid schedule does not appear to be strongly associated with the number of non-work trips. 
Fully remote workers make somewhat fewer non-work trips, while fully in-person workers make the fewest non-work trips of all.
This may be due to their ability to go shopping, run errands or conduct social activities during work breaks or on the return trip from work.
Looking at non-work trips by third place use also indicates that people who spend more time at third places also make more non-work trips. 

With respect to mode choice, the overall trends are similar to the commuting trips although the two are not directly comparable due to the different question formats. 
More than 50\% of the population drives to conduct non-work trips at least five times per week, while more than 75\% of the population rarely or never choose transit, walking or biking for non-work trips.
These shares are somewhat different depending on the home location; urban residents are more likely to use non-car modes for non-work trips than those who live in rural areas. 

\subsection{Departure time}

Flexibility related to the physical location of work has also translated into flexibility with regards to the temporal aspect of work. 
The freedom to work at home allows remote workers to start working in the morning, then leave travel to the office or a third place after peak congestion has subsided. 
This section reviews the distribution of work trip departure times, how those departure times have changed since 2019, and how they vary by work location including third places.

\subsubsection{Commuting departure times}

\begin{figure}[ht!]
    \centering
    \includegraphics[width=0.9\textwidth]{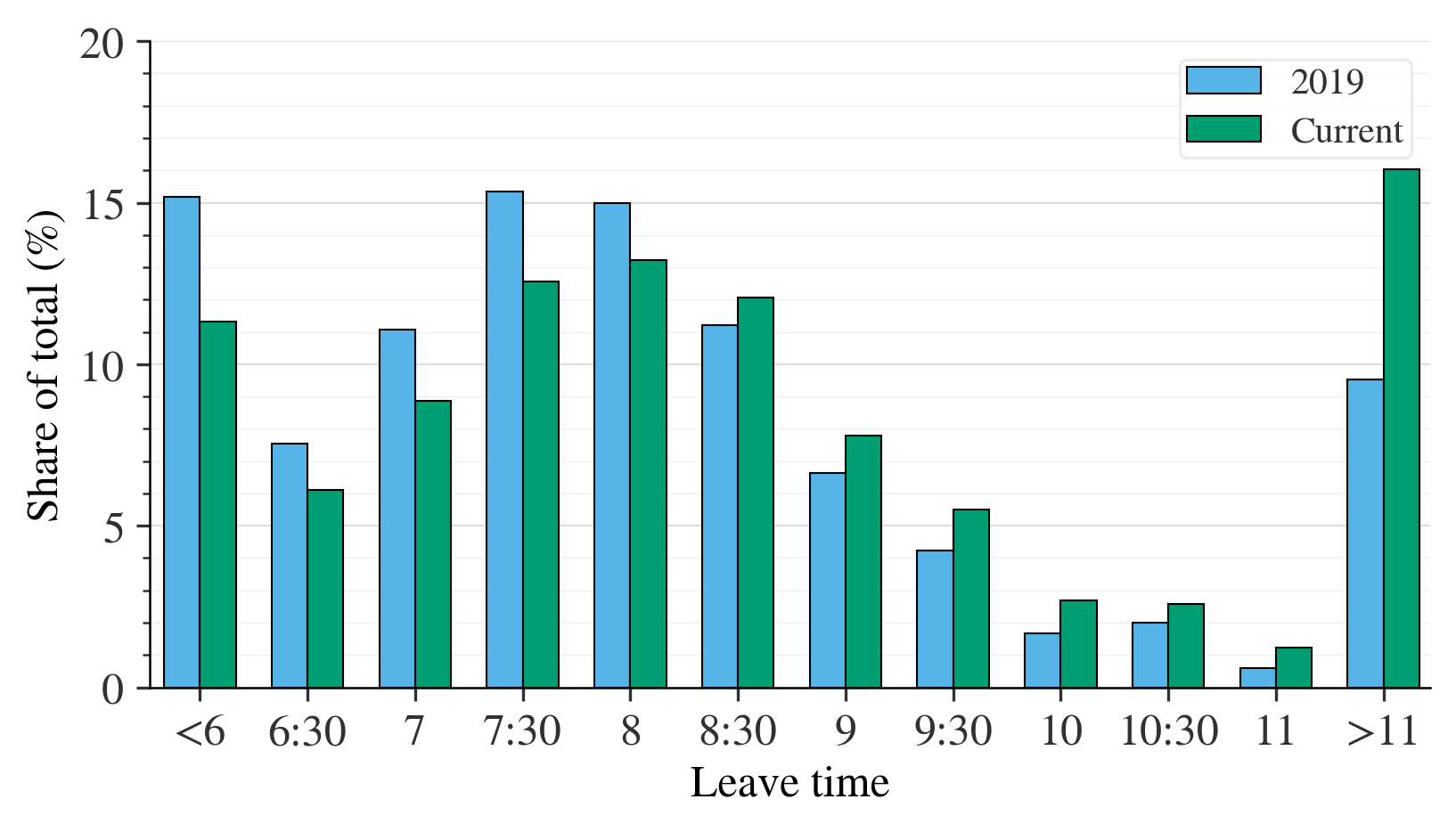}
    \begin{minipage}{0.74\textwidth}
    \footnotesize Nov 2021 - Feb 2022, N = 16,723. 
    \end{minipage}
    \caption{Work trip departure time changes from 2019 to current}
    \label{fig:deptime}
\end{figure}

Typical departure times have shifted later in the day relative to 2019, as shown in Figure~\ref{fig:deptime}.
All departure times from 8:30 AM onward have become more popular, while all earlier departure times have become less popular.
Notably the portion of the population reporting departure times after 11:00AM has increased from 9.5\% to 16\%, likely reflecting an increase in the use of third places for remote work. 

\subsubsection{Third places}

\begin{figure}[ht!]
    \centering
    \includegraphics[width=0.9\textwidth]{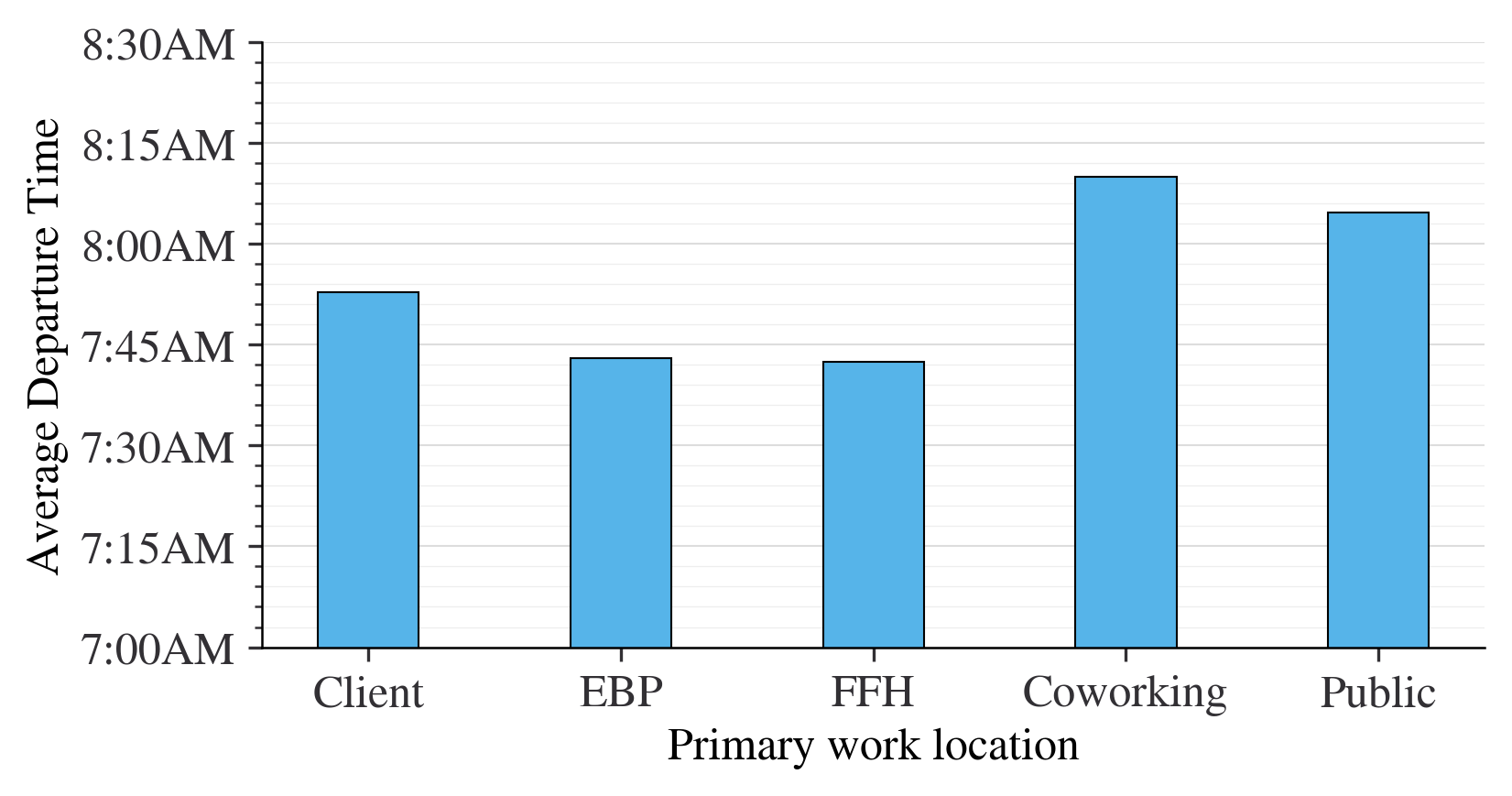}
    \begin{minipage}{0.60\textwidth}
    \footnotesize Nov 2021 - Feb 2022, N = 16,723. 
    \end{minipage}
    \caption{Departure times by primary work location}
    \label{fig:deptime_third}
\end{figure}

Departure times can also be differentiated based on the respondent's primary work location.
As shown in Figure~\ref{fig:deptime_third}, departure times by work location reflect a similar trend to third place trip duration.
Departure times for trips to coworking spaces specifically are similar to departure times for traditional commutes.
Departure times for trips to public spaces and FFH, conversely, occur much later in the day on average. 

\subsection{Productivity and travel}

One of the unique benefits of the SWAA questionnaire is that it includes a comprehensive set of questions related to employment and productivity in addition to the travel questions. 
This section reviews how the travel choices described above are related to perceptions of productivity.
These results connect the often disconnected fields of organizational behavior and transportation planning, demonstrating how remote work is inherently an interdisciplinary topic with complex trade-offs. 

\begin{figure}[ht!]
    \centering
    \includegraphics[width=0.9\textwidth]{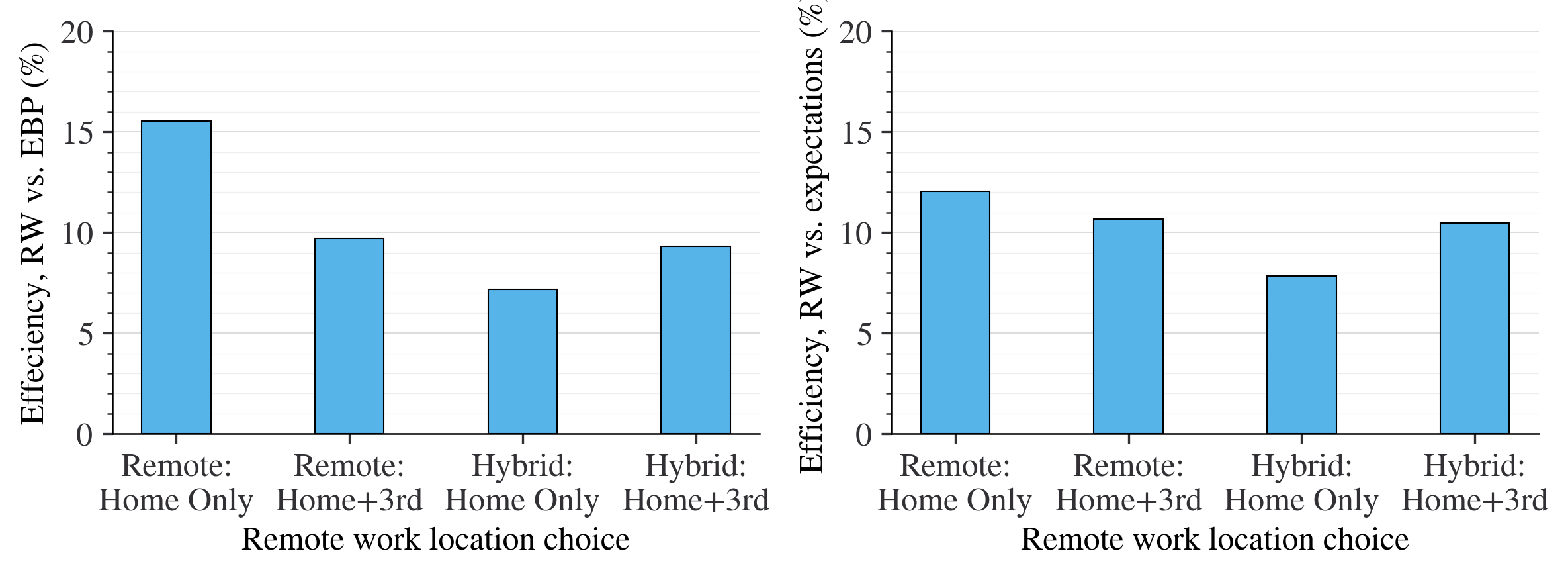}
    \begin{minipage}{0.60\textwidth}
    \footnotesize Nov 2021 - Apr 2022, N = 9,012. 
    \end{minipage}
    \caption{Perceived change in efficiency during remote work relative to working at EBP (a) and relative to expectations (b)}
    \label{fig:productivity}
\end{figure}

Figure~\ref{fig:productivity} shows how perceptions of personal efficiency during remote work is closely associated with the choice of work location.
In Figure~\ref{fig:productivity}(a), we observe that people who work fully remote and spend all of their remote work time at home (``Remote: Home Only'') have the highest self-reported efficiency relative to working at their employer's business premises.
People who work 100\% remotely but choose to spend at least some of their remote work hours working at a third place (``Remote: Home+3rd'') still have positive perceptions of their remote work productivity, but less than that of people who work entirely at home.
This may be because the choice to work at a third place is associated with a somewhat unproductive home work environment, meaning that the actual third place work is not necessarily the cause of the lower perceived efficiency.
This trend may also be a result of people making work location choices based on factors other than maximizing productivity; third places can offer a more social environment, new networking opportunities or more comfortable surroundings. 

The opposite trend is can be observed for remote workers with a hybrid remote work schedule where some of their time is spent working at their employer's business premises. 
Hybrid workers who do not use third places for remote work perceive themselves to be less productive than people who split their working time between their employer's business premises, home and third places.
This suggests that providing easily accessible third places for hybrid workers, who are the largest cohort of workers in the economy, could be associated with a small rise in work efficiency. 
Similar trends can be observed for efficiency relative to expectations in 2019, albeit with less variation. 

\section{Discussion} \label{sec:discussion}

% Ramble about remote work and the uncertain future
The future of working arrangements and work-related travel patterns remains uncertain. 
Demand for remote work rose very quickly and appears durable in the medium term.
New infrastructure, technology and services to support remote work are only just beginning to emerge and may increase the demand by making remote work more accessible and convenient for a wider range of tasks.
Alternatively, a future economic downturn may shift the balance of power to employers, who typically favor less remote work than their employees would want.
What is clear, however, is that remote work has a strong influence on travel behavior, including destination choice, mode choice, and departure time. 
Furthermore, widespread remote work, by providing flexibility about where and when to work, connects travel demand with employment characteristics and attitudes to a much greater degree than ever before. 

% Briefly recap the notable and surprising results from this paper
The SWAA results demonstrate a number of notable findings relative to the travel behavior of remote workers. 
Conducting remote work at a third place has become relatively popular and now accounts for over a third of all commuting trips. 
The characteristics of these trips depends on the type of third place; trips to co-working spaces are fairly similar to traditional commutes with respect to travel time, duration and departure time, but trips to FFH and public spaces are altogether different.
Sustainable travel modes including walking, carpooling, public transit and cycling have lost mode share since 2019, but third place commuting trips are more likely to use these modes than the average commuting trip.
Departure times for commuting trips have shifted later in the day overall, and the total number of non-work trips is likely to have grown given that hybrid and fully remote workers conduct more non-work trips than people who do not work remotely. 

% Describe the policy implications of this research
This research has many important policy implications. 
The first is that it is essential to consider third place commutes when estimating overall travel demand. 
Many people assume that ``working from home'' and ``remote work'' are synonymous, but this research shows that remote work often happens outside of the home, inducing a commuting trip with social externalities that differ from traditional commutes. 
Preferences for third places vary considerably, not just by the typical demographic and household characteristics that are often included in household travel surveys, but also by employer characteristics, task characteristics, employer remote work policies, coordination between colleagues, and attitudes towards remote work.
Policy makers must consider these factors and how they might change over time when making investments in transportation infrastructure. 
Public transit agencies can also use these insights to identify where third place trips are occurring and offer new services to make third places more accessible. 

The SWAA findings also show that third place commutes are generally shorter, are less likely to take place during peak hours, and have a more sustainable mode share than typical commutes.
Moreover, perceptions of work efficiency are greater among hybrid workers who use third places relative to those who only conduct remote work from home.
These trends are despite the fact that land use, transportation systems and third place operators have not fully adapted to the rapid and unexpected rise in remote work.
As shown in Figure~\ref{fig:third_traveltime}, the average commuting trip to a third place remains relatively long. 
By making third places more accessible and evenly distributed across urban areas, policy makers can make third place commutes even shorter and thus more accessible by walking and cycling in the future. 
This could be especially beneficial for suburban areas, rural areas, and communities where the existing housing stock might not be conducive to working at home. 

This paper explores an ongoing source of comprehensive data on travel, remote work, and employment. 
There are many future directions for this research.
Developing statistical models to quantify the effects of different independent variables on travel behavior would be helpful in calibrating travel demand forecasting tools. 
Supplementing this aggregate longitudinal survey with travel diaries could also help to provide detailed information about third place trips, such as travel distance and carbon emissions.
Further research is also needed into the factors that affect the choice of a specific third place for remote work.
These are likely to include traditional factors such as travel time and accessibility by various modes, but also the types of amenities available, the work task to be accomplished, and the average occupancy.
The data and code used for this analysis are freely available at \url{https://github.com/jtl-transit/swaa} for anyone interested in pursuing future research in this area. 

\section{Acknowledgements}
The authors are extremely grateful to Professors Jos\'e Barrero, Nick Bloom and Steven J. Davis for establishing the SWAA survey, for supporting the introduction of new travel-related questions, and for their continuing enthusiastic collaboration on remote work and travel research.
In addition, the authors thank their colleagues Bhuvan Alturi and Jim Aloisi for providing feedback on an early version of this paper. 
The authors also thank the MIT Mobility Initiative, the MIT Energy Initiative and the MIT Task Force on the Work of the Future for funding this research. 

\section{Author Contribution Statement}
The authors confirm contribution to the paper as follows: study conception and design: N.S. Caros, X. Guo, Y. Zheng, J. Zhao; data collection: N.S. Caros, X. Guo, Y. Zheng; analysis and interpretation of results: N.S. Caros, X. Guo, Y. Zheng, J. Zhao; draft manuscript preparation: N.S. Caros, X. Guo, Y. Zheng, J. Zhao. All of the authors reviewed the results and approved the final version of the manuscript. 
The authors do not have any conflicts of interest to declare.

\newpage
% \nolinenumbers
\bibliographystyle{apalike}
\bibliography{references}

\newpage
\input{appendix}
\end{document}

%% file: appendix.tex
\clearpage
\pagenumbering{arabic}% resets `page` counter to 1
\renewcommand*{\thepage}{\Alph{section}-\arabic{page}}

\appendix

\section{Remote work trends, plans and preferences}
\renewcommand{\thesubsection}{\Alph{section}\arabic{subsection}}

This section presents supplementary results from the SWAA related to current remote work trends, employee preferences for remote work in the future and employer plans for remote work in the future.

\subsection{Current remote work shares}

First, we look at the observed trends in remote work. 
The results are based on a question asking respondents what percent of paid full days in the past week were conducted remotely. 
Responses for people who were not employed at the time of the survey are excluded.

\begin{figure}[ht!]
    \centering
    \includegraphics[width=\textwidth]{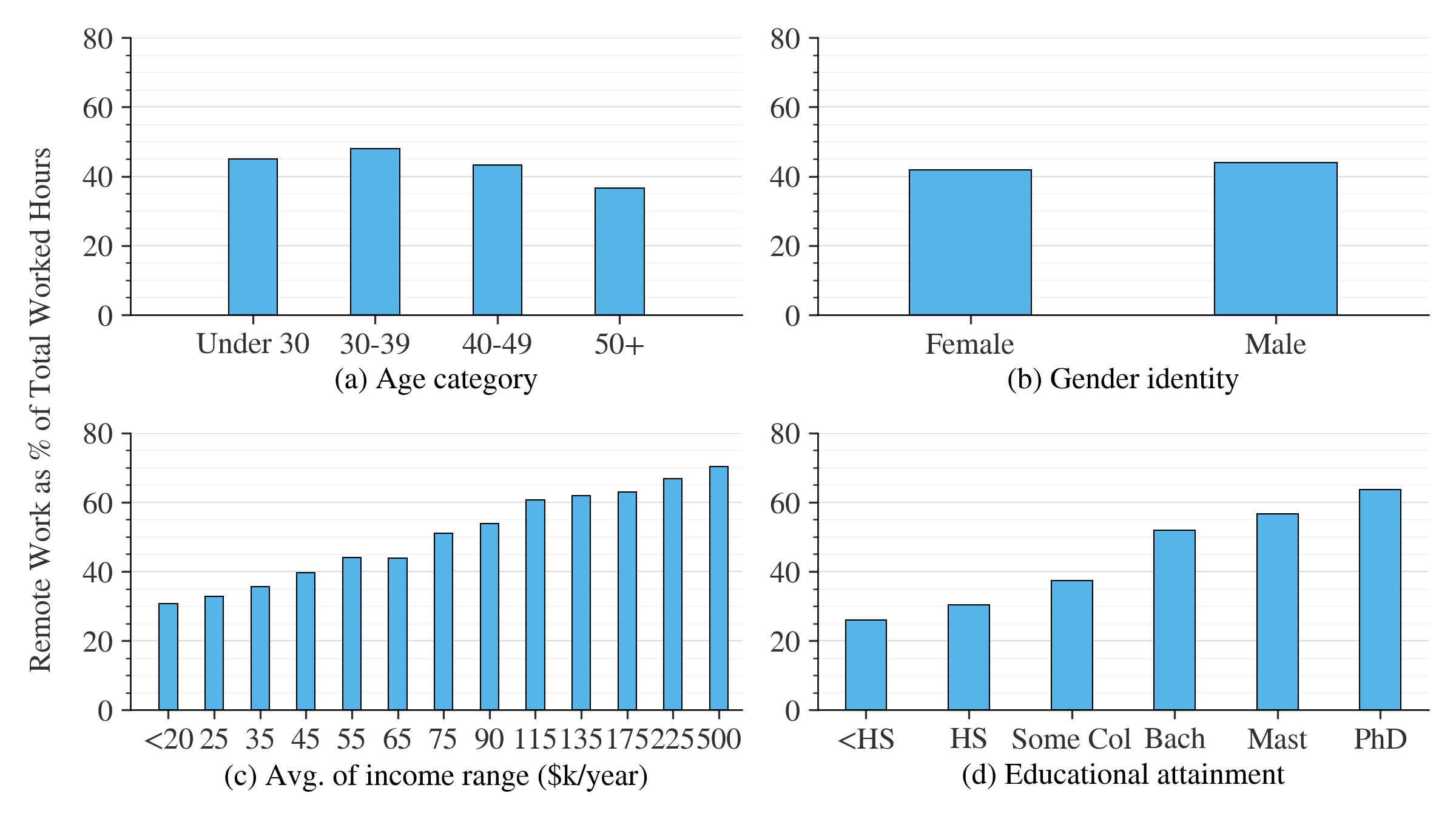}
    \begin{minipage}{0.84\textwidth}
    \footnotesize All: May 2020 - Jan 2023, N = 108,804.
    \end{minipage}
    \caption{Current remote work share by demographic group}
    \label{fig:current_demographics}
\end{figure}

Remote work shares by demographic group are shown in Figure~\ref{fig:current_demographics}. 
Consistent with previous studies, the SWAA finds that younger people, men, and those with higher incomes and education generally engage in more remote work than the average.

\begin{figure}[!ht]
    \centering
    \includegraphics[width=\textwidth]{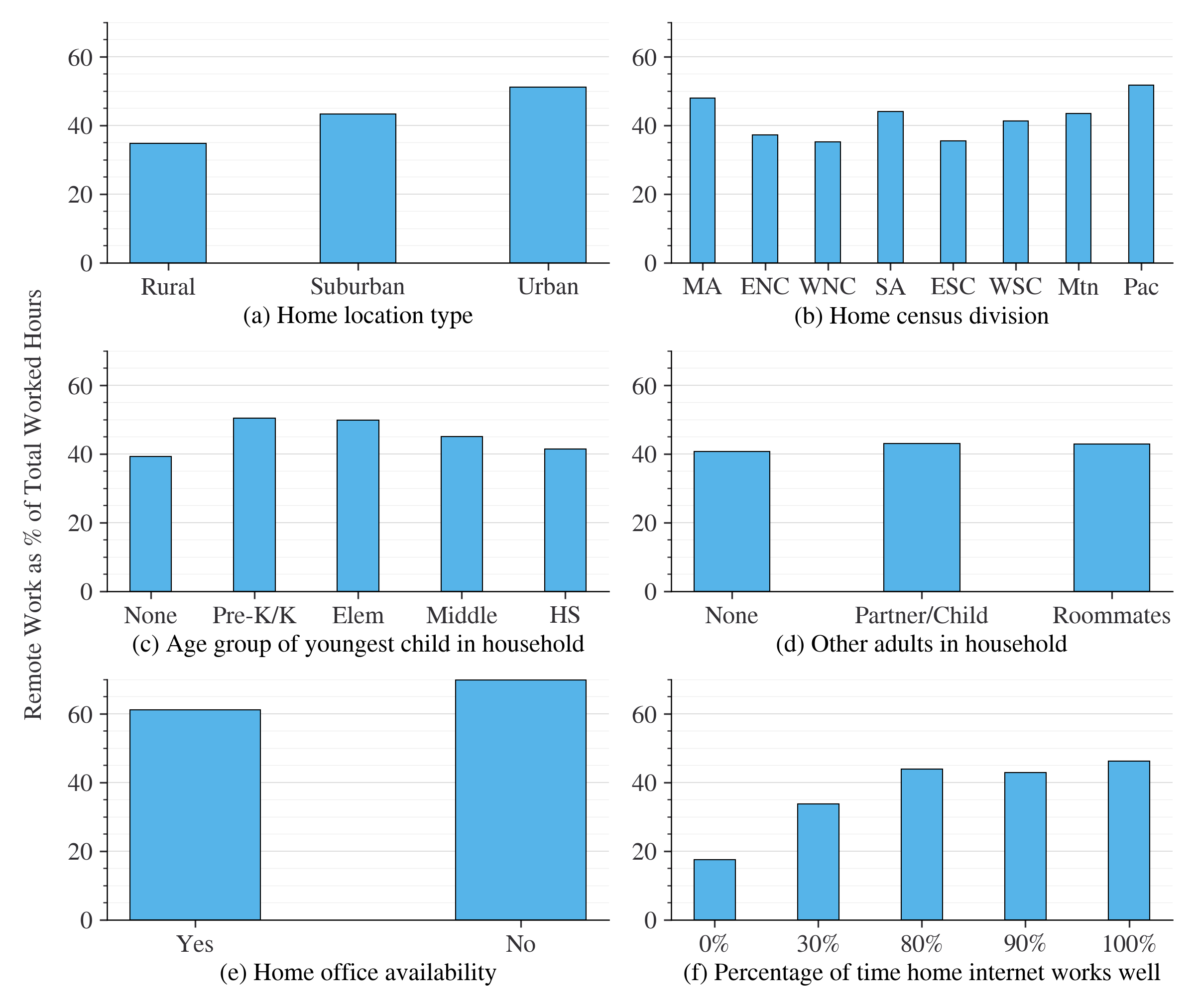}
    \begin{minipage}{0.84\textwidth}
    \footnotesize (a): Aug 2020 - Jan 2023, N = 100,572; (b): May 2020 - Jan 2023, N = 108,801; (c): Jul 2020 - Oct 2022, N = 91,809; (d): Jul 2020 - Jan 2023, N = 101,045; (e): May 2020 - Dec 2021, N = 39,873; (f): May 2020 - Jan 2023, N = 65,705. 
    \end{minipage}
    \caption{Current remote work share by household characteristics}
    \label{fig:current_household}
\end{figure}

Current remote work shares by household characteristic are shown in Figure~\ref{fig:current_household}. 
Urban dwellers are much more likely to be working remotely. 
There is significant variation by census division, with Mid-Atlantic, South Atlantic, Mountain and Pacific engaged in the highest amount of remote work.
Remote work shares are also differentiated by children's age and internet quality in an intuitive sense. 
Those without a home office are actually more likely to prefer remote work, but that trend is thought to be a result of home offices being correlated with other characteristics such as age and home location. 

\begin{figure}[!h]
    \centering
    \includegraphics[width=\textwidth]{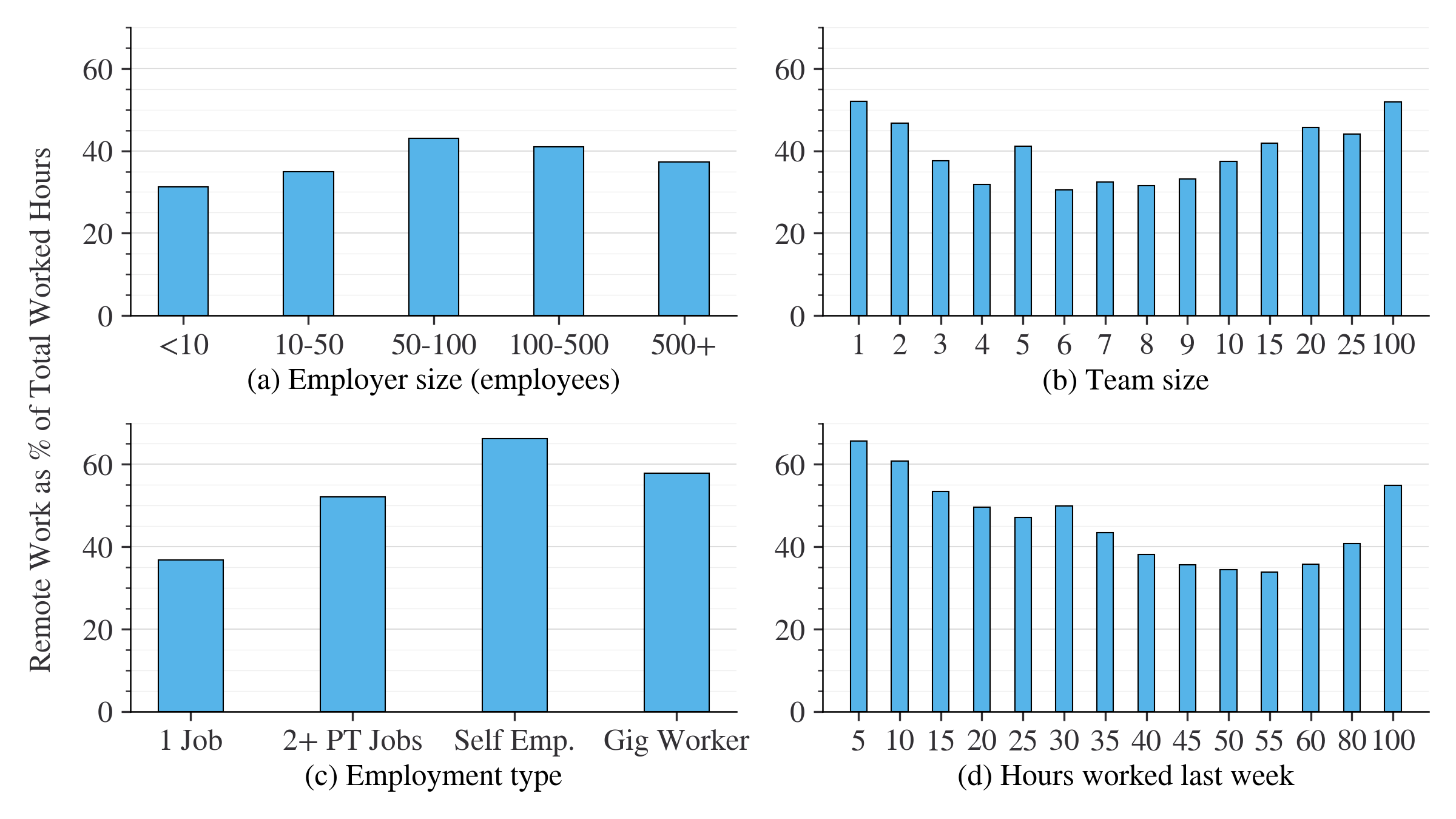}
    \begin{minipage}{0.84\textwidth}
    \footnotesize (a): Aug 2021 - Jan 2023, N = 65,730; (b): Sep 2021 - Oct 2021, N = 6,028; (c): Feb 2021 - Jan 2023, N = 89,886; (d): Jul 2020 - Jan 2023, N = 105,873.    \end{minipage}
    \caption{Current remote work share by employment characteristics}
    \label{fig:current_employment}
\end{figure}

Current remote work shares by employment characteristics are shown in Figure~\ref{fig:current_employment}. 
Team size and number of hours worked both appear to have higher remote work shares among people with the lowest and highest response values. 
There is limited variation in current remote work by employer size.

\begin{figure}[!h]
    \centering
    \includegraphics[width=\textwidth]{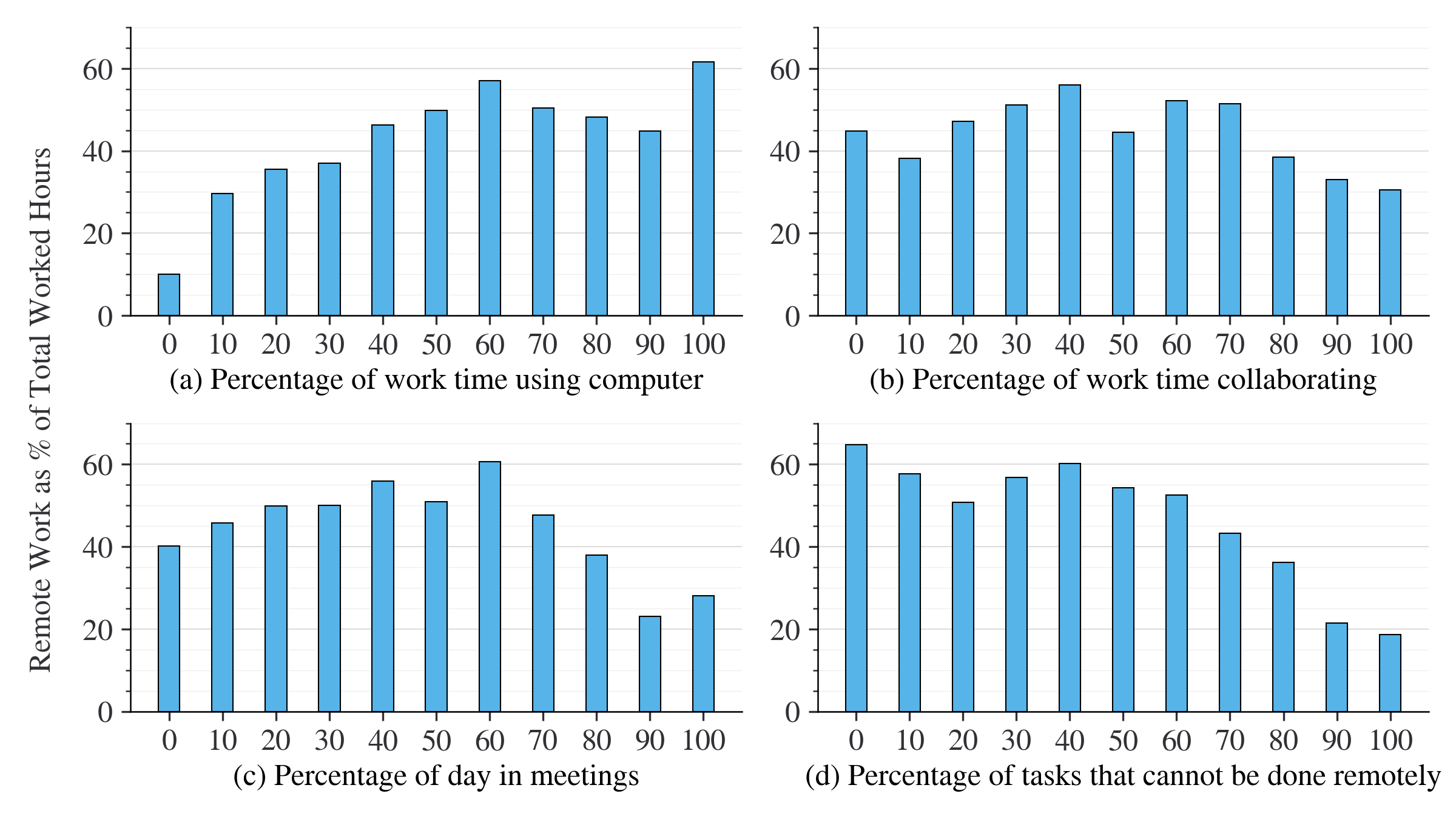}
    \begin{minipage}{0.84\textwidth}
    \footnotesize (a): Apr 2022 - Sep 2022, N = 22,804; (b): Sep 2021 - Oct 2021, N = 6,659; (c): May 2022 - Jun 2022, N = 7,416; (d): Mar 2022 - Apr 2022, N = 5,780.
    \end{minipage}
    \caption{Current remote work share by task characteristics}
    \label{fig:current_tasks}
\end{figure}

Current remote work shares by job task characteristics are shown in Figure~\ref{fig:current_tasks}. 
As expected, a general trend can be observed that people who use a computer less often and people who spend more time in meetings or engaged in collaborative tasks are less likely to conduct their work remotely. 

\begin{figure}[!h]
    \centering
    \includegraphics[width=\textwidth]{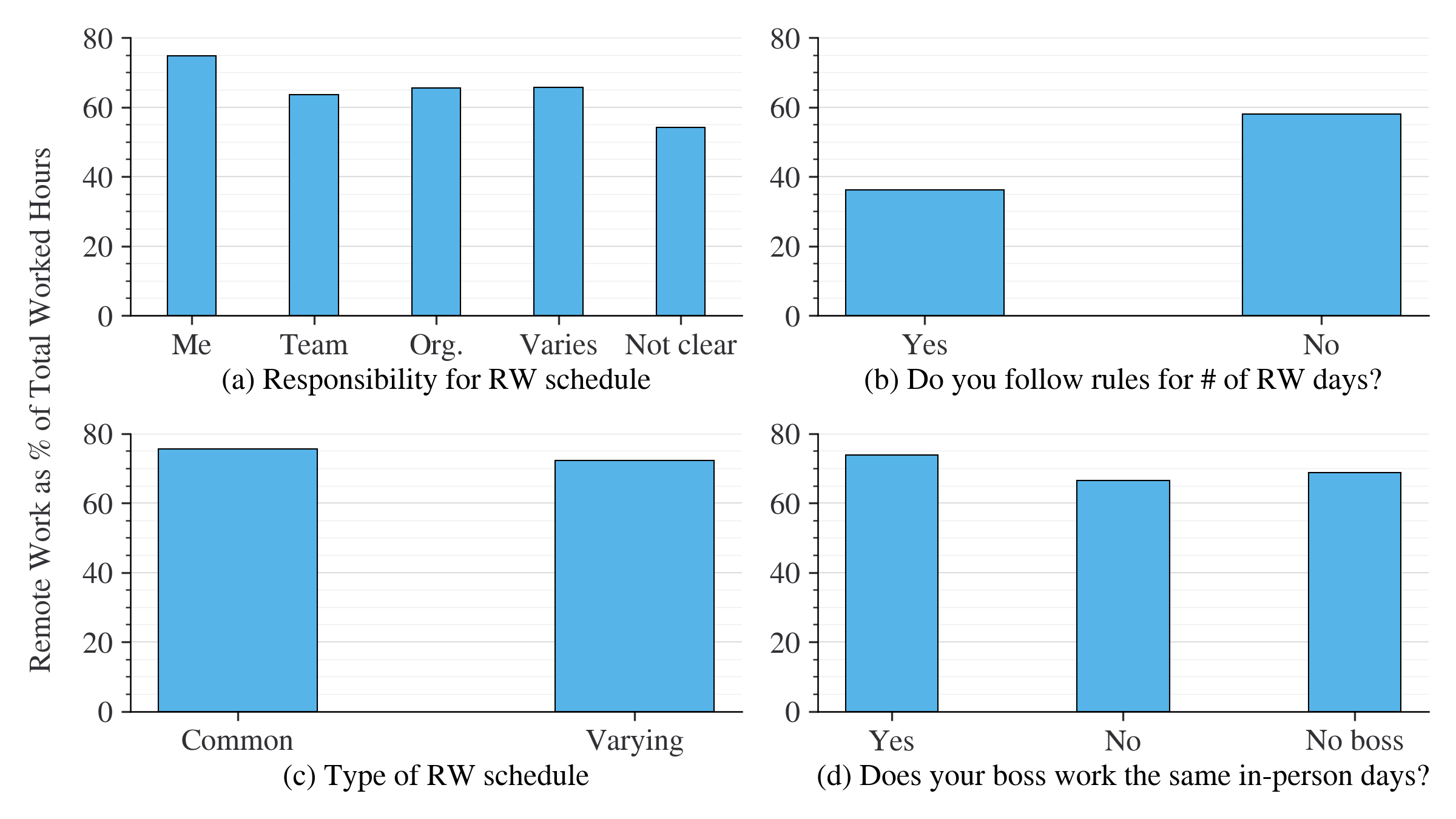}
    \begin{minipage}{0.84\textwidth}
    \footnotesize (a): Jul 2021 - Jan 2022, N = 2,896; (b): May 2022 - Nov 2022, N = 29,657; (c): Jan 2022 - Sep 2022, N = 6,681; (d): Oct 2021 - Sep 2022, N = 10,064.
    \end{minipage}
    \caption{Current remote work share by remote work policy}
    \label{fig:current_wfhpolicy}
\end{figure}

Current remote work shares by remote work policy types are shown in Figure~\ref{fig:current_wfhpolicy}. 
People who get to set their own remote work schedule are generally conducting more remote work, as are those who choose not to follow their employer's guidelines for number of remote work days.
The difference in current remote work share between those whose employers set a common remote work schedule and those whose remote work schedule varies is relatively small. 

\begin{figure}[!h]
    \centering
    \includegraphics[width=\textwidth]{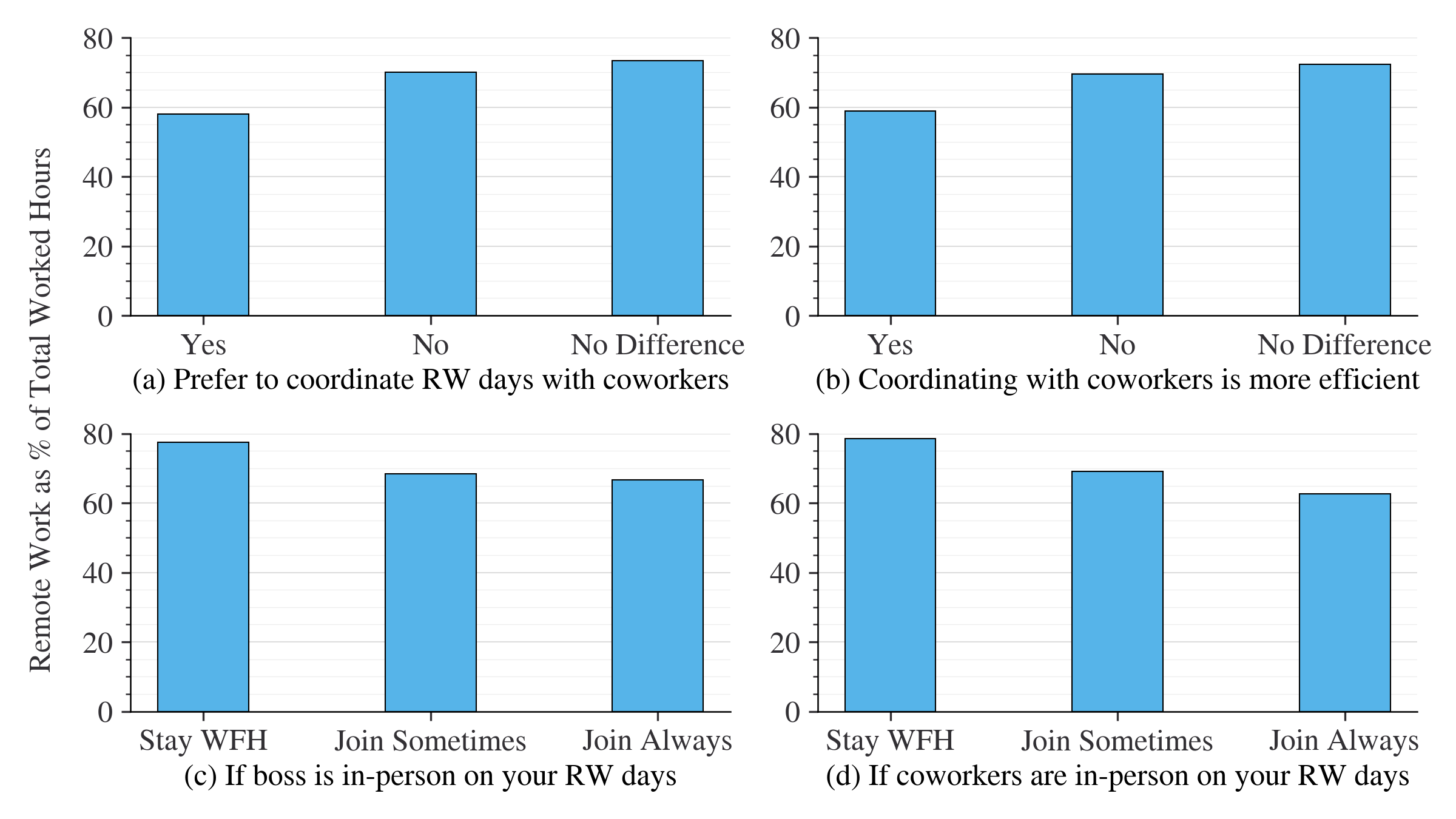}
    \begin{minipage}{0.84\textwidth}
    \footnotesize (a): Feb 2022, N = 2,627; (b): Same as (a); (c): Oct 2021 - Sep 2022, N = 7,416; (d): Oct 2021 - Sep 2022, N = 7,416.
    \end{minipage}
    \caption{Current remote work share by attitudes towards coordinating with colleagues}
    \label{fig:current_coordination}
\end{figure}

Current remote work shares by attitudes towards remote work coordination with colleagues are shown in Figure~\ref{fig:current_coordination}. 
Those who are less interested in coordination across all four questions generally conduct more remote work than those who value coordination. 

\begin{figure}[!h]
    \centering
    \includegraphics[width=\textwidth]{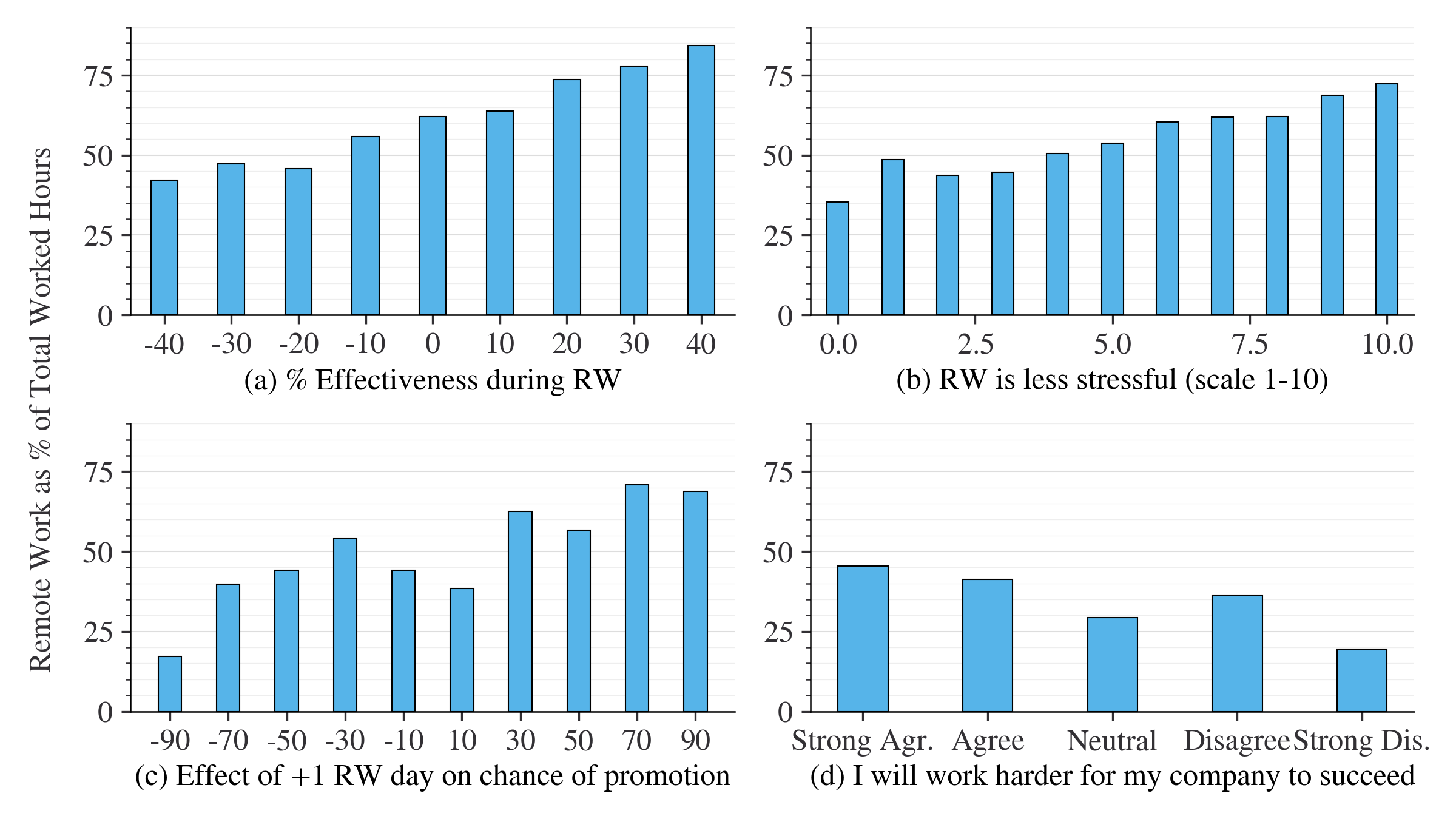}
    \begin{minipage}{0.84\textwidth}
    \footnotesize (a): Jul 2020 - Jan 2023, N = 76,685 (b): Oct 2021, N = 2,622; (c): May 2021 - Aug 2022, N = 5,641; (d): Jul 2022, N = 2,743.
    \end{minipage}
    \caption{Current remote work share by attitudes towards remote work}
    \label{fig:current_wfhattitudes}
\end{figure}

Current remote work shares by attitudes towards remote work more generally are shown in Figure~\ref{fig:current_wfhattitudes}. 
Those who find remote work to increase their effectiveness or less stressful are broadly more likely to be conducting higher levels of remote work. 
Similarly, people who think remote work increases their chances of promotion have higher levels of remote work. 
Results for the survey question about whether respondents are willing to work harder than expected to help their organization succeed are somewhat mixed.
As with all questions, the effects are likely to be bi-directional, so those who are allowed to do more remote work may be perceive the additional remote work as a personal benefit and therefore have a stronger positive attitude towards their employer. 

\begin{figure}[!h]
    \centering
    \includegraphics[width=\textwidth]{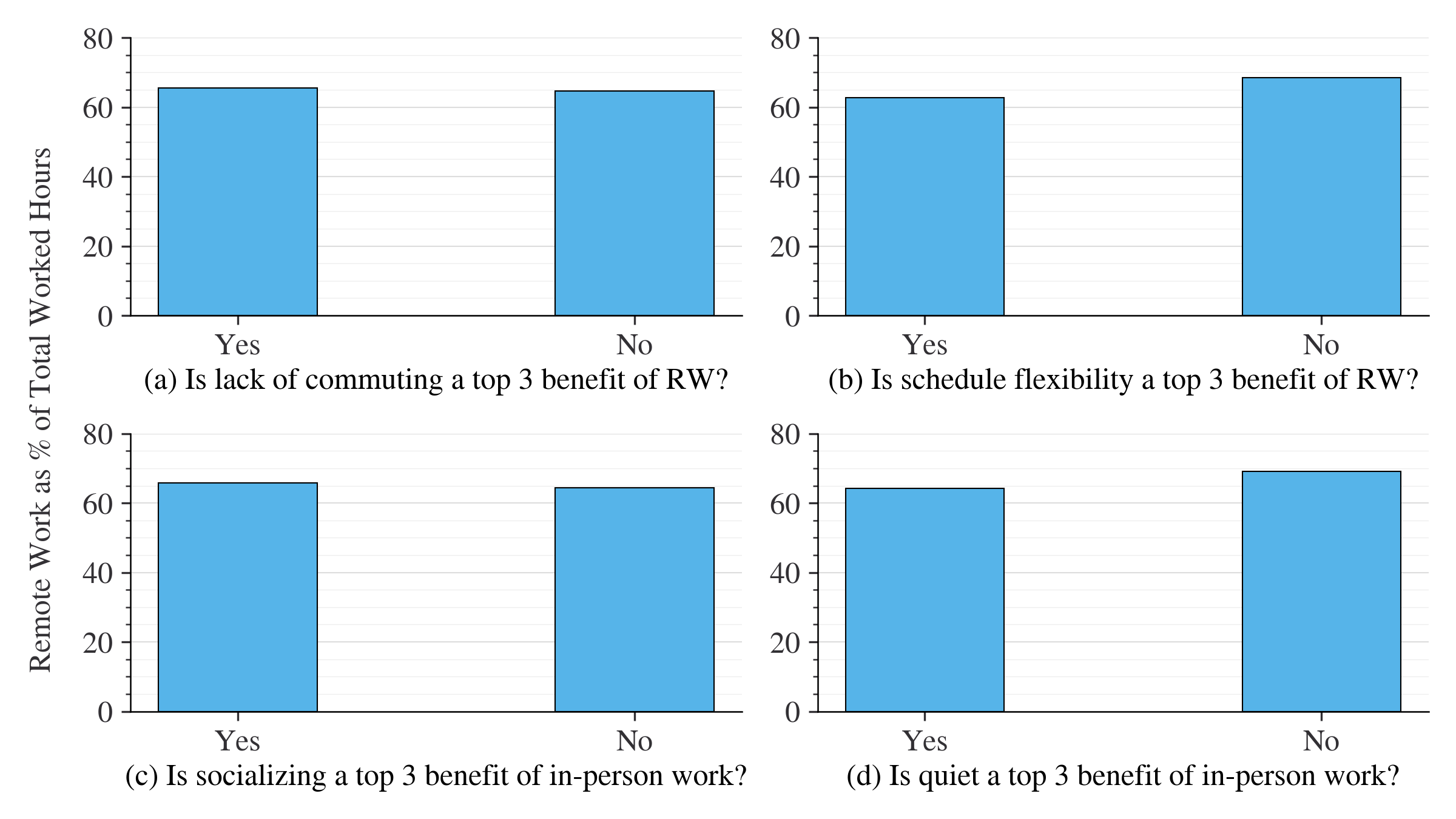}
    \begin{minipage}{0.84\textwidth}
    \footnotesize All: Feb 2022 - Sep 2022, N = 24,685.
    \end{minipage}
    \caption{Current remote work share by perceived benefits of remote work}
    \label{fig:current_benefits}
\end{figure}

Current remote work shares by the perceived benefits of remote work are shown in Figure~\ref{fig:current_benefits}. 
There is very little variation among response groups with respect to current shares of remote work. 
Greater variation is observed with respect to remote work preferences which are presented in the next section. 

\begin{figure}[!h]
    \centering
    \includegraphics[width=\textwidth]{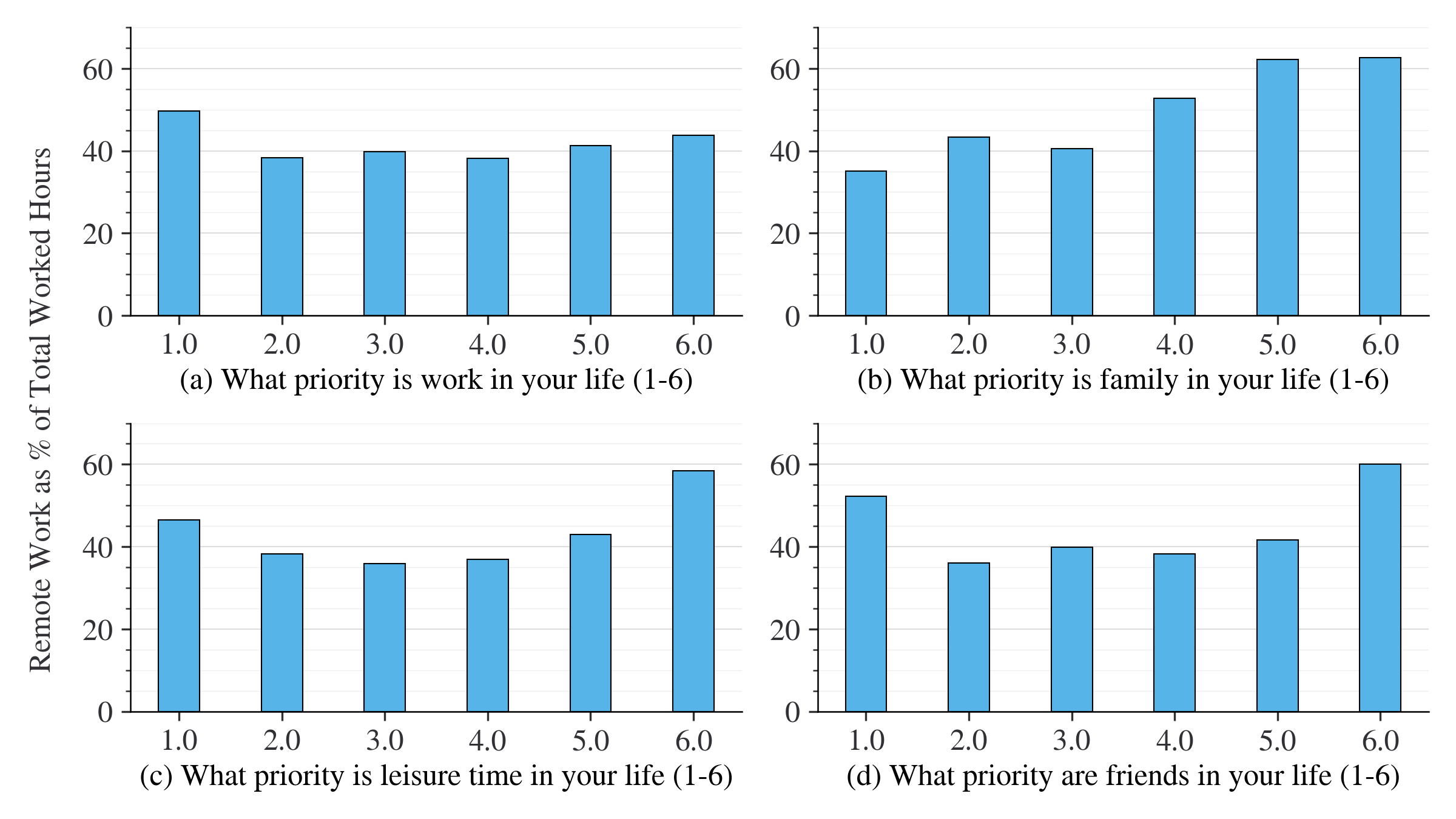}
    \begin{minipage}{0.84\textwidth}
    \footnotesize All: Jul 2022, N = 2,636.
    \end{minipage}
    \caption{Current remote work share by life priorities}
    \label{fig:current_priorities}
\end{figure}

Finally, current remote work shares by life priorities are shown in Figure~\ref{fig:current_priorities}. 
Interestingly, those who value leisure and friends highest and lowest are participating in the most amount of remote work.
A high value of work and low value of family are both positively associated with additional remote work, which is somewhat counter-intuitive. 

\subsection{Preferred remote work shares}

Second, we can explore employee preferences for remote work. 
The results are based on a question asking for respondents' preferences for remote work in future, assuming remote work did not affect their pay. 

\begin{figure}[ht!]
    \centering
    \includegraphics[width=\textwidth]{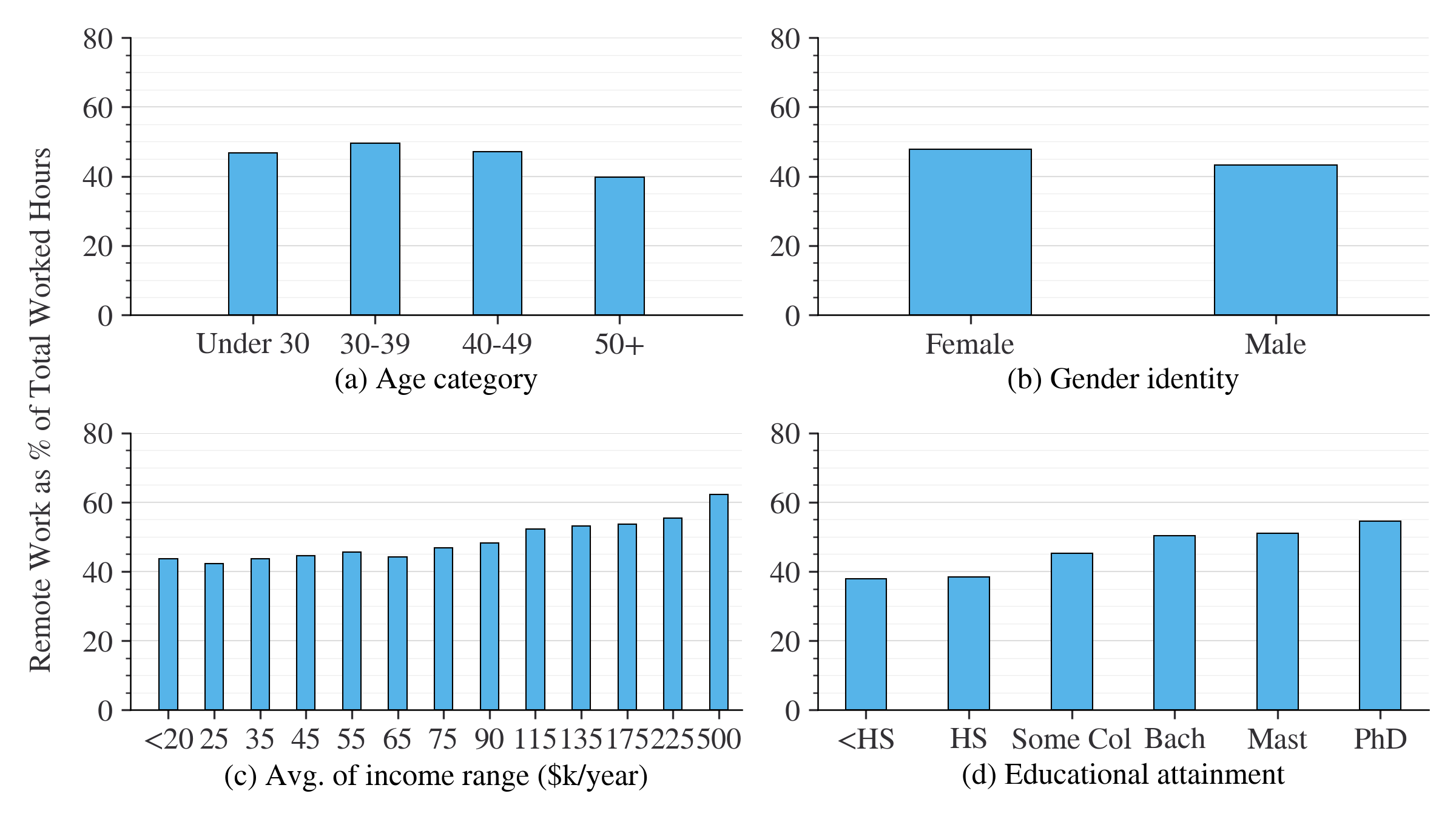}
    \begin{minipage}{0.84\textwidth}
    \footnotesize All: May 2020 - Jan 2023, N = 139,465.
    \end{minipage}
    \caption{Current remote work share by demographic group}
    \label{fig:preferred_demographics}
\end{figure}

Remote work preferences by demographic group are shown in Figure~\ref{fig:preferred_demographics}. 
Age, income and education largely track with observed remote work shares.
Men prefer less remote work than women, despite conducting remote work more often in their current positions. 

\begin{figure}[!ht]
    \centering
    \includegraphics[width=\textwidth]{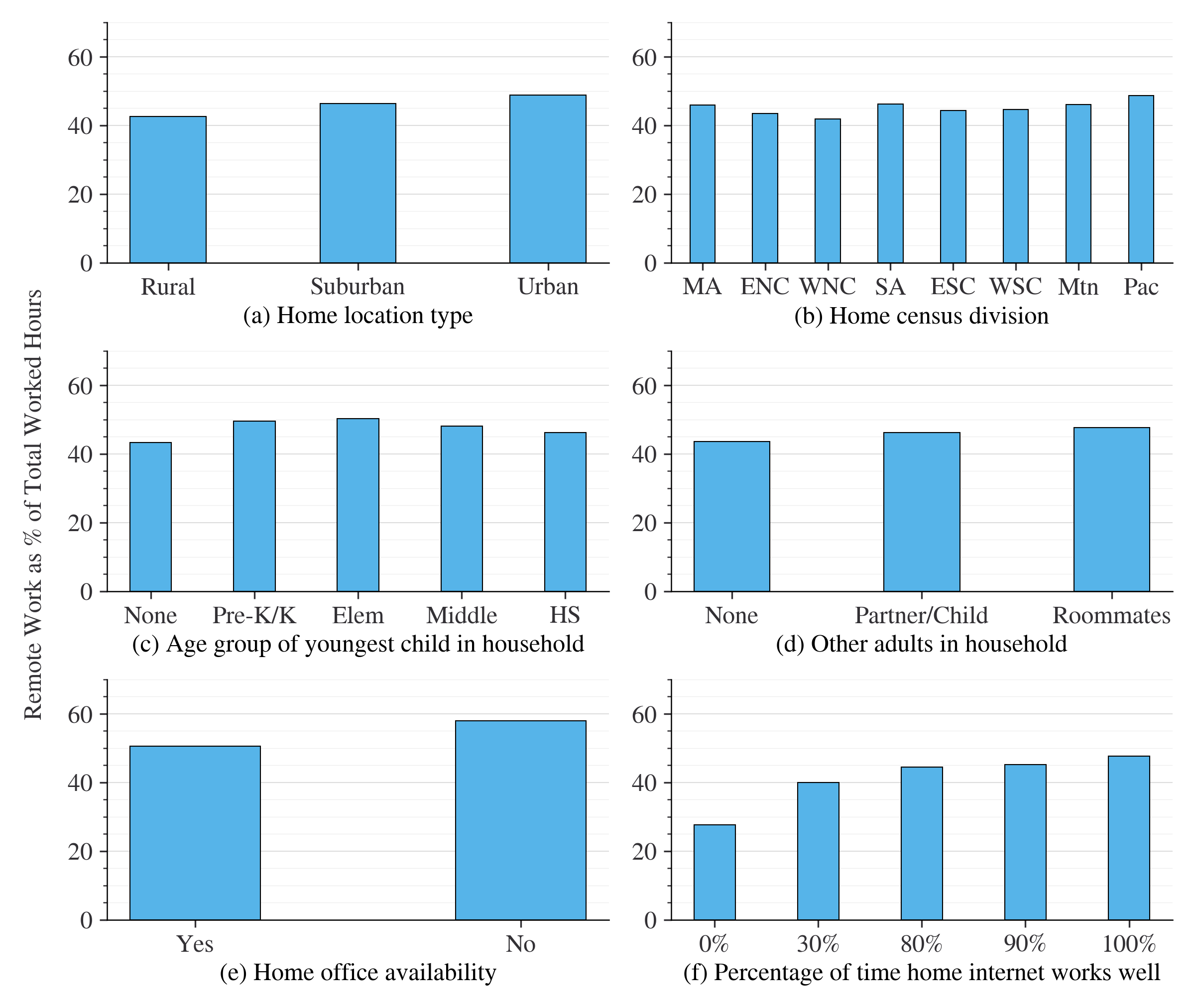}
    \begin{minipage}{0.84\textwidth}
    \footnotesize (a): Aug 2021 - Jan 2023, N = 129,138; (b): May 2020 - Jan 2023, N = 139,460; (c): Jul 2020 - Oct 2022, N = 117,719; (d): Jul 2020 - Jan 2023, N = 129,405; (e): May 2020 - Dec 2021, N = 47,903; (f): May 2020 - Jan 2023, N = 84,336. 
    \end{minipage}
    \caption{Current remote work share by household characteristics}
    \label{fig:preferred_household}
\end{figure}

Remote work preferences by household characteristic are shown in Figure~\ref{fig:preferred_household}. 
Preferences by home location type and census division are quite even. 
The presence of children or other adults in the home also does not seem to be associated with different preferences for remote work.
Internet quality, however, is somewhat correlated with higher preferences. 

\begin{figure}[!h]
    \centering
    \includegraphics[width=\textwidth]{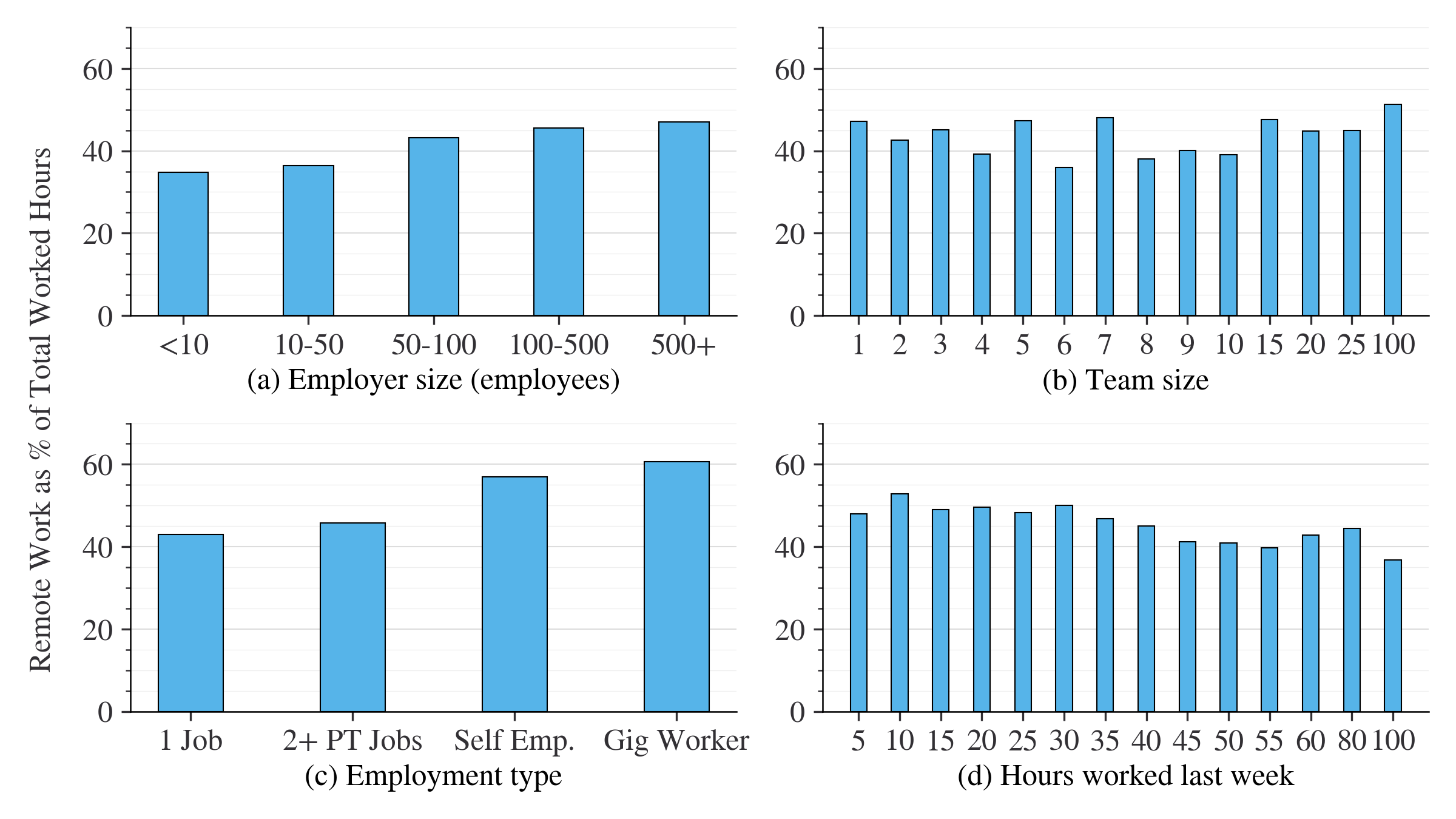}
    \begin{minipage}{0.84\textwidth}
    \footnotesize (a): Aug 2021 - Jan 2023, N = 69,436; (b): Sep 2021 - Oct 2021, N = 6,439; (c): Feb 2021 - Jan 2023, N = 96,001; (d): Jul 2020 - Jan 2023, N = 106,513.    \end{minipage}
    \caption{Current remote work share by employment characteristics}
    \label{fig:preferred_employment}
\end{figure}

Remote work preferences by employment characteristics are shown in Figure~\ref{fig:preferred_employment}. 
Team size and number of hours worked have less variation with respect to preferences than the observed remote work shares shown in Figure~\ref{fig:current_employment}.
Employees of large companies and those who are self-employed or gig workers are more likely to prefer high levels of remote work. 

\begin{figure}[!h]
    \centering
    \includegraphics[width=\textwidth]{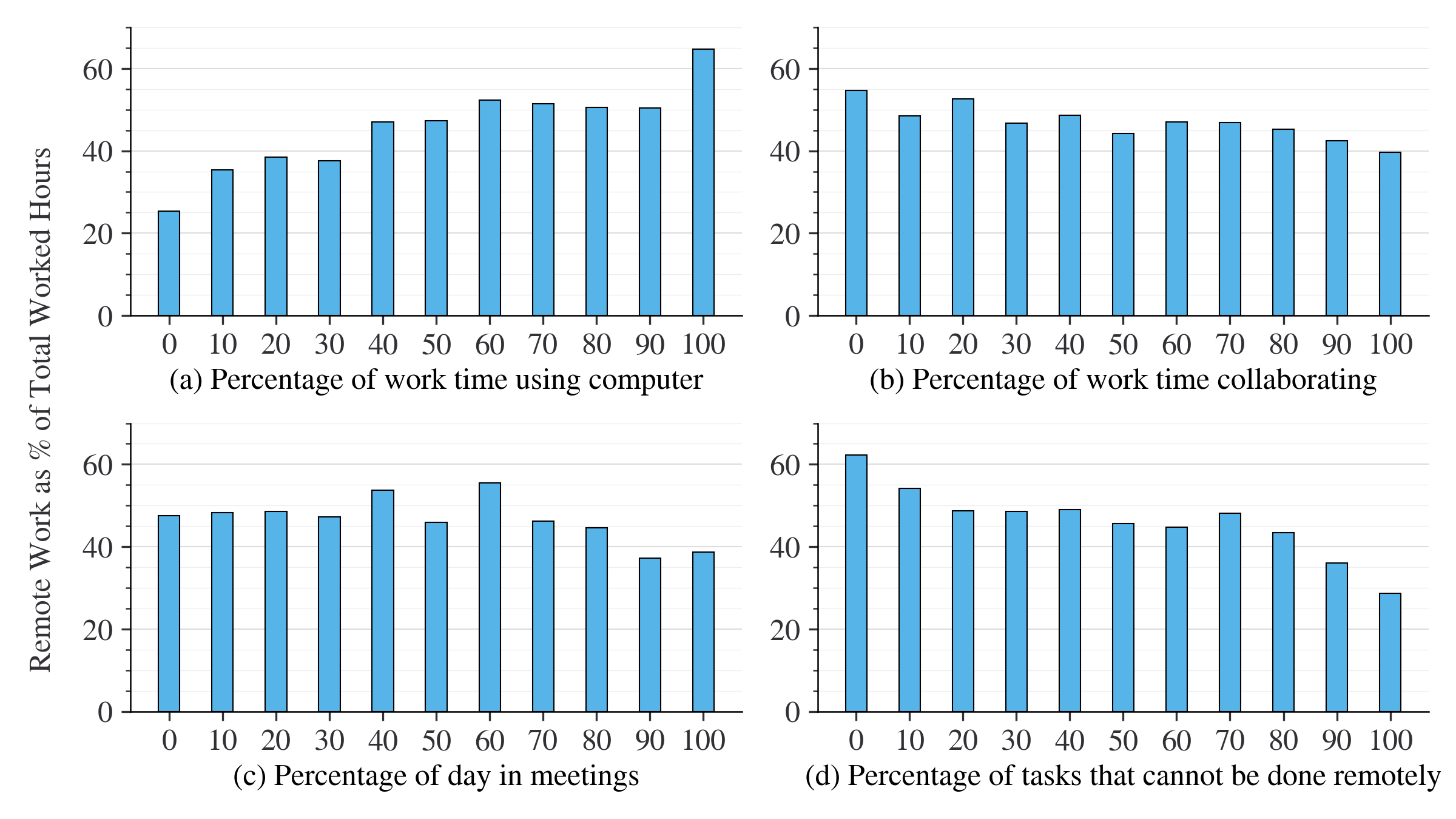}
    \begin{minipage}{0.84\textwidth}
    \footnotesize (a): Apr 2022 - Sep 2022, N = 24,157; (b): Sep 2021 - Oct 2021, N = 7,117; (c): May 2022 - Jun 2022, N = 8,199; (d): Mar 2022 - Apr 2022, N = 6,323.
    \end{minipage}
    \caption{Current remote work share by task characteristics}
    \label{fig:preferred_tasks}
\end{figure}

Remote work preferences by job task characteristics are shown in Figure~\ref{fig:preferred_tasks}. 
Percentage of time using the computer and percentage of tasks that can be done remotely appear to have a stronger correlation with remote work preferences than percentage of work time spent in meetings or on collaborative tasks.
The average person who works 100\% of the time on a computer and all of whose tasks can be done remotely prefers about 60\% remote work, which is high relative to other groups but still includes a substantial amount of in-person work. 

\begin{figure}[!h]
    \centering
    \includegraphics[width=\textwidth]{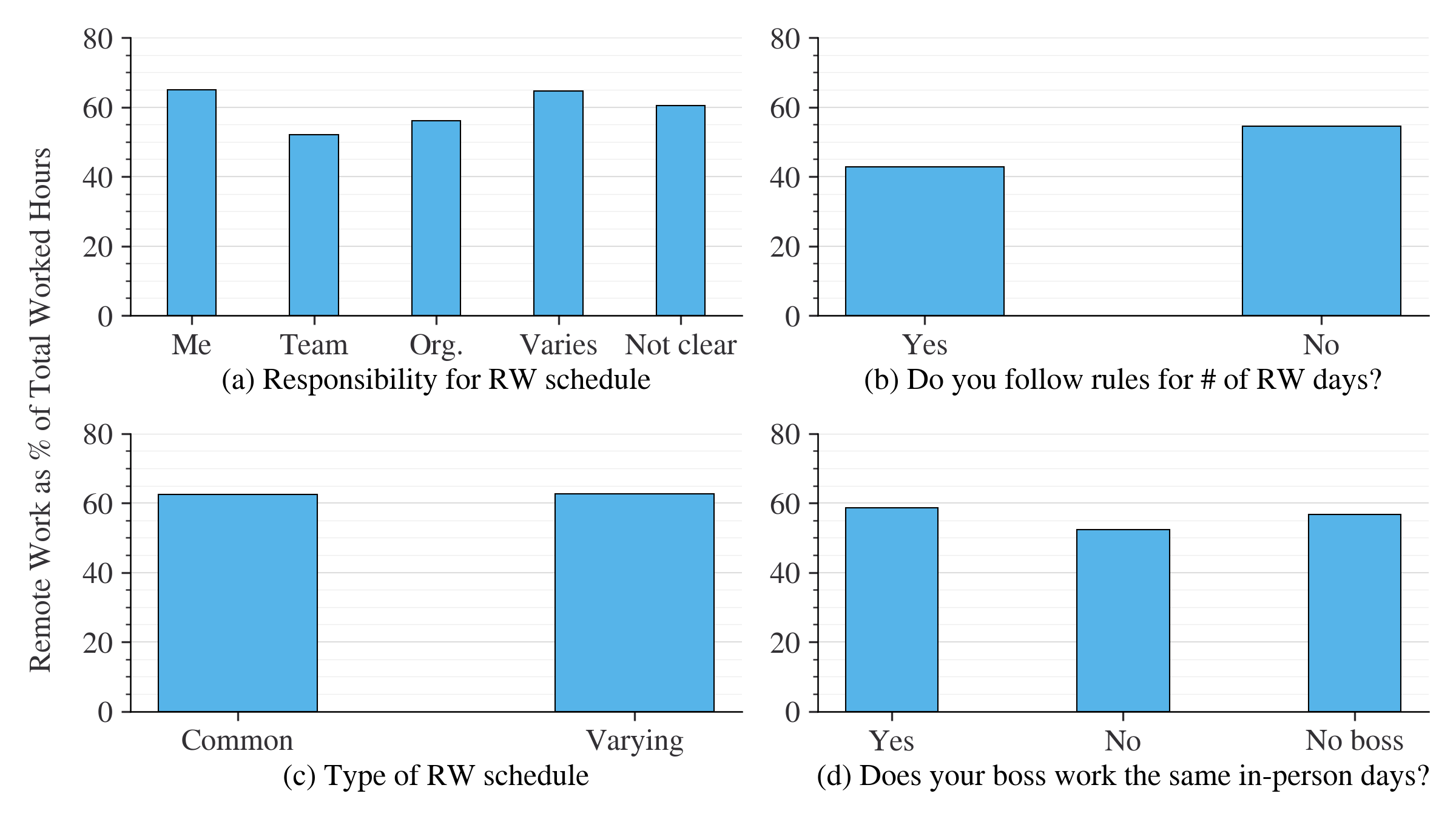}
    \begin{minipage}{0.84\textwidth}
    \footnotesize (a): Jul 2021 - Jan 2022, N = 3,253; (b): May 2022 - Nov 2022, N = 29,657; (c): Jan 2022 - Sep 2022, N = 6,963; (d): Oct 2021 - Sep 2022, N = 10,673.
    \end{minipage}
    \caption{Current remote work share by remote work policy}
    \label{fig:preferred_wfhpolicy}
\end{figure}

Remote work preferences by remote work policy types are shown in Figure~\ref{fig:preferred_wfhpolicy}. 
People who prefer higher levels of remote work are those who set their own remote work schedule and those who do not follow their employer's guidelines for remote work. 
Working the same in-person days as one's boss is associated with slightly greater preferences for remote work, suggesting that face-to-face encounters during in-person days allows people to feel more comfortable with additional remote work time. 

\begin{figure}[!h]
    \centering
    \includegraphics[width=\textwidth]{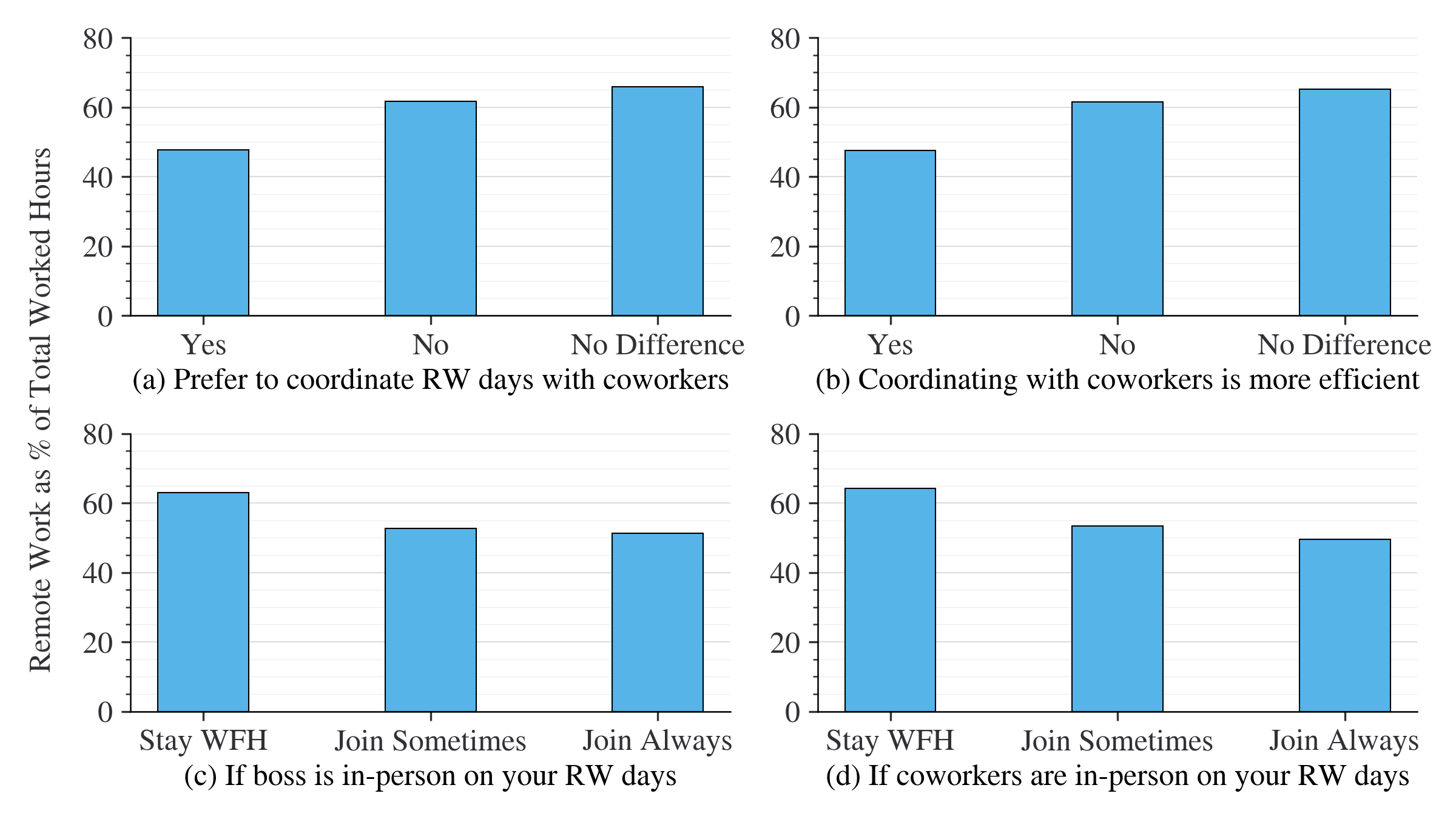}
    \begin{minipage}{0.84\textwidth}
    \footnotesize (a): Feb 2022, N = 3,122; (b): Same as (a); (c): Oct 2021 - Sep 2022, N = 7,796; (d): Oct 2021 - Sep 2022, N = 8,022.
    \end{minipage}
    \caption{Current remote work share by attitudes towards coordinating with colleagues}
    \label{fig:preferred_coordination}
\end{figure}

Remote work preferences by attitudes towards remote work coordination with colleagues are shown in Figure~\ref{fig:preferred_coordination}. 
There is a substantial difference in preferences between those who prefer to coordinate with their colleagues and those who do not (or those who do not perceive any difference). 
This result is intuitive and highlights how attitudes and employment factors are now important to consider when estimating travel preferences.

\begin{figure}[!h]
    \centering
    \includegraphics[width=\textwidth]{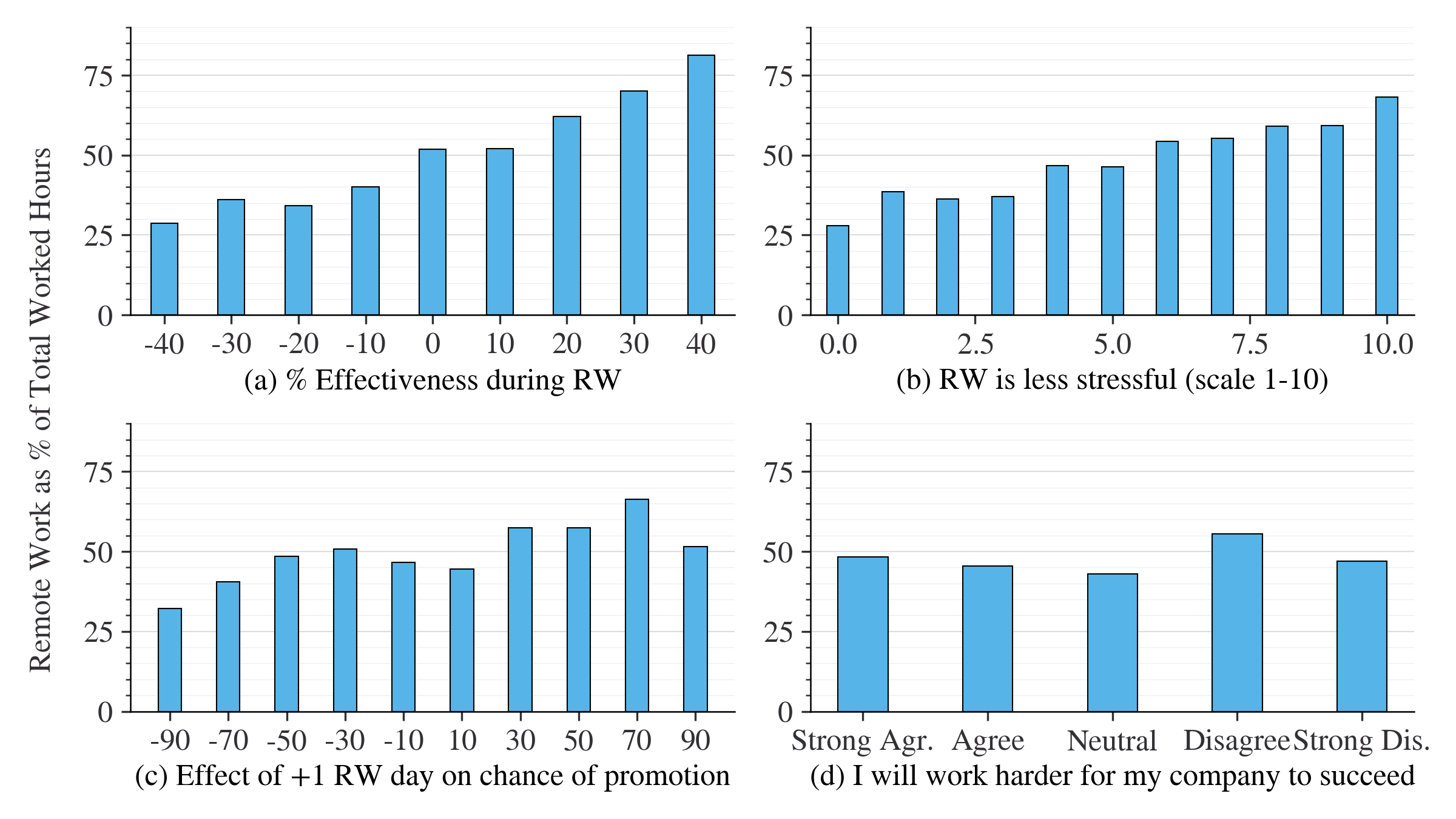}
    \begin{minipage}{0.84\textwidth}
    \footnotesize (a): Jul 2020 - Jan 2023, N = 90,213 (b): Oct 2021, N = 2,780 (c): May 2021 - Aug 2022, N = 5,843; (d): Jul 2022, N = 3,452.
    \end{minipage}
    \caption{Current remote work share by attitudes towards remote work}
    \label{fig:preferred_wfhattitudes}
\end{figure}

Remote work preferences by attitudes towards remote work more generally are shown in Figure~\ref{fig:preferred_wfhattitudes}. 
Unsurprisingly, those who find remote work to increase their effectiveness or less stressful prefer more remote work. 
Remote work preferences in relation to the question about the degree to which remote work affects chances of a promotion have a bi-modal distribution.
People who expect remote work to increase their chances of a promotion have an understandable preference for higher levels of remote work. 
The second peak is those who expect that remote work lowers their chance of a promotion by 30 to 50 percent, but prefer remote work regardless.
That peak might also be reflective of the dynamics shown by the ``disagree'' respondents in Figure~\ref{fig:preferred_wfhattitudes}(d), who prefer remote work at higher levels and are not particularly interested in working harder in order for their organization to succeed.

\begin{figure}[!h]
    \centering
    \includegraphics[width=\textwidth]{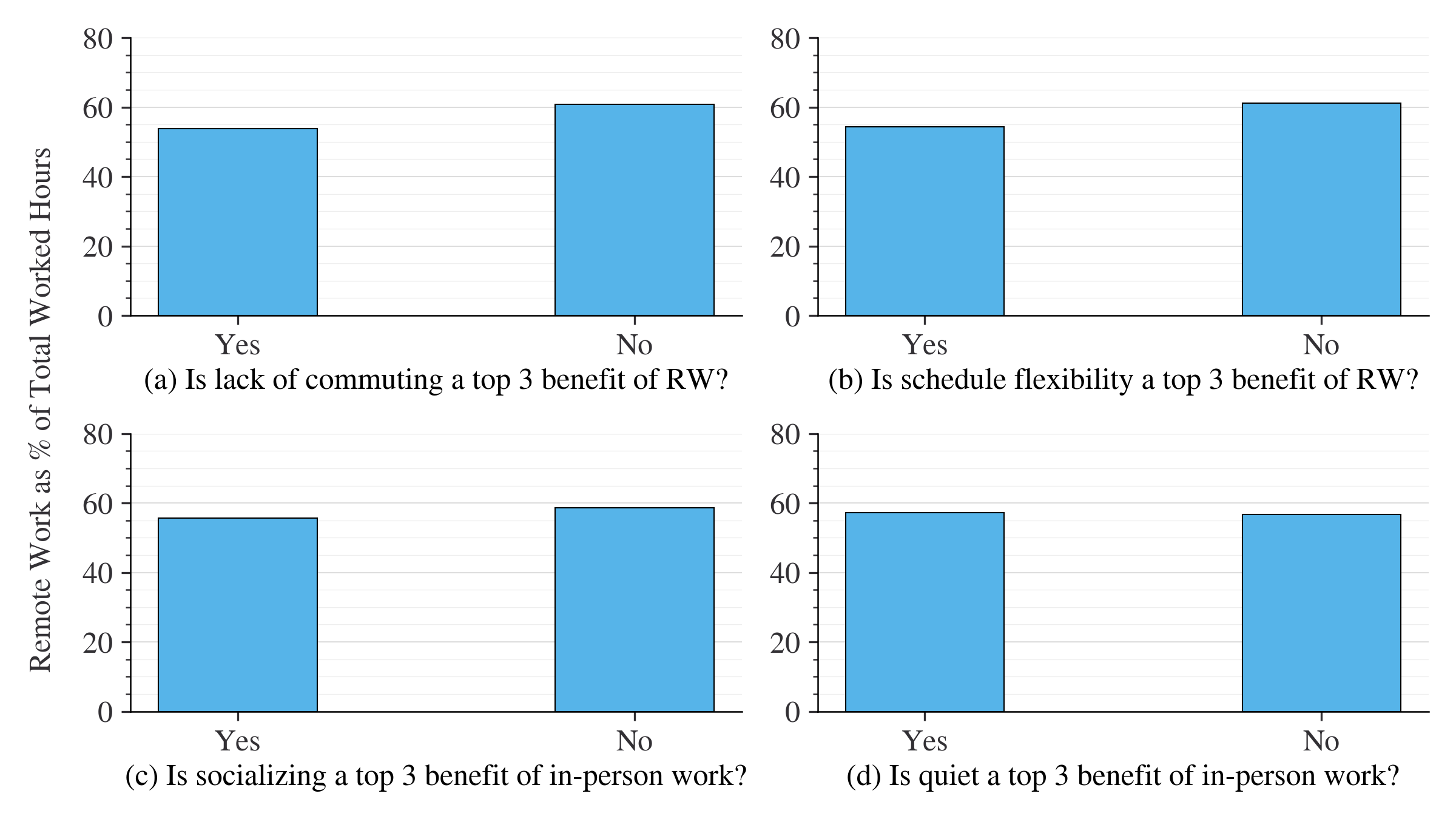}
    \begin{minipage}{0.84\textwidth}
    \footnotesize All: Feb 2022 - Sep 2022, N = 28,836.
    \end{minipage}
    \caption{Current remote work share by perceived benefits of remote work}
    \label{fig:preferred_benefits}
\end{figure}

Remote work preferences by the perceived benefits of remote work are shown in Figure~\ref{fig:preferred_benefits}. 
People who enjoy remote work due to the commuting and schedule flexibility benefits are actually less likely to prefer remote work. 
Those who enjoy the social benefits of in-person work prefer slightly less remote work than those who do not. 

\begin{figure}[!h]
    \centering
    \includegraphics[width=\textwidth]{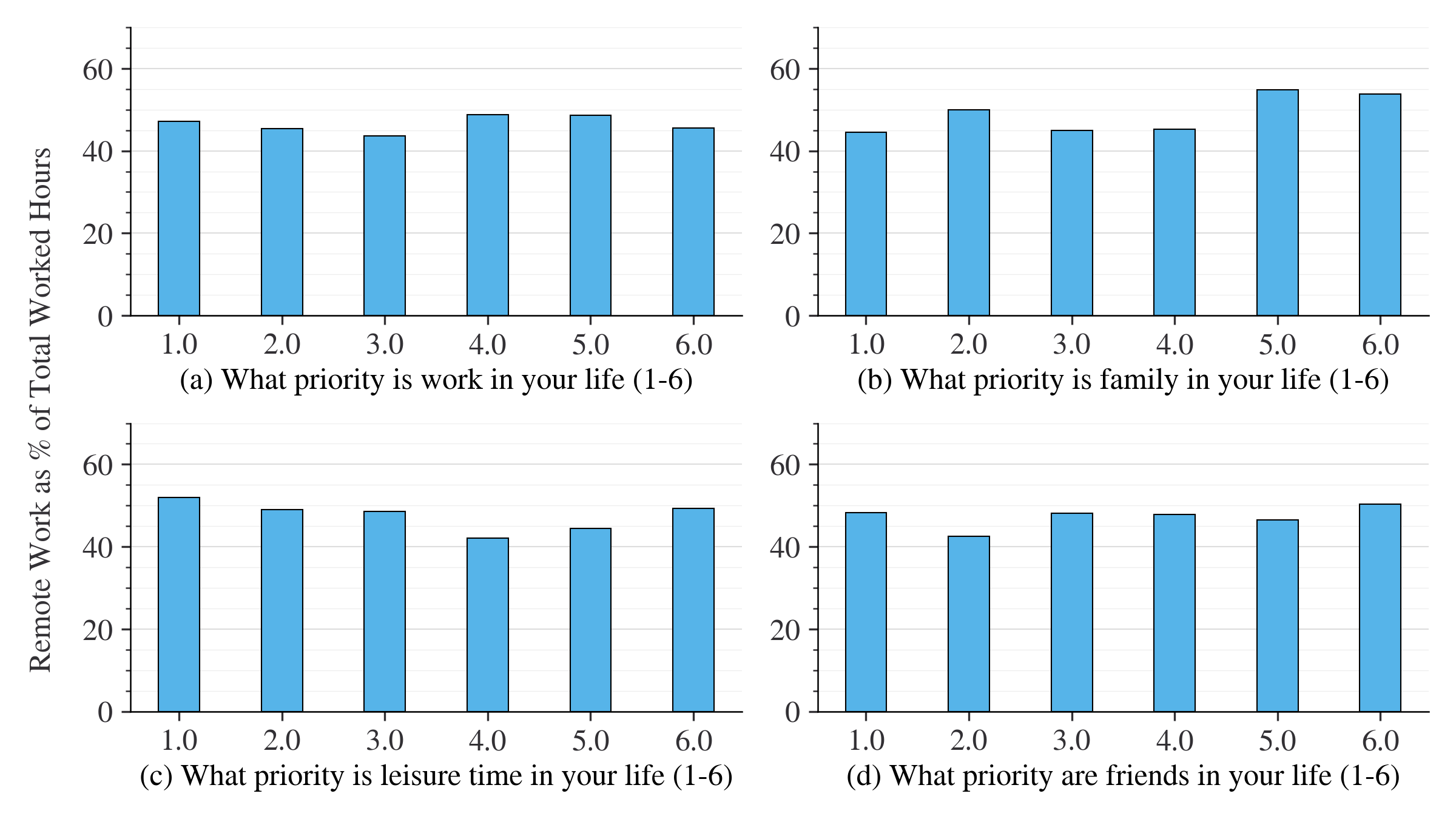}
    \begin{minipage}{0.84\textwidth}
    \footnotesize All: Jul 2022, N = 3,309.
    \end{minipage}
    \caption{Current remote work share by life priorities}
    \label{fig:preferred_priorities}
\end{figure}

Finally, preferred remote work shares by life priorities are shown in Figure~\ref{fig:preferred_priorities}. 
Unlike the observed remote work share results shown in Figure~\ref{fig:current_priorities}, preferences are relatively constant across each of the life priority questions. 
This is somewhat of a surprising result that suggest remote work is not widely perceived as a means towards fulfilling any of the life priorities presented in the survey.

\subsection{Employer planned remote work shares}

Lastly, we can explore employee's perceptions of their employer's plans for remote work in the future. 
The aggregate results are generally lower than the results for preferences, confirming that there is a gap between employer plans for remote work and employee preferences. 

\begin{figure}[ht!]
    \centering
    \includegraphics[width=\textwidth]{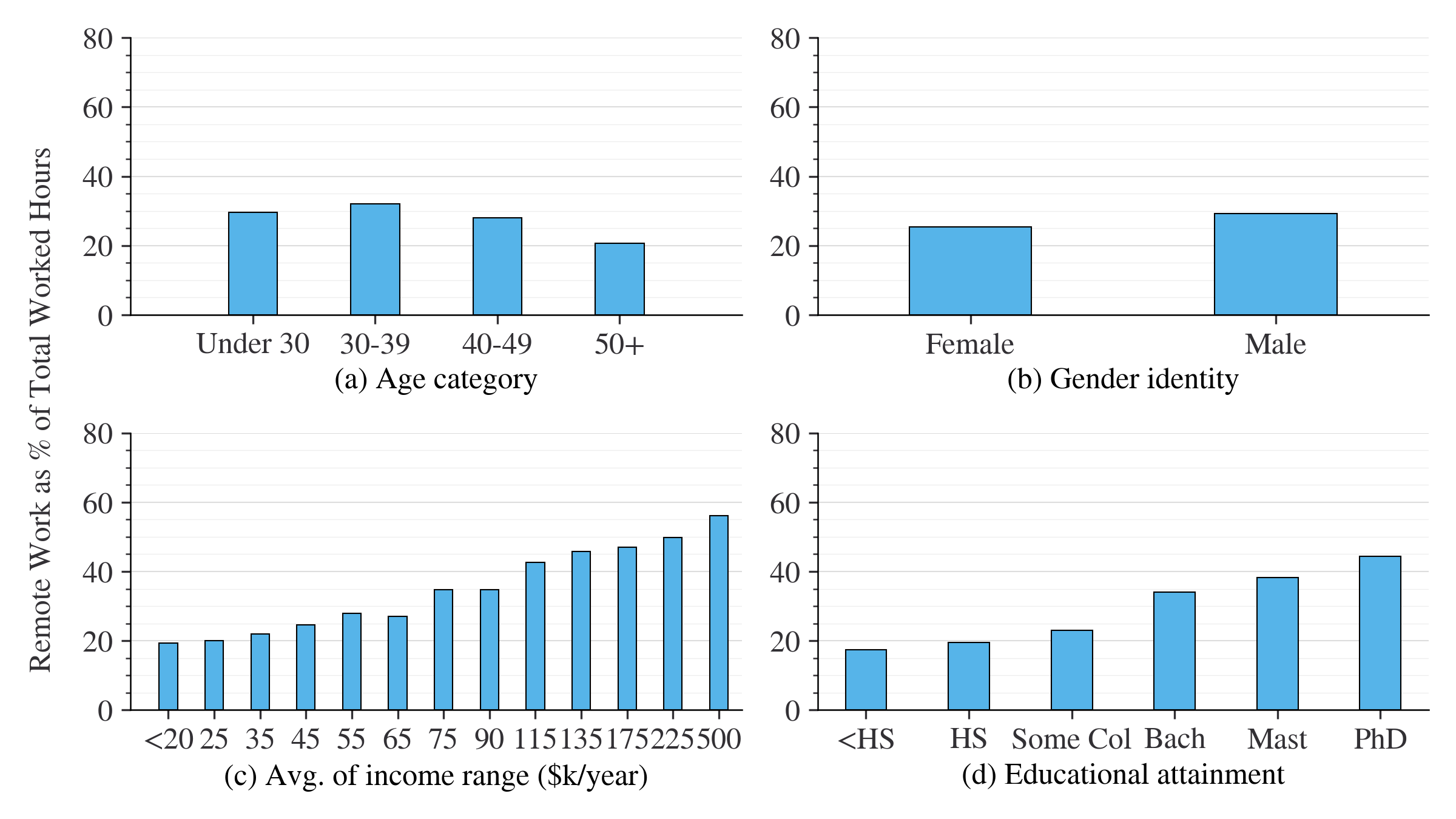}
    \begin{minipage}{0.84\textwidth}
    \footnotesize All: May 2020 - Jan 2023, N = 139,465.
    \end{minipage}
    \caption{Current remote work share by demographic group}
    \label{fig:planned_demographics}
\end{figure}

Remote work plans by demographic group are shown in Figure~\ref{fig:planned_demographics}. 
The trends are similar to the current remote work share results in Figure~\ref{fig:current_demographics}, albeit at a lower overall level of remote work than currently worked. 
People in the lowest income categories, those with a high school education or less, and people aged 50 and above expect their employers to allow them to work remotely for 20\% of their hours on average, equivalent to one day per week for a typical full time job. 

\begin{figure}[!ht]
    \centering
    \includegraphics[width=\textwidth]{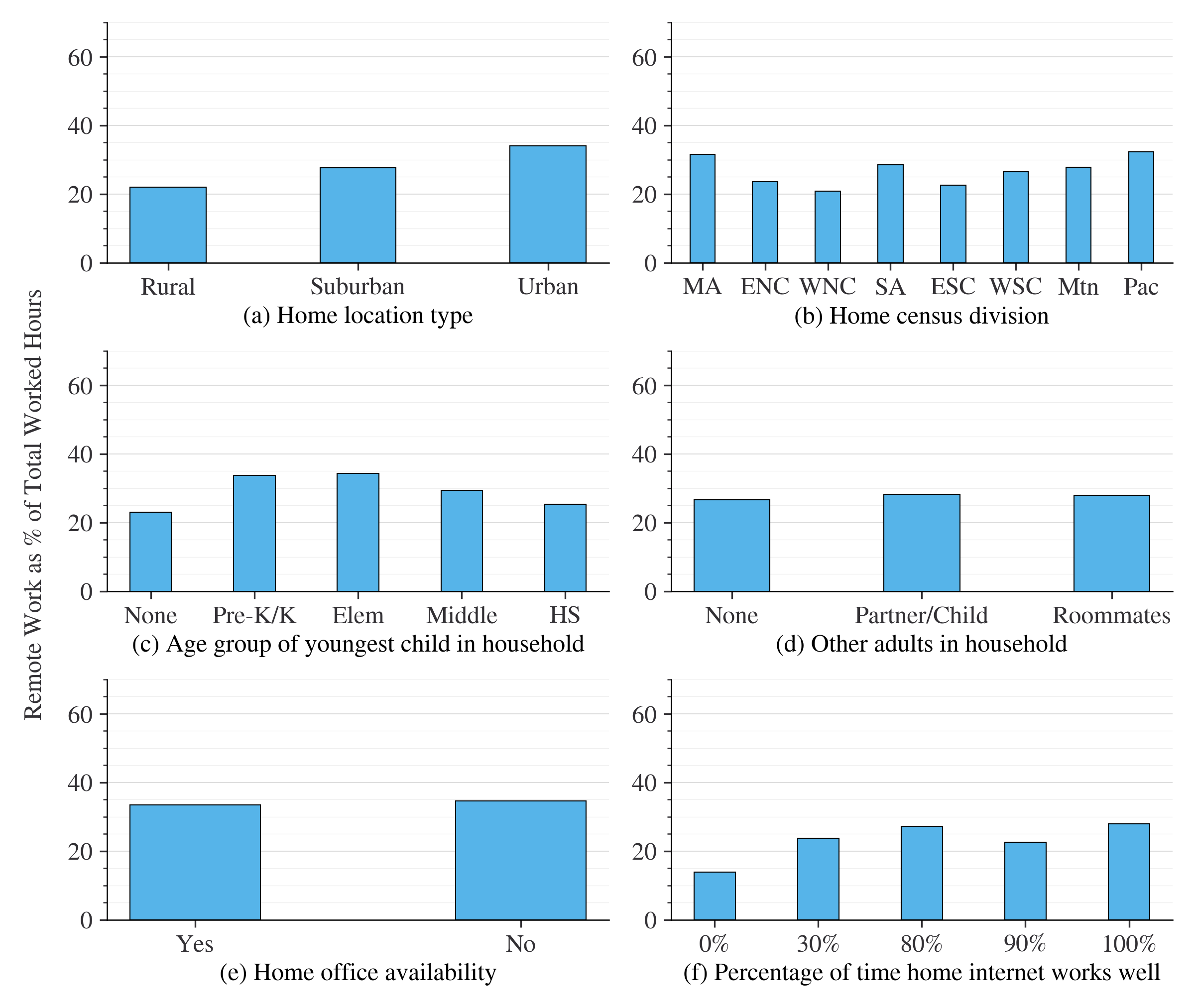}
    \begin{minipage}{0.84\textwidth}
    \footnotesize (a): Aug 2021 - Jan 2023, N = 129,138; (b): May 2020 - Jan 2023, N = 139,460; (c): Jul 2020 - Oct 2022, N = 117,719; (d): Jul 2020 - Jan 2023, N = 129,405; (e): May 2020 - Dec 2021, N = 47,903; (f): May 2020 - Jan 2023, N = 84,336. 
    \end{minipage}
    \caption{Current remote work share by household characteristics}
    \label{fig:planned_household}
\end{figure}

Employer planned remote work shares by household characteristic are shown in Figure~\ref{fig:planned_household}. 
Home location type and census division appear to have a stronger association with employer plans than with employee preferences.  
The availability of a home office and the presence of other adults in the household do not appear to be correlated with employer plans for remote work, but people with young children expect to be granted more remote work than those who have older children or no children at all. 

\begin{figure}[!h]
    \centering
    \includegraphics[width=\textwidth]{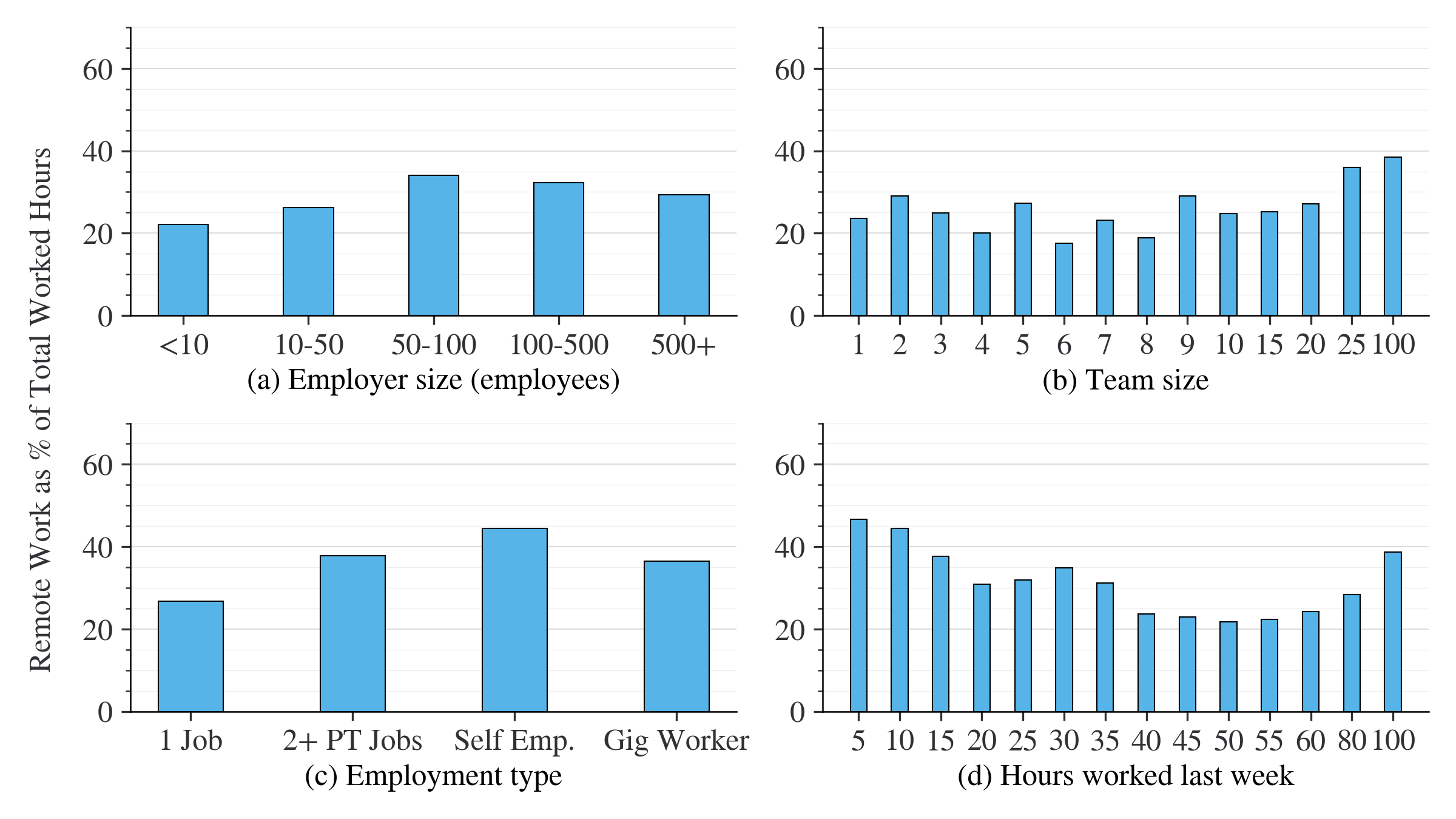}
    \begin{minipage}{0.84\textwidth}
    \footnotesize (a): Aug 2021 - Jan 2023, N = 69,436; (b): Sep 2021 - Oct 2021, N = 6,439; (c): Feb 2021 - Jan 2023, N = 96,001; (d): Jul 2020 - Jan 2023, N = 106,513.    \end{minipage}
    \caption{Current remote work share by employment characteristics}
    \label{fig:planned_employment}
\end{figure}

Employer planned remote work shares by employment characteristics are shown in Figure~\ref{fig:planned_employment}. 
People in very large teams appear to expect higher remote work plans than those with medium-sized teams. 
Interestingly, people who work relatively few hours or very many hours also expect to be granted more remote work by their employers than those who work 40 to 60 hours per week.
There is limited variation in planned remote work by employer size, with medium-sized employers expected to offer more remote work than very large or small employers.

\begin{figure}[!h]
    \centering
    \includegraphics[width=\textwidth]{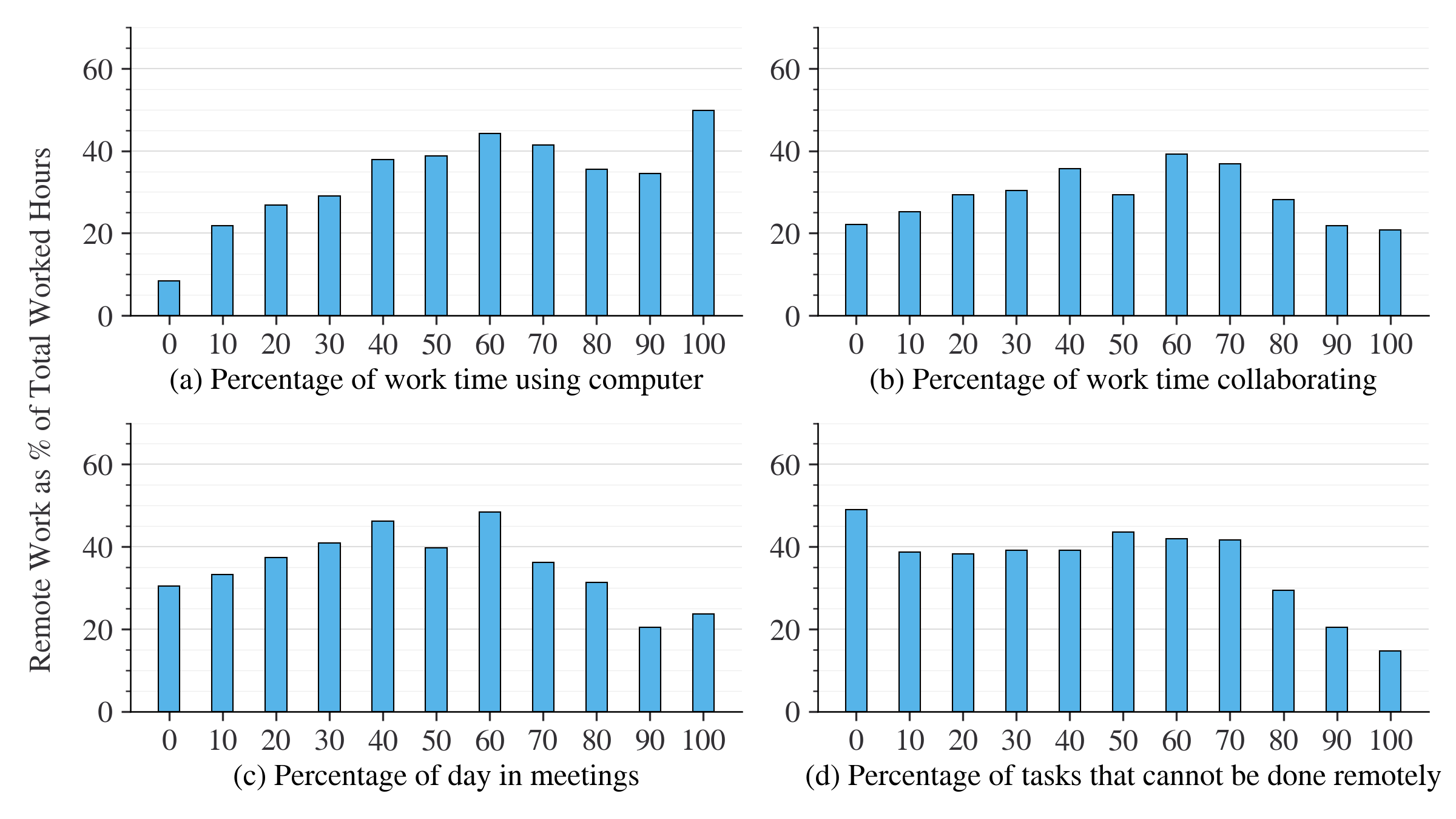}
    \begin{minipage}{0.84\textwidth}
    \footnotesize (a): Apr 2022 - Sep 2022, N = 24,157; (b): Sep 2021 - Oct 2021, N = 7,117; (c): May 2022 - Jun 2022, N = 8,199; (d): Mar 2022 - Apr 2022, N = 6,323.
    \end{minipage}
    \caption{Current remote work share by task characteristics}
    \label{fig:planned_tasks}
\end{figure}

Employer planned remote work shares by job task characteristics are shown in Figure~\ref{fig:planned_tasks}. 
Interestingly, the people whose employers are planning the highest levels of remote work are those whose jobs involve meetings and collaborative tasks 60\% of the time. 
Employer plans also appear to change little for people for whom 10\% to 70\% percent of tasks cannot be done remotely, but plans are quite low above the 70\% threshold. 

\begin{figure}[!h]
    \centering
    \includegraphics[width=\textwidth]{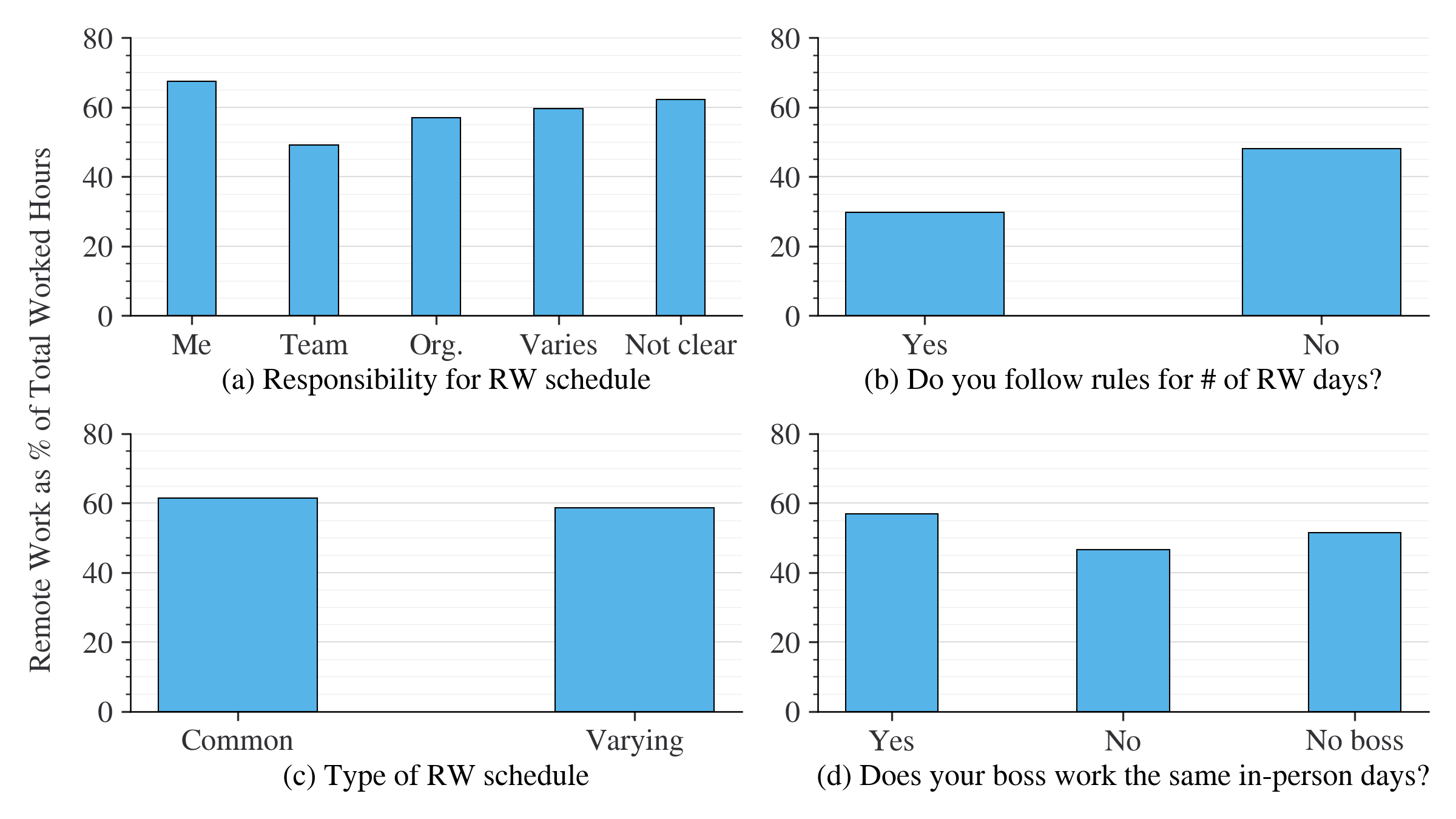}
    \begin{minipage}{0.84\textwidth}
    \footnotesize (a): Jul 2021 - Jan 2022, N = 3,253; (b): May 2022 - Nov 2022, N = 29,657; (c): Jan 2022 - Sep 2022, N = 6,963; (d): Oct 2021 - Sep 2022, N = 10,673.
    \end{minipage}
    \caption{Current remote work share by remote work policy}
    \label{fig:planned_wfhpolicy}
\end{figure}

Employer planned remote work shares by remote work policy types are shown in Figure~\ref{fig:planned_wfhpolicy}. 
It seems reasonable that people who get to set their own remote work schedule are expecting to be afforded greater levels of remote work.
Remote workers who work the same in-person days as their boss also expect to be allowed more remote work in the future.

\begin{figure}[!h]
    \centering
    \includegraphics[width=\textwidth]{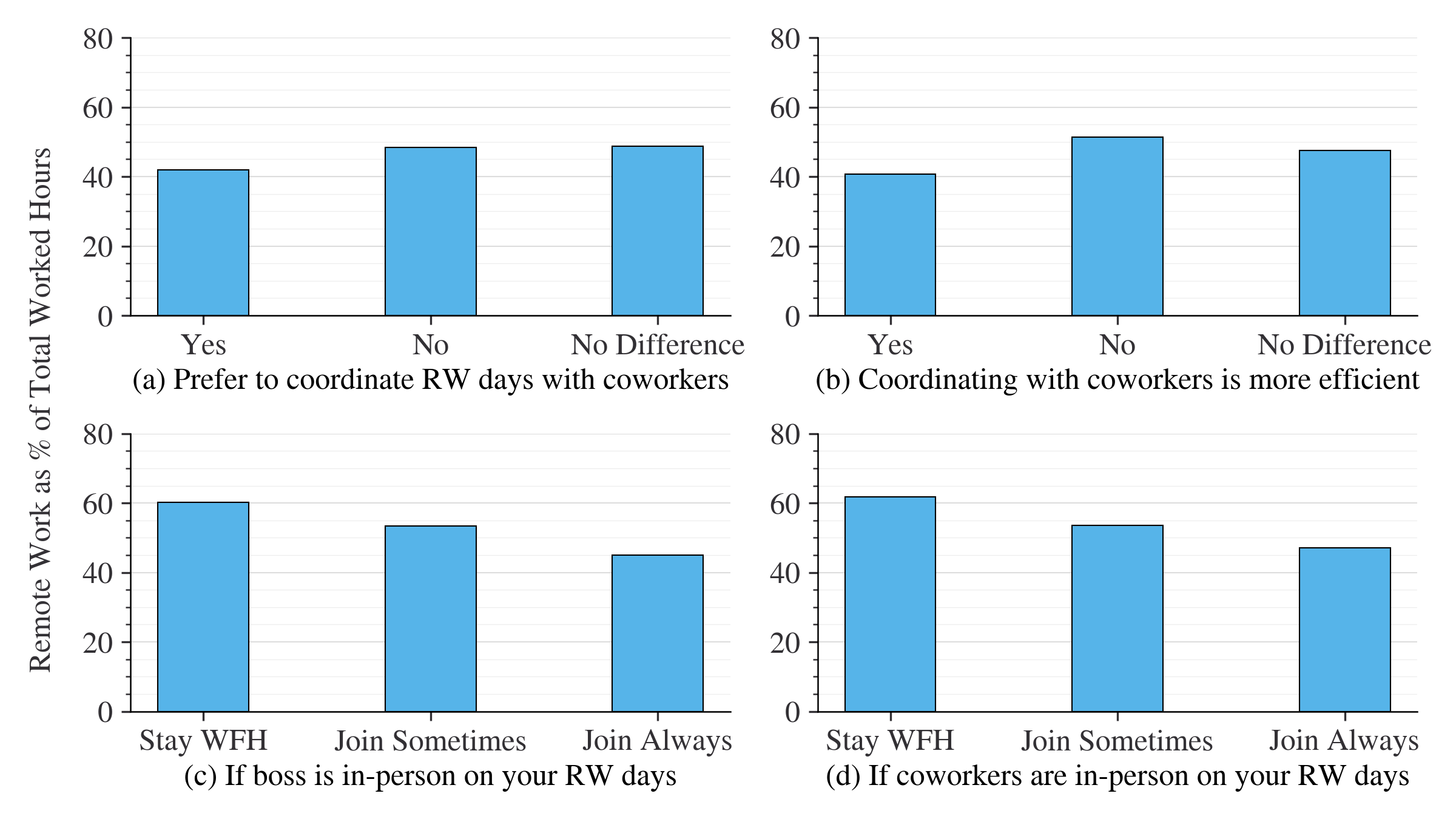}
    \begin{minipage}{0.84\textwidth}
    \footnotesize (a): Feb 2022, N = 3,122; (b): Same as (a); (c): Oct 2021 - Sep 2022, N = 7,796; (d): Oct 2021 - Sep 2022, N = 8,022.
    \end{minipage}
    \caption{Current remote work share by attitudes towards coordinating with colleagues}
    \label{fig:planned_coordination}
\end{figure}

Employer planned remote work shares by attitudes towards remote work coordination with colleagues are shown in Figure~\ref{fig:planned_coordination}. 
The difference in employer plans between those who prefer to coordinate with colleagues are less pronounced than differences in remote work preferences for the same groups. 
Interestingly, people who join their boss or colleagues working in-person expect to receive less remote work in the future, suggesting expectations of a wider return-to-office shift within their organization. 

\begin{figure}[!h]
    \centering
    \includegraphics[width=\textwidth]{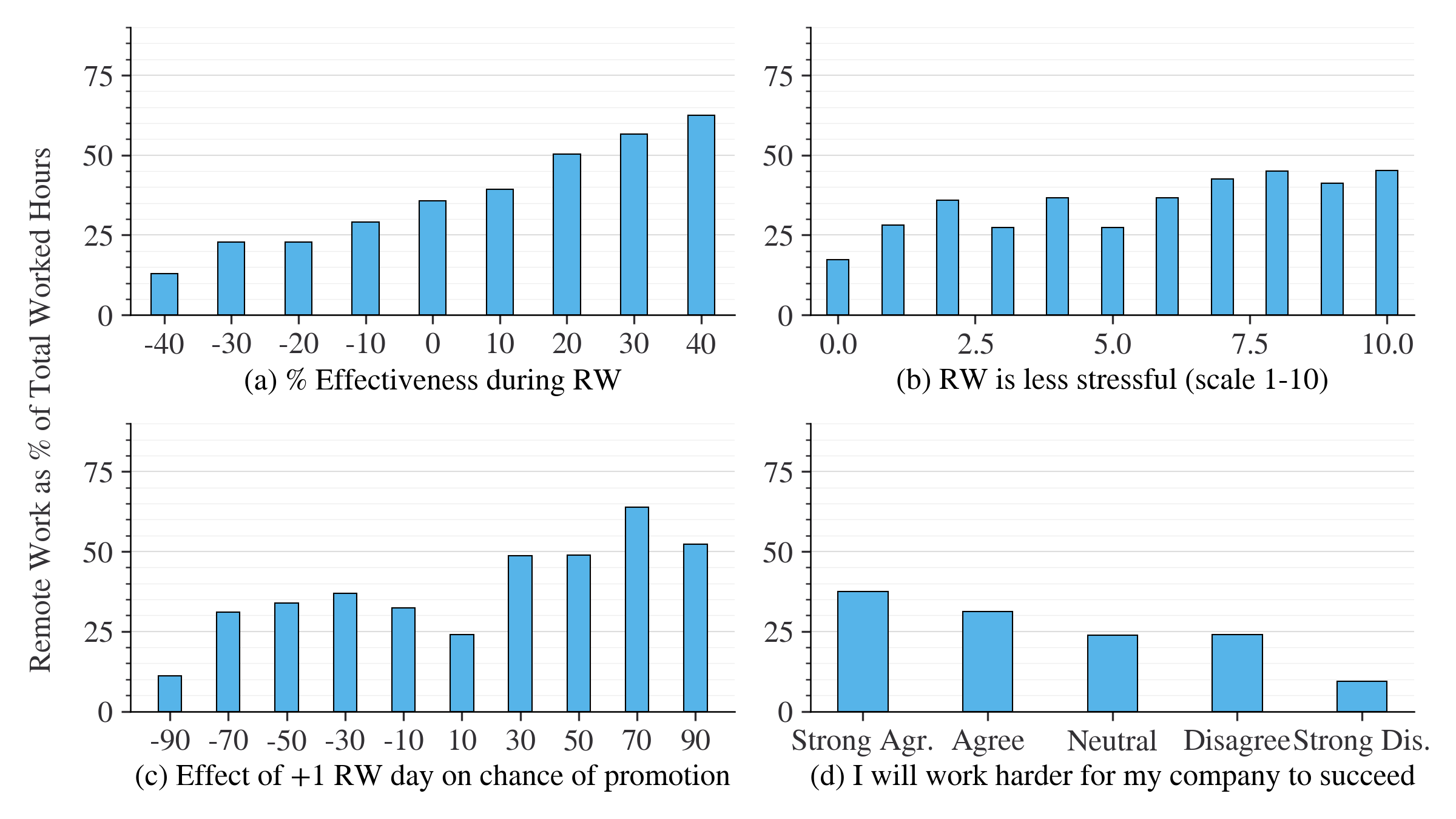}
    \begin{minipage}{0.84\textwidth}
    \footnotesize (a): Jul 2020 - Jan 2023, N = 90,213 (b): Oct 2021, N = 2,780 (c): May 2021 - Aug 2022, N = 5,843; (d): Jul 2022, N = 3,452.
    \end{minipage}
    \caption{Current remote work share by attitudes towards remote work}
    \label{fig:planned_wfhattitudes}
\end{figure}

Employer planned remote work shares by attitudes towards remote work more generally are shown in Figure~\ref{fig:planned_wfhattitudes}. 
Those who find remote work to increase their effectiveness appear to expect their employers to respond to this increased effectiveness by offering greater levels of remote work, although the effect is almost certainly bi-directional.
In contrast, people do not expect their employer to consider their stress levels when setting remote work plans, as there is no obvious trend across groups.
Those who are particularly eager to work harder to support their organization's success do expect more remote work, and those with the opposite attitude expect their employer to allow them to work remotely for less than 10\% of their hours. 

\begin{figure}[!h]
    \centering
    \includegraphics[width=\textwidth]{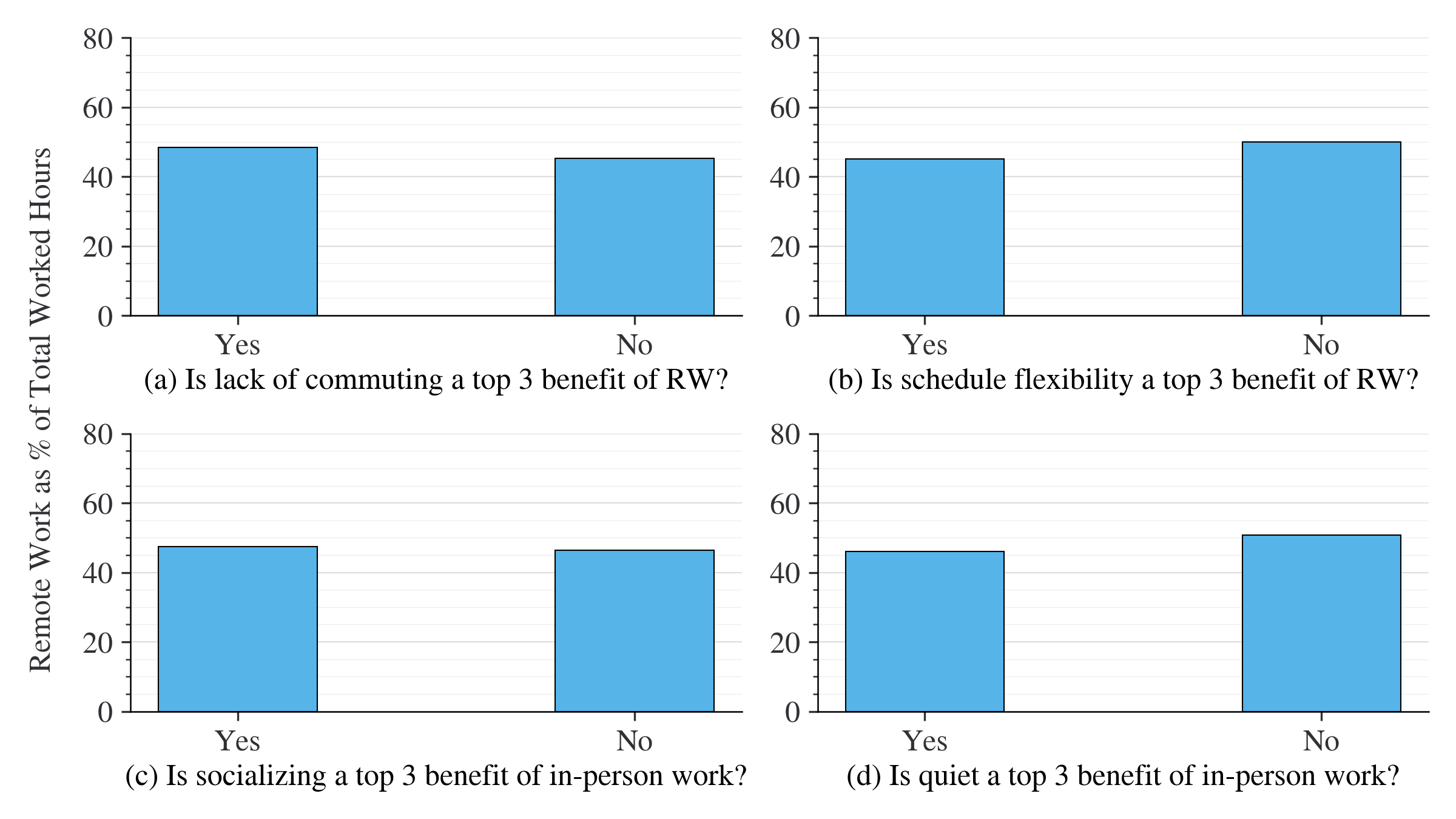}
    \begin{minipage}{0.84\textwidth}
    \footnotesize All: Feb 2022 - Sep 2022, N = 28,836.
    \end{minipage}
    \caption{Current remote work share by perceived benefits of remote work}
    \label{fig:planned_benefits}
\end{figure}

Employer planned remote work shares by the perceived benefits of remote work are shown in Figure~\ref{fig:planned_benefits}. 
There is little variation among the response groups with respect to employer planned shares of remote work.

\begin{figure}[!h]
    \centering
    \includegraphics[width=\textwidth]{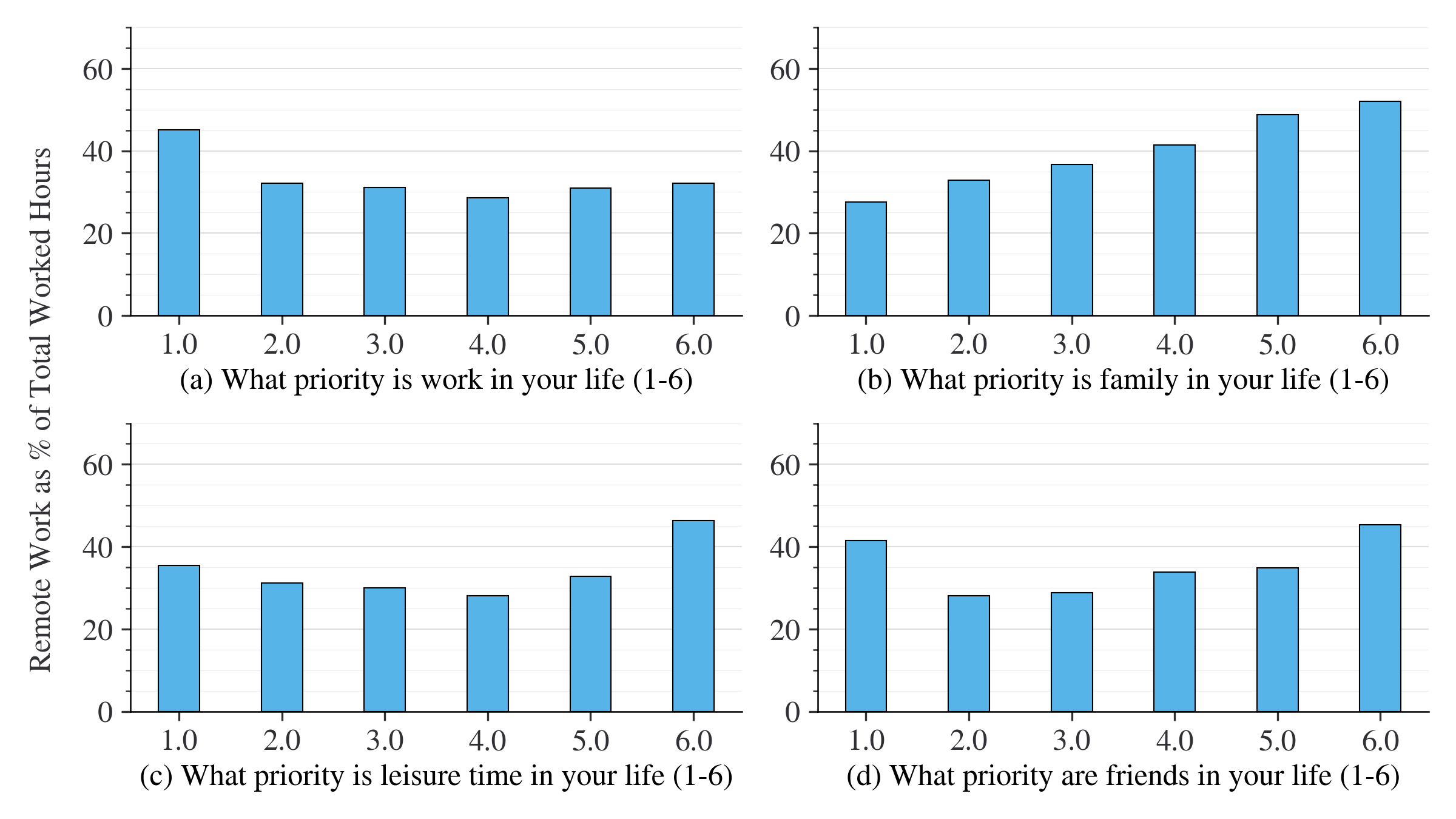}
    \begin{minipage}{0.84\textwidth}
    \footnotesize All: Jul 2022, N = 3,309.
    \end{minipage}
    \caption{Current remote work share by life priorities}
    \label{fig:planned_priorities}
\end{figure}

Finally, planned remote work shares by life priorities are shown in Figure~\ref{fig:planned_priorities}. 
As with the observed shares of remote work in Figure~\ref{fig:current_priorities}, there are some notable differences between response groups. 
Placing a higher priority on work and a lower priority on family is associated with higher levels of remote work. 
Employer plans for high levels of remote work are also associated with the extremes for prioirty of leisure and friends. 

